\documentclass[11pt]{article} 

\usepackage[showcomments]{ajtex}

\title{Effective field theory for non-relativistic hydrodynamics}

\author{Akash Jain}\email{ajain@uvic.ca}

\affiliation{Department of Physics \& Astronomy, University of Victoria, PO
  Box 1700 STN CSC, Victoria, BC, V8W 2Y2, Canada}

\abstract{We write down a Schwinger-Keldysh effective field theory for
  non-relativistic (Galilean) hydrodynamics. We use the null background
  construction to covariantly couple Galilean field theories to a set of
  background sources. In this language, Galilean hydrodynamics gets recast as
  relativistic hydrodynamics formulated on a one dimension higher spacetime
  admitting a null Killing vector. This allows us to import the existing field
  theoretic techniques for relativistic hydrodynamics into the Galilean setting,
  with minor modifications to include the additional background vector field. We
  use this formulation to work out an interacting field theory describing
  stochastic fluctuations of energy, momentum, and density modes around thermal
  equilibrium. We also present a translation of our results to the more
  conventional Newton-Cartan language, and discuss how the same can be derived
  via a non-relativistic limit of the effective field theory for relativistic
  hydrodynamics.}

\usepackage[bbgreekl]{mathbbol}

\def\bbn{\mathbb{n}}
\def\bbh{\mathbb{h}}

\def\bbg{\mathbb{g}}

\def\rel{\text{rel}}
\def\kB{k_{\text B}}

\def\SK{Schwinger-Keldysh\xspace}

\setlist[itemize]{itemsep=0pt,topsep=5pt}
\setlist[enumerate]{itemsep=0pt,topsep=5pt}

\usepackage{tikz}
\usetikzlibrary{decorations.markings,decorations.pathmorphing,arrows.meta}

\tikzset{aux/.style={decorate,decoration={snake,segment length=1.5mm,
      amplitude=0.4mm}}}
\tikzset{left/.style={arrows={Stealth[scale length=0.5, scale width=1.5]-}}}
\tikzset{right/.style={arrows={-[sep=-2pt]Stealth[scale length=0.5, scale
      width=1.5]}}}
\tikzset{leftright/.style={arrows={Stealth[scale length=0.5, scale
      width=1.5]-Stealth[scale length=0.5, scale width=1.5]}}}

\definecolor{lightblue}{RGB}{200,225,255}
\definecolor{lightblueborder}{RGB}{30,100,255}

\begin{document}

\maketitle 

\section{Introduction and overview} 

Hydrodynamics is an effective framework that aims to describe low-energy
dynamics of many body systems near thermal equilibrium. Classically,
hydrodynamics is posed as a set of conservation equations associated with the
global symmetries that the system enjoys. The underlying rationale is that
non-conserved quantities are expected to decay and thermalise at much smaller
distance and time scales compared to conserved quantities, which need to be
transported out to infinity to thermalise. Hence, one expects that if one waits
long enough and is only interested in macroscopic phenomenon, the low-energy
behaviour of a thermal system would be entirely governed by the dynamics of its
conserved operators. However, this expectation is only true up to a leading
approximation. As we start to take into account interactions between various
hydrodynamic modes, the inevitable interaction between hydrodynamic and the
ignored high-energy degrees of freedom also starts to play an important role.  A
concrete realisation of this phenomenon are the so-called long-time
tails~\cite{Martin:1973zz, Pomeau:1974hg, 1974Phy....75....1D}, which signal the
breakdown of the hydrodynamic derivative expansion at late times. To account for
such effects, one needs to pass to the more sophisticated formalism of
stochastic hydrodynamics (see e.g.~\cite{Kovtun:2012rj}). Here, the collective
high-energy excitations are modelled by random small-scale noise sources in the
hydrodynamic equations~\cite{Landau:1980mil, landauHydroFluctuations1957,
  Hohenberg:1977ym, DeDominicis:1977fw, Khalatnikov:1983ak}, followed by
statistical averaging over all possible noise configurations. The noisy
interactions are chosen in a way that they reproduce the classical hydrodynamic
results in the appropriate limit. This formalism is still highly
phenomenological and minimalistic and does not account for the most generic
structure of stochastic interactions nor quantum fluctuations of the
hydrodynamic modes themselves.

Ideally, one would like to not base the hydrodynamic framework on a set of
conservation equations, but rather write down a bona fide effective field theory
(EFT) starting from certain fundamental degrees of freedom and symmetries. This
would systematically account for the high-energy stochastic interactions without
having to introduce them by hand. Thanks to a series of collaborated efforts
over the last decade, there now exists a complete EFT description for stochastic
hydrodynamics based on the Schwinger-Keldysh (SK) formalism of thermal field
theories~\cite{Dubovsky:2011sj, Grozdanov:2013dba, Crossley:2015evo,
  Glorioso:2017fpd, Glorioso:2016gsa, Gao:2017bqf, Glorioso:2017lcn,
  Gao:2018bxz, Jensen:2017kzi, Jensen:2018hse, Haehl:2018lcu};
see~\cite{Glorioso:2018wxw} for a review. One of the major achievements of the
new framework is an understanding of the second law of thermodynamics as a
macroscopic manifestation of the microscopic KMS (Kubo–Martin–Schwinger)
condition. The KMS condition also ensures that the hydrodynamic correlators
predicted by the effective theory satisfy non-linear fluctuation-dissipation
theorems~\cite{Wang:1998wg}. So far, such field theoretic tools have been
developed for a handful of low-energy condensed matter systems including
relativistic fluids and superfluids~\cite{Crossley:2015evo, Glorioso:2017fpd},
and Abelian and non-Abelian diffusive fluctuations~\cite{Chen-Lin:2018kfl,
  Glorioso:2020loc}. See also,~\cite{Landry:2019iel, Glorioso:2018kcp,
  Gralla:2018kif} for similar tools in non-dissipative relativistic solids,
supersolids, liquid crystals, relativistic magnetohydrodynamics, and force-free
electromagnetism. Nonetheless, an effective field theory approach for
non-relativistic (Galilean) hydrodynamics is still largely missing. This is an
obvious limitation because most physical situations around us where such a
theory could be useful are non-relativistic. The goal of this paper is to remedy
this situation and write down a complete \SK effective field theory for
dissipative Galilean hydrodynamics.

A primary ingredient in the EFT framework are a set of background fields coupled
to various conserved operators that make up the spectrum of hydrodynamics. These
act as sources in the associated field theory generating functional and can be
used to obtain correlation functions of the hydrodynamic operators. For example,
in a relativistic setting it is widely known that the energy momentum tensor
$T^{\mu\nu}$ and charge (particle number) current $J^\mu$ can be coupled to a
background spacetime metric $g_{\mu\nu}$ and gauge field $A_\mu$
respectively. The situation is slightly more tricky in the Galilean case. The
hydrodynamic observables are the mass (particle number) density $\rho^t$, mass
(particle number) flux/momentum density $\rho^i$, energy density $\epsilon^t$,
energy flux $\epsilon^i$, and stress tensor $\tau^{ij}$. These need to be
coupled to the so-called Newton-Cartan (NC) structure: mass gauge field
$(A_t,A_i)$, clock form $(n_t,n_i)$, and spatial metric $h_{ij}$ respectively,
that describes a curved non-relativistic spacetime~\cite{Cartan:1924yea,
  Cartan:1923zea, Jensen:2014aia, Jensen:2014ama, Jensen:2014wha,
  Jensen:2014hqa, Festuccia:2016awg, Bergshoeff:2017dqq, Bergshoeff:2014uea,
  Christensen:2013rfa, Christensen:2013lma, Hartong:2016nyx, Hartong:2014pma,
  Geracie:2015xfa, Geracie:2016dpu, Duval:2009vt,
  Banerjee:2017rch,Banerjee:2016laq}; see~\cite{Jensen:2014aia} for a
review.\footnote{Here we have chosen the Newton-Cartan Galilean frame velocity
  $v^i = 0$. A more formal discussion on Newton-Cartan backgrounds is given in
  \cref{sec:null-redn}.}  However, Galilean boost symmetry (also referred to as
Milne boosts in the NC literature) acts on these quantities quite non-trivially
and is quite tedious to implement in an effective theory written in the NC
language. It is this symmetry that is responsible for setting the momentum
density equal to the mass flux.

Fortunately, there is another manifestly covariant formulation for coupling
Galilean field theories to background sources, using the so-called \emph{null
  backgrounds}~\cite{Banerjee:2015hra}. The idea is that by introducing a
fiducial null coordinate $x^-$, Galilean background sources can be arranged into
a one-dimensional higher relativistic metric
$g_{\sM\sN} \df x^\sM \df x^\sN = -2\lb \df x^- - A_t \df t - A_i \df x^i \rb
\lb n_t \df t+ n_i \df x^i \rb + h_{ij} \df x^i \df x^j$. Analogously, the
Galilean hydrodynamic observables also arrange themselves neatly into a
one-dimensional higher relativistic energy-momentum tensor $T^{\sM\sN}$ coupled
to $g_{\sM\sN}$. Galilean symmetries, including boosts, act on these quantities
naturally as one-dimensional higher Poincar\'e transformations that leave the
null vector $V^\sM\dow_\sM = \dow_-$ invariant. Posed in these terms, Galilean
physics gets dressed as one-dimensional higher relativistic physics with a null
Killing vector $V^\sM$. The NC results can always be obtained from here by
reducing the theory over the fiducial null coordinate, formally known as null
reduction~\cite{Maldacena:2008wh, Adams:2008wt, Herzog:2008wg}. Not only does
this formalism makes all the symmetries of the Galilean theory manifest, it also
allows one to directly import a plethora of well-developed relativistic results
directly into the Galilean setting with only minor modifications to accommodate
the null Killing vector; see e.g.~\cite{Duval:1984cj, Duval:1990hj,
  Julia:1994bs, Banerjee:2015hra, Jain:2015jla, Banerjee:2015uta,
  Banerjee:2016qxf, Banerjee:2017ouw, Jain:2018jxj, Hassaine:1999hn,
  Horvathy:2009kz}. Along the same lines, for our current goal of interest, we
can use the null background framework to write down an effective field theory
for Galilean hydrodynamics using the techniques already developed for
relativistic hydrodynamics.

For the benefit of the interested readers who might not want to get their hands
dirty with all the technicalities, we summarise the main results of the paper
here in as non-technical terms as we can manage. The fundamental degrees of
freedom in the EFT for hydrodynamics, relativistic or Galilean, are the comoving
coordinates of the fluid elements $\sigma^i(x)$, comoving time coordinate
$\sigma^\tau(x)$, mass/particle number phase $\varphi_r(x)$. These are
accompanied by associated stochastic noise fields $X^i_a(x)$, $X^t_a(x)$, and
$\varphi_a(x)$ capturing our statistical ignorance for the respective
quantities. Here $(x^\mu) = (t,x^i)$ collectively denotes physical time and
space coordinates. The theory is coupled to two copies of background sources in
spirit with the SK formalism; for the Galilean case these are the ``difference''
sources $A_{a\mu}(x)$, $n_{a\mu}(x)$, and $h_{a\mu\nu}(x)$ (with
$h_{a\mu t}(x) = 0$) that probe the physical mass density-flux $\rho^\mu_r$,
energy density-flux $\epsilon^\mu_r$, and stress tensor $\tau^{\mu\nu}_r$
(normalised as $\tau^{\mu\nu}_r n_{r\nu} = 0$) operators respectively, while the
``average'' sources $A_{r\mu}(x)$, $n_{r\mu}(x)$, and $h_{r\mu\nu}(x)$ (with
$h_{r\mu t}(x) = 0$) probe the respective stochastic noise operators
$\rho^\mu_a$, $\epsilon^\mu_a$, and $\tau^{\mu\nu}_a$ (normalised as
$\tau^{\mu\nu}_a n_{r\nu} = 0$). This ``$r/a$'' usage for background fields and
associated operators is standard in the SK literature; see
e.g.~\cite{Bellac:2011kqa}. The ``classical'' hydrodynamic fields are defined in
terms of these as
\begin{gather}
  \text{Velocity}: \quad
  u^\mu = \frac{\beta^\mu}{n_{r\mu} \beta^\mu}, \qquad
  \text{Temperature}:\quad
  \kB T = \frac{1}{n_{r\mu} \beta^\mu}, \nn\\
  \text{Mass chemical potential}:\quad
  \mu = \frac{\Lambda_\beta + \beta^\nu A_{r\nu}}{n_{r\mu} \beta^\mu}
  + \half \frac{h_{r\nu\rho} \beta^\nu \beta^\rho}{(n_{r\mu}\beta^\mu)^2},
\end{gather}
where $\beta^\mu(x) = \beta_0\dow x^\mu(\sigma(x))/\dow\sigma^\tau(x)$ is called
the thermal vector and
$\Lambda_\beta(x) = \beta_0\mu_0 + \beta^\mu(x)\dow_\mu\varphi_r(x)$ the
chemical shift field, with $x^\mu(\sigma)$ being the inverse of
$\sigma^\alpha(x)$. Furthermore, $\beta_0 = (\kB T_0)^{-1}$ and $\mu_0$ are the
constant inverse temperature and chemical potential of the global thermal state
respectively. Note that the fluid velocity $u^\mu$ is normalised as
$u^\mu n_{r\mu} = 1$ and only has $d$ independent components in $d$ spatial
dimensions. It is also convenient to introduce a Galilean frame velocity
$v^\mu = \delta^\mu_t/n_{rt}$, obeying $v^\mu n_{r\mu} = 1$. We can define the
inverse spatial metric as $h_r^{\mu\nu}$ with components
$h^{tt}_r = h_{r}^{ij}n_{ri}n_{rj}/n_{rt}^2$,
$h^{ti}_r = -h_{r}^{ij}n_{rj}/n_{rt}$, and $h^{ij}_r = (h_r^{-1})^{ij}$,
satisfying $h_r^{\mu\nu} n_{r\mu} = 0$ and
$h_r^{\mu\lambda} h_{\lambda\nu} + v^\mu n_{r\nu} = \delta^\mu_\nu$. Note that
$h_{r}^{\mu\nu}$ is \emph{not} the inverse of $h_{r\mu\nu}$.

At ideal order, a Galilean fluid is characterised by the grand-canonical
equation of state, i.e. its thermodynamic pressure $p(T,\mu)$ expressed as a
function of $T$ and $\mu$. Its derivatives define the thermodynamic densities:
mass density $\rho(T,\mu)$, internal energy density $\varepsilon(T,\mu)$, and
entropy density $s(T,\mu)$ via the Gibbs-Duhem relation
$\df p = s\,\df T+\rho\,\df\mu$ and Euler relation $\varepsilon+p =
Ts+\mu\rho$. Note that these two imply the local first law of thermodynamics
$\df \varepsilon = T\, \df s + \mu\, \df\rho$.  One-derivative dissipative
corrections to ideal hydrodynamics are parametrised by three non-negative
transport coefficients: shear viscosity $\eta(T,\mu)$, bulk viscosity
$\zeta(T,\mu)$, and thermal conductivity $\kappa(T,\mu)$. For a parity-violating
fluid, there are further non-dissipative corrections that can be found later in
the main text. In $d$ spatial dimensions, the fluid is described by the
effective action
\begin{align}
  S
  &= \int \df t\, \df^dx\, n_{rt} \sqrt{\det h_r} \bigg[
    \rho\, u^\mu \lb A_{a\mu} + \dow_{\mu}\varphi_a + \lie_{X_a} A_{r\mu} \rb \nn\\
  &\qquad
    - \Big( \lb \varepsilon + p
    + {\textstyle\half} \rho\, h_{r\rho\sigma}u^\rho u^\sigma \rb u^\mu
    - p\, v^\mu \Big)
    \lb n_{a\mu} + \lie_{X_a}n_{r\mu} \rb
    + \frac12 \lb \rho\, u^\mu u^\nu + p\, h_r^{\mu\nu} \rb
    \lb h_{a\mu\nu} + \lie_{X_a} h_{r\mu\nu} \rb \nn\\
  &\qquad
    + \frac{i\kB T}{4}\lb 2\eta\, h_r^{\mu(\rho} h_r^{\sigma)\nu}
    + \lb \zeta - {\textstyle\frac{2}{d}}\eta \rb h_r^{\mu\nu}h_r^{\rho\sigma} \rb
    \lb h_{a\mu\nu} + \lie_{X_a}h_{r\mu\nu}
    - 2\lb n_{a(\mu} + \lie_{X_a}n_{r(\mu}\rb h_{r\nu)\lambda} u^\lambda \rb \nn\\
  &\qquad\qquad\qquad
    \lb h_{a\rho\sigma} 
    + \lie_{(X_a+i\beta)}h_{r\rho\sigma}
    - 2\lb n_{a(\mu} + \lie_{(X_a+i\beta)}n_{r(\mu}\rb h_{r\nu)\lambda} u^\lambda \rb \nn\\
  &\qquad
    + i\kB T^2\kappa\, h^{\mu\nu}_r \lb n_{a\mu} + \lie_{X_a}n_{r\mu} \rb
    \lb n_{a\nu} + \lie_{(X_a + i\beta)}n_{r\nu} \rb \bigg].
    \label{eq:1der-action-NC-intro}
\end{align}
Here $\lie_{X_a}$ denotes a Lie derivative along $X_a^\mu$, while
$\lie_{(X_a + i\beta)}$ along $X^\mu_a + i u^\mu/T$. Varying with respect to the
difference sources leads to the well-known constitutive relations of
one-derivative Galilean hydrodynamics, modified with stochastic noise
contributions. Upon setting the sources to trivial:
$A_{a\mu}=n_{a\mu}=h_{a\mu\nu}=0$, $A_{r\mu} = 0$, $n_{r\mu} = \delta^t_\mu$,
and $h_{r\mu\nu} = \delta^i_{\mu}\delta^j_{\nu}\delta_{ij}$, these are reduced
to
\begin{gather}
  \rho^t_r = \rho, \qquad
  \rho^i_r = \rho\,u^i, \qquad
  \epsilon^t_r = \varepsilon + \half\rho \vec u^2, \nn\\
  \epsilon^i_r = \lb \varepsilon + p + \half\rho \vec u^2 \rb u^i
  - 2\eta\,u_j \dow^{(i} u^{j)}
  - \lb \zeta - {\textstyle\frac{2}{d}}\eta\rb u^i \dow_k u^k
  - \kappa\, \dow^i T
  + i\epsilon^i_{\text{noise}}, \nn\\
  \tau^{ij}_r = \rho\, u^i u^j + p\,\delta^{ij}
  - 2\eta\,\dow^{(i} u^{j)} - \zeta\,\delta^{ij} \dow_k u^k
  + i\tau_{\text{noise}}^{ij},
  \label{eq:intro-consti}
\end{gather}
where $\epsilon^i_{\text{noise}}$ and $\tau_{\text{noise}}^{ij}$ are the stochastic
noise contributions given by
\begin{align}
  \epsilon^i_{\text{noise}}
  &= 4\kB T\eta\, u_j\lb \dow^{(i} X_a^{j)} - u^{(i} \dow^{j)}X_a^t \rb
    + 2\kB T \lb \zeta - {\textstyle\frac{2}{d}}\eta\rb u^i
    \lb \dow_k X_a^k - u^k \dow_k X^t_a \rb
    + 2\kB T^2\kappa \dow^i X_a^t, \nn\\
  \tau_{\text{noise}}^{ij}
  &= 4\kB T\eta \lb \dow^{(i} X_a^{j)} - u^{(i} \dow^{j)}X_a^t \rb
    + 2\kB T \lb \zeta - {\textstyle\frac{2}{d}}\eta\rb \delta^{ij}
    \lb \dow_k X_a^k - u^k \dow_k X^t_a \rb
    + 2\kB T^2\kappa \dow^i X_a^t.
\end{align}
Varying with respect to the average sources, on the other hand, one can also
obtain the partner noise operators: $\rho^t_a$, $\rho^i_a$, $\epsilon^t_a$,
$\epsilon^i_a$, and $\tau^{ij}_a$. These are fairly complicated, so we do not
present them here. In the absence of sources, the effective action itself turns
into
\begin{align}
  S
  &= \int \df t\, \df^3x \bigg[
    \rho^t \dow_t \varphi_a
    + \rho^i \dow_i \varphi_a
    - \epsilon^t \dow_t  X^t_{a}
    - \epsilon^i \dow_i  X^t_{a}
    + \rho^i \dow_t X_{ai} 
    + \tau^{ij} \dow_i X_{aj} \nn\\
  &\qquad
    + 2i\kB T \eta \Big( \dow_{(i}X_{aj)} - u_{(i} \dow_{j)}X^t_a \Big)
    \Big( \dow^{(i}X_{a}^{j)} - u^{(i} \dow^{j)}X^t_a \Big)
    + i\kB T \lb \zeta - {\textstyle\frac{2}{d}}\eta\rb
    \lb \dow_k X_{a}^k - u^k \dow_k X^t_a \rb^2 \nn\\
  &\qquad
    + i \kB T^2 \kappa\, \dow_i X^t_a \dow^i X^t_a \bigg].
    \label{eq:1der-action-flat-intro}
\end{align}
Here $\rho^t$, $\rho^i$, $\epsilon^t$, $\epsilon^i$, and $\tau^{ij}$ (without
any subscript) are the classical constitutive relations, obtained from
\cref{eq:intro-consti} by turning off the noise contributions. One can verify
that varying with respect to $\varphi_a$, $X^t_a$, and $X_a^i$ leads to the
well-known conservation equations of Galilean hydrodynamics in the absence of
noise: $\dow_t \rho^t + \dow_i \rho^i = 0$,
$\dow_t\epsilon^t + \dow_i\epsilon^i = 0$, and
$\dow_t\rho^i + \dow_j\tau^{ij} = 0$. Note that the effective action has a
non-zero imaginary part coupled to dissipative transport coefficients $\eta$,
$\zeta$, and $\kappa$. This captures the fact that the effective action is
dissipative.

To study the effects of stochastic fluctuations around an equilibrium
configuration, we can expand \cref{eq:1der-action-flat-intro} order-by-order in
fluctuations around the state $\sigma^i(x) = x^i$, $\sigma^\tau(x) = t$,
$\varphi_r(x) = 0$, and $X^i_a(x) = X^t_a(x) = \varphi_a(x) = 0$. Truncating the
expansion at the desired order in interactions, depending on the sensitivity
required, we can set up a diagrammatic analysis for computing the hydrodynamic
correlation functions. Detailed expressions for the effective action truncated
at three-point interactions are given in \cref{sec:1derGal}. A similar analysis
was performed by the authors of~\cite{Chen-Lin:2018kfl}, but restricted to just
energy fluctuations, to compute non-analytic long-time tail behaviour in the
energy-energy retarded two-point function. By contrast, the effective action
\eqref{eq:1der-action-flat-intro} is suitable for describing all energy,
momentum, and density fluctuations.

Given the effective action coupled to sources, we can define the SK generating
functional for the EFT via the path integral
$\exp W[\phi_r,\phi_a] = \int\mathcal{D}\sigma\mathcal{D}
\varphi_r\mathcal{D}X_a\mathcal{D}\varphi_a\, \exp(iS)$, which is a functional
of the background sources
$\phi_{r,a} = (A_{r,at}, A_{r,ai}, -n_{r,at}, -n_{r,ai}, h_{r,aij})$. This can
be used to compute the correlation functions of the hydrodynamic operators
$\mathcal{O}_{r,a} = (\rho^t_{r,a}, \rho^i_{r,a}, \epsilon^t_{r,a},
\epsilon^i_{r,a}, \tau^{ij}_{r,a})$ using the variational formulae
\begin{equation}
  G_{\mathcal{O}_\alpha\ldots}(x,\ldots)
  = i^{n_a} \lb \frac{-i\delta}{\delta \phi_{\bar\alpha}(x)} \ldots \rb
  W[\phi_r,\phi_a],
  \label{eq:intro-correlators}
\end{equation}
where $\alpha=r,a$ and $\bar\alpha = a,r$ denotes its conjugate, while $n_a$ is
the number of $a$ type fields in the correlator on the left. Following the
general SK machinery (see e.g.~\cite{Wang:1998wg}), the correlators of the type
``$ra\ldots a$'' are interpreted as the fully retarded correlation functions,
``$a\ldots ar$'' as the fully advanced, ``$rr\ldots r$'' as the symmetric
correlation functions, and various other time-ordering schemes in between. For
instance, $G_{\mathcal{O}_r\mathcal{O}_a}$, $G_{\mathcal{O}_a\mathcal{O}_r}$,
$G_{\mathcal{O}_r\mathcal{O}_r}$ are the retarded, advanced, and symmetric
propagators of the hydrodynamic observables. The SK generating functional
describing an out-of-equilibrium thermal system is expected to satisfy a number
of consistency conditions. These are reflected as a set of constraints in the
effective action as proposed by~\cite{Crossley:2015evo}. \textbf{(1)} The
effective action should map to minus its complex conjugate when we flip the
signs of all the ``$a$'' type fields. This results in
$W[\phi_r,\phi_a] = W^*[\phi_r,-\phi_a]$ and ensures that all the correlations
functions are real (in position space). \textbf{(2)} The action should vanish
when we switch off all the ``$a$'' type fields, which ensures that
$W[\phi_r,\phi_a=0]=0$ and all the correlation functions of the type
``$aa\ldots a$'' identically vanish.\footnote{Originally, it was believed that
  one needs to introduce additional BRST ghost fields and emergent supersymmetry
  into the hydrodynamic setup to ensure that this condition is satisfied at all
  loop orders~\cite{Crossley:2015evo}. However, it is now understood that one
  can avoid these ghost fields altogether by employing a consistent
  regularisation procedure for frequency loop
  integrals~\cite{Gao:2018bxz}.\label{foot:noghosts}} \textbf{(3)} The imaginary
part of the effective action should be positive semi-definite, leading to
$\mathrm{Re}\,W[\phi_r,\phi_a]\leq 0$, which ensures that variously ordered
correlation functions have the appropriate singularity structure in the momentum
space. \textbf{(4)} Lastly, the effective action should be invariant under the
discrete KMS symmetry given by\footnote{Defining
  $\phi_{1,2} = \phi_r \pm \hbar/2\,\phi_a$, background field transformations in
  \cref{eq:KMS-intro} are merely the small $\hbar$ limit of the KMS
  transformation $\phi_1(x) \to \phi_1(-x)$ and
  $\phi_2(x) \to \phi_2(-t-i\hbar\beta_0,-\vec x)$.}
\begin{gather}
  \phi_r(x) \to \phi_r(-x), \qquad
  \phi_a(x) \to \phi_a(-x) - i\beta_0\dow_t\phi_r(-x), \nn\\
  \beta^\mu(x) \to \beta^\mu(-x), \qquad
  \Lambda_\beta(x) \to \Lambda_\beta(-x), \nn\\
  X^\mu_a(x) \to -X^\mu_a(-x) - i\beta^\mu(-x) + i\beta_0\delta^\mu_t, \qquad
  \varphi_a(x) \to - \varphi_a(-x) - i\Lambda_\beta(-x).
  \label{eq:KMS-intro}
\end{gather}
This ensures that the correlations functions \eqref{eq:intro-correlators}
satisfy (non-linear) fluctuation-dissipation theorems~\cite{Wang:1998wg}. These
conditions are together responsible for an emergent local second law of
thermodynamics and Onsager's reciprocity relations~\cite{Glorioso:2016gsa}. In
addition to the SK constraints, the effective action for hydrodynamics has to be
invariant under \textbf{(5)} reparametrisations of the comoving spacetime
coordinates associated with the fluid elements and, of course, for Galilean
hydrodynamics under \textbf{(6)} local Galilean transformations of the
background sources, including Milne boosts. It can be checked that the effective
action \eqref{eq:1der-action-NC-intro} obeys all these requirements. In the main
text, we look at these symmetries and conditions in more detail and outline the
construction of the most generic effective action for Galilean hydrodynamics
allowed by the EFT framework.

The organisation of the remainder of the paper is as follows. We start
\cref{sec:ClassicalGalilean} with a lightening review of classical Galilean
hydrodynamics and its coupling to null background sources. We introduce the
local second law of thermodynamics and derive the Galilean adiabaticity equation
encapsulating the second law constraints to arbitrary high orders in the
derivative expansion. We also spend some time discussing the subtle issue of
hydrodynamic frame transformations. \Cref{sec:GalSK} makes up the bulk of the
paper. Here we give a detailed construction of the EFT framework for Galilean
hydrodynamics based on symmetry principles and SK formalism of thermal field
theory. Also in this section we construct the most generic effective action
consistent with the framework, in small $\hbar$ limit, and discuss its
connection to classical hydrodynamics. A proof of the local second law of
thermodynamics, within the context of the effective theory, also appears
here. Leaving formalities aside, we discuss the explicit effective action for
one-derivative Galilean hydrodynamics in \cref{sec:1derGal}. We linearise this
action around a fluid configuration at rest on a flat background and setup a
perturbative theory for stochastic fluctuations, keeping terms up to three-point
interactions. In \cref{sec:NC-formulation}, we translate our results to the
Newton-Cartan language; the Milne boost symmetry is no longer manifest but the
results can be presented without introducing an auxiliary spacetime
direction. Finally, we close the paper with some discussion in
\cref{sec:discussion}. The paper has two appendices. In \cref{sec:second}, we
give a brief derivation of the second law constraints in one-derivative Galilean
hydrodynamics. In \cref{sec:nonrelLimit}, we illustrate how the effective action
for Galilean hydrodynamics arises through a large speed of light expansion of
its relativistic cousin.

Factors of $\hbar$, $c$, and $\kB$ have been kept explicit throughout this
paper. Similarly, constant global temperature $T_0$ and chemical potential
$\mu_0$ appear explicitly. While we generically work in $d$ spatial dimensions,
the parity-violating results are presented only for $d=3$.

\section{Review of classical Galilean hydrodynamics}
\label{sec:ClassicalGalilean}

Before we pursue an EFT for Galilean hydrodynamics, let us take a step back and
review the essential features of classical Galilean hydrodynamics. We initiate
our discussion with Galilean symmetries and associated conservation equations
and derive the constitutive relations of a dissipative Galilean fluid in
\cref{sec:flatGalHydro}. We mainly follow the textbook approach
of~\cite{landau1959fluid}, with a few novel features derived
from~\cite{Jensen:2014ama} for parity-violating fluids. We review the null fluid
framework of~\cite{Banerjee:2015hra} to couple Galilean hydrodynamics to
background sources in \cref{sec:source-coupling-null}. The discussion in terms
of the more widely used Newton-Cartan sources has been deferred to
\cref{sec:NC-formulation}. Subtleties regarding frame transformations in
hydrodynamics have been discussed in \cref{sec:frame}. Finally, in
\cref{sec:GalAdiabaticity}, we outline an all-order argument to derive the
Galilean hydrodynamic constitutive relations in accordance with the second law
of thermodynamics using the offshell
formalism~\cite{Jain:2018jxj,Loganayagam:2011mu}.

\subsection{Dissipative hydrodynamics on flat background}
\label{sec:flatGalHydro}

Classical hydrodynamics is a theory of locally conserved quantities. One sets
out by enumerating the complete set of Noether currents associated with any
global symmetries that the system might enjoy, and expresses various ``fluxes''
in terms of the conserved ``densities'', arranged in a perturbative expansion in
derivatives. For a given set of such ``constitutive relations'', the time
evolution of the conserved densities is determined by their respective
conservation equations. In Galilean (non-relativistic) hydrodynamics, the
conserved densities of interest are the energy, momentum, and mass/particle
number density of the fluid. The associated fluxes are the energy flux and
stress tensor; mass flux is the same as momentum density due to Galilean
symmetries. Up to one-derivative order, the constitutive relations are specified
in terms of a thermodynamic equation of state, along with number of
one-derivative transport coefficients such as viscosities and conductivity.

\subsubsection{Conservation equations}
\label{sec:conservation-eqs}

Let us consider a field theory in $d$ spatial dimensions with coordinates $t$,
$x^i$ with $i=1,2,\ldots d$. For the field theory to qualify as Galilean, the
dynamical evolution must be invariant under global Galilean coordinate
transformations: translations ($a^i$), time-translation ($a^t$), $\SO(d)$
rotations $(\Lambda^i{}_j)$, and Galilean boost ($v^i$) defined as
\begin{equation}
  t \to t' = t + a^t, \qquad
  x^i \to x'^i = \Lambda^i{}_j\lb x^j + a^j + v^j t \rb.
  \label{eq:Gal-trans}
\end{equation}
The Noether equations associated with spacetime translations imply the well-known
conservation of energy and momentum
\begin{subequations}
\begin{align}
  \text{Energy conservation:}
  &\qquad
    \dow_t \epsilon^t
    + \dow_i \epsilon^i = 0, \nn\\
  \text{Momentum conservation:}
  &\qquad
    \dow_t \pi^i + \dow_j \tau^{ij} = 0.
\end{align}
Here $\epsilon^t$ is the energy density, $\epsilon^i$ energy flux, $\pi^i$
momentum density, and $\tau^{ij}$ is the stress tensor. Typically, a Galilean
theory also features a conserved mass density $\rho^t$ and the associated flux
$\rho^i$, leading to the continuity equation
\begin{align}
  \text{Continuity equation:}
  &\qquad
    \dow_t \rho^t + \dow_i \rho^i = 0.
\end{align}%
\label{eq:Gal-conservation}%
\end{subequations}
The Noether equations associated with rotations and boost lead to angular
momentum and centre-of-mass conservation respectively, which further enforce
momentum density to equal the mass flux, $\pi^i = \rho^i$, and stress tensor to
be symmetric, $\tau^{ij} = \tau^{ji}$.\footnote{The angular momentum and
  centre-of-mass conservation equations are respectively given as
  \begin{equation}
    \dow_t \lb \pi^{i} x^{j} - \pi^j x^i \rb
    + \dow_k \lb \tau^{ki}x^{j} - \tau^{kj}x^i \rb
    = \tau^{ji} - \tau^{ij} = 0, \qquad
    \dow_t \lb \rho^t x^{i} - \pi^i t \rb
    + \dow_k \lb \rho^k x^{i} - \tau^{ki} t \rb
    = \rho^i - \pi^i = 0.
  \end{equation}
} \Cref{eq:Gal-conservation} are the generic conservation equations in a
Galilean-invariant field theory. They are left invariant under the Galilean
transformations \eqref{eq:Gal-trans} provided that various quantities transform
according to
\begin{gather}
  \rho^t \to \rho^t, \qquad
  \rho^i  \to \Lambda^i{}_j\lb \rho^j + \rho^t v^j \rb, \qquad
  \tau^{ij} \to \Lambda^i{}_k \Lambda^j{}_l \lb \tau^{kl}
  + 2 v^{(k}\rho^{l)} + \rho^t v^k v^l \rb, \nn\\
  \epsilon^t \to \epsilon^t + \half \rho^t \vec v^2 + v_i \rho^i , \qquad
  \epsilon^i \to \Lambda^i{}_j\lb \epsilon^j + \epsilon^t v^j
  + \half \vec v^2 \lb \rho^j + \rho^t v^j \rb
  + \lb \tau^{jk} + v^{j} \rho^k \rb v_k \rb.
  \label{eq:Gal-quantity-trans}
\end{gather}

To recast above equations in a more familiar form, we can define a velocity
associated with the flow of mass according to $u^i = \rho^i/\rho^t$. It
transforms as $u^i \to \Lambda^i{}_j( u^j + v^j)$. We also define thermodynamic
mass density $\rho$, internal/comoving thermodynamic energy density
$\varepsilon$, comoving heat flux $q^i$, and comoving stress tensor $t^{ij}$ in
the local rest frame according to\footnote{The distinction between the
  thermodynamic mass density $\rho$ and physical mass density $\rho^t$ is purely
  notational at this point. However when coupled to background sources, this
  relationship gets modified with a red-shift factor. Furthermore, the
  definition of thermodynamic mass density is also sensitive to hydrodynamic
  frame redefinitions, as we will later diagnose in \cref{sec:frame}. For
  readers familiar with relativistic hydrodynamics, this distinction is similar
  to the distinction between thermodynamic energy density $\epsilon$ and
  physical energy density $T^{tt}$ in the relativistic case.}
\begin{gather}
  \rho^t = \rho, \qquad
  \epsilon^t = \varepsilon + \half \rho\, u^i u_i, \nn\\
  \epsilon^i
  = \lb \varepsilon
  + \half \rho\, u^i u_i \rb u^i
  + t^{ij} u_j + q^i, \qquad
  \tau^{ij} = \rho\, u^i u^j + t^{ij}.
  \label{eq:internal-observables}
\end{gather}
The benefit of working with these quantities is that they are invariant under
boosts. In terms of these definitions, the conservation equations turn into
\begin{align}
  \lb \dow_t + u^i \dow_i\rb \rho
  &= - \rho\, \dow_i u^i,  \nn\\
  \lb \dow_t + u^i \dow_i \rb \varepsilon
  &= - \varepsilon\, \dow_i u^i
    - t^{ij} \dow_i u_j - \dow_i q^i,  \nn\\
  \lb \dow_t + u^j \dow_j \rb u^i
  &= - \frac{1}{\rho} \dow_j t^{ij}.
    \label{eq:Gal-dynamics}
\end{align}

\Cref{eq:Gal-dynamics} are a feature of any Galilean-invariant field theory. The
departure point of hydrodynamics, however, is the assumption that in a
near-equilibrium state at long-enough distance and time scales and low-enough
energy and momentum scales, the dynamics of the system is entirely governed by
the conserved operators: mass density $\rho^t$, energy density $\epsilon^t$, and
momentum density $\pi^i$.\footnote{For this assumption to be applicable, one
  requires a mass-gap in the spectrum of the theory, which gaps out all the
  irrelevant massive degrees of freedom at low-enough energies, leaving us only
  with the conserved operators. In presence of massless degrees of freedom,
  however, one can modify the hydrodynamic equations to consistently account for
  the additional low energy dynamics, e.g. superfluids~\cite{Banerjee:2016qxf}.}
It is convenient to reshuffle these fields and instead work with the
thermodynamic mass density $\rho$, internal energy density $\varepsilon$, and
``fluid'' velocity $u^i$. Hydrodynamics is then characterised by the most
generic expressions for the comoving heat current and stress tensor in terms of
the chosen variables and their spatial derivatives, i.e.
\begin{equation}
  q^i[\rho,\varepsilon,u^i,\dow_i], \qquad
  t^{ij}[\rho,\varepsilon,u^i,\dow_i],
\end{equation}
consistent with Galilean symmetries. These are known as the \emph{hydrodynamic
  constitutive relations}. Note that the temporal derivatives of various
quantities are determined by \cref{eq:Gal-dynamics} and hence are not
independent. Our assumption of \emph{near-equilibrium} allows us to arrange the
constitutive relations in a \emph{derivative expansion}, truncated at a given
order in derivatives according to the phenomenological sensitivity required. At
any given order in the derivative expansion, the constitutive relations contain
all the possible tensor structures made out of derivatives of $\rho$,
$\varepsilon$, and $u^i$, appended with arbitrary \emph{transport coefficients}
as a function of $\rho$ and $\varepsilon$.

\subsubsection{Constitutive relations}

Up until this point, writing down the constitutive relations is a purely
combinatorial exercise. The physical content comes in the form of certain
phenomenological constraints that have to be imposed on these constitutive
relations. Most important of these is the ``local second law of thermodynamics''
that requires that there must exist an entropy density $s^t$ and an associated
flux $s^i$ such that
\begin{equation}
  \dow_t s^t + \dow_i s^i \geq 0.
\end{equation}
At zero derivative order, entropy density is given by an arbitrary function of
$\rho$ and $\varepsilon$, i.e. $s^t = s(\varepsilon,\rho)$. Let us define the
intensive thermodynamic variables: temperature $T(\varepsilon,\rho)$, mass
chemical potential $\mu(\varepsilon,\rho)$,\footnote{Note that $\mu$ here is the
  chemical potential associated with mass density, as used by Landau
  in~\cite{landau1959fluid}). For a single species fluid with mass per unit
  particle $m$, the particle number chemical potential is given by $m\mu$.} and
pressure $p(\varepsilon,\rho)$ via the thermodynamic relations: local first law
of thermodynamics and Euler relation respectively
\begin{equation}
  T\df s = \df \varepsilon - \mu\, \df \rho,  \qquad
  p = Ts+\mu\rho  - \varepsilon.
  \label{thermodynamics}
\end{equation}
It is easy to check that
\begin{equation}
  \partial_t s^t + \dow_i s^i
  = - \frac1T \lb t^{ij} - p\, \delta^{ij} \rb \dow_{i} u_{j}
  - \frac{1}{T^2} q^i \dow_i T,
  \label{eq:2ndLaw-non-cov}
\end{equation}
where we have identified the entropy flux as $s^i = s\, u^i + q^i/T$.
Therefore, at leading order in derivatives, the second law constraint simply
requires that
\begin{equation}
  t^{ij} = p\,\delta^{ij} + \mathcal{O}(\dow), \qquad
  q^i = \mathcal{O}(\dow),
\end{equation}
We note that the comoving heat flux is zero at this order, while the comoving
stress tensor is just given by the thermodynamic pressure of the fluid.  Note
also that entropy is conserved at this order, $\dow_t s^t + \dow_i s^i = 0$,
hence these fluids are known as ``ideal/perfect fluids''. Inserting these into
\cref{eq:Gal-dynamics}, one recovers the well-known Euler equations of Galilean
ideal hydrodynamics.
% \begin{align}
%   \lb \dow_t + u^i \dow_i\rb \rho
%   &= - \rho\, \dow_i u^i,  \nn\\
%   \lb \dow_t + u^i \dow_i \rb \varepsilon
%   &= - \lb\varepsilon + p\rb\, \dow_i u^i + \mathcal{O}(\dow^2),  \nn\\
%   \lb \dow_t + u^j \dow_j \rb u^i
%   &= - \frac{1}{\rho} \dow^ip  + \mathcal{O}(\dow^2).
%     \label{eq:ideal-hydro-cons}
% \end{align}
% The last one of these is the well-known Navier-Stokes equation.

The relation $s = s(\varepsilon,\rho)$, or equivalently
$\varepsilon = \varepsilon(s,\rho)$, is known as the micro-canonical equation of
state of the fluid and completely characterises its constitutive relations at
ideal order through the thermodynamic relations
\eqref{thermodynamics}.\footnote{As a well-known example, the ideal gas
  ``equation of state'' is typically expressed as
  $p(\rho,T) = \rho/m\, k_{\text B}T$, where $m$ is the mass per particle. This,
  however, does not uniquely specify the thermodynamic state of the fluid; one
  also needs to provide $\varepsilon(\rho,T) = c_v \rho/m\, k_{\text B}T$, where
  $c_v$ is the dimensionless specific heat capacity. Both of these can,
  nonetheless, be combined into a micro-canonical equation of state
  $\varepsilon(s,\rho) = \Lambda_0 c_v (\rho/m)^{(c_v+1)/c_v}
  \exp\lb\frac{ms}{c_v\rho k_{\text B}} - \frac{c_v+1}{c_v} \rb$ or a
  grand-canonical equation of state
  $p(T,\mu) = \Lambda_0^{-c_v}(k_{\text B}T)^{c_v+1}\exp\bfrac{m\mu}{kT}$ for an
  arbitrary energy scale $\Lambda_0$.} We note, however, that hydrodynamics as a
physical system is better defined in the grand canonical ensemble, because a
fluid element is allowed to freely exchange mass and energy with its
surroundings. Keeping this in mind, we can take the fundamental dynamical fields
to be $T$ and $\mu$ instead of $\varepsilon$ and $\rho$. In this case, the
equation of state is given in terms of $p(T,\mu)$ instead of
$\varepsilon(s,\rho)$ with the thermodynamic relations: Gibbs-Duhem relation and
Euler relation respectively
\begin{equation}
  \df p = s\, \df T + \rho\, \df\mu, \qquad
  \varepsilon = Ts + \mu\rho - p.
  \label{eq:GCthermo}
\end{equation}
These define $\varepsilon$, $\rho$, and $s$ in terms of $T$ and $\mu$. 

The second law can also be used to constrain the admissible derivative
corrections to the hydrodynamic equations. One finds that the entropy density
remains unperturbed by one-derivative parity-preserving corrections, but is
sensitive to parity-violating corrections. Focusing on $d=3$ spatial dimensions,
one finds (see~\cite{Jensen:2014ama} or \cref{sec:second} for more
details)\footnote{In fairness, implementation of the second law of
  thermodynamics on flat space allows for the $K_2T$ term
  \cref{eq:Gal-EC-non-cov} to be replaced with an arbitrary function of $T$. The
  present form is enforced by implementing the second law in the presence of
  background sources, or requiring the fluid to admit an equilibrium partition
  function~\cite{Banerjee:2015hra}.}
\begin{align}
  s^t &= s + K_0\epsilon^{ijk} u_i \dow_j u_k, \nn\\
  s^i &= s\, u^i + \frac1T q^i
        + K_0 \epsilon^{ijk} u_j \lb \dow_t u_k + \half \dow_k \vec u^2 \rb
        + \lb 2 K_0 \mu + \half K_0 \vec u^2 + K_2T \rb
        \epsilon^{ijk}\dow_j u_k,
        \label{eq:Gal-EC-non-cov}
\end{align}
where $K_0$, $K_2$ are constants. Requiring the second law to hold in generality
leads to\footnote{Note the identity
  $\epsilon^{ljk} u^k\dow_k u_l \dow_j u_k = \half \epsilon^{ijk} \dow_i \vec
  u^2 \dow_j u_k$.}
\begin{align}
  \partial_t s^t + \dow_i s^i 
  = - \frac1T \lb t^{ij} - p\, \delta^{ij} \rb \dow_{(i} u_{j)}
    - \frac{1}{T^2} \lb q^i - \xi_\Omega \epsilon^{ijk}\dow_j u_k \rb
  \partial_i T
  \geq 0,
  \label{eq:2ndLaw-ideal-odd}
\end{align}
where $\xi_\Omega = T^2K_2 + 2\mu T K_0 - ((\varepsilon+p)/\rho) 2TK_0$ is a
parity-odd transport coefficient. In obtaining this equation, we have used the
ideal order Navier-Stokes equation from \cref{eq:Gal-dynamics}. This results in
the constitutive relations
\begin{equation}
  t^{ij} = p\, \delta^{ij}
  - \eta\, \sigma^{ij}
  - \zeta\, \delta^{ij}  \dow_k u^k + \cO(\dow^2), \qquad
  q^i = - \kappa\, \dow^i T + \xi_\Omega \epsilon^{ijk}\dow_j u_k + \cO(\dow^2).
  \label{eq:Gal-consti-noncov}
\end{equation}
where
$\sigma^{ij} = \dow^{i} u^{j} + \dow^j u^i - \frac{2}{d} \delta^{ij} \dow_k u^k$
is the fluid shear tensor. The ``dissipative'' transport coefficients
$\eta(T,\mu)$ shear viscosity, $\zeta(T,\mu)$ bulk viscosity, and
$\kappa(T,\mu)$ thermal conductivity contribute to entropy production and are
required to be positive semi-definite on the account of the second law. On the
other hand, $\xi_\Omega(T,\mu)$ is an ``adiabatic'' transport coefficient
that does not lead to entropy production. Note that the parity-preserving part
in \cref{eq:Gal-consti-noncov} is applicable for any $d$, and is standard in
Galilean hydrodynamics; see e.g.~\cite{landau1959fluid}.

% The dissipative hydrodynamic equations follow from \cref{eq:Gal-dynamics} as
% \begin{align}
%   \lb \dow_t + u^i \dow_i\rb \rho
%   &= - \rho\, \dow_i u^i,  \nn\\
%   \lb \dow_t + u^i \dow_i \rb \varepsilon
%   &= - \lb \varepsilon + p\rb \dow_i u^i
%     + \half \eta\, \sigma^{ij} \sigma_{ij}
%     + \zeta (\dow_i u^i)^2
%     + \dow_i \lb \kappa\, \dow^i T \rb
%     - \epsilon^{ijk} \dow_i \lambda_\Omega \dow_j u_k
%     + \mathcal{O}(\dow^3),  \nn\\
%   \lb \dow_t + u^j \dow_j \rb u^i
%   &= - \frac{1}{\rho} \dow^i p
%     - \frac{1}{\rho} \dow_j \lb
%     \eta\, \sigma^{ij}
%     + \zeta\, \delta^{ij}  \dow_k u^k \rb
%     + \mathcal{O}(\dow^3),
%     \label{eq:Gal-dynamics-1der}
% \end{align}
% where the parity-odd effects are specific to $d=3$ spatial dimensions, while the
% parity-even terms are correct for any $d$.

We can similarly work out the constraints following from the second law of
thermodynamics at arbitrarily high orders in the derivative expansion;
see~\cite{Jain:2018jxj}. We can also impose additional phenomenological
constraints such as the microscopic time-reversal invariance (Onsager's
relations), PT invariance, or the existence of an equilibrium partition
function~\cite{Banerjee:2015uta}.

\subsection{Coupling to sources}
\label{sec:source-coupling-null}

We would like to introduce a set of background sources into the Galilean
conservation equations \eqref{eq:Gal-dynamics} covariantly coupled to the
conserved operators. To this end, we have a choice between two frameworks:
Newton-Cartan (NC) backgrounds~\cite{Jensen:2014aia} or null
backgrounds~\cite{Banerjee:2015hra}. While the former framework is much widely
used to couple to Galilean field theories~\cite{Cartan:1924yea, Cartan:1923zea,
  Jensen:2014aia, Jensen:2014ama, Jensen:2014wha, Jensen:2014hqa,
  Festuccia:2016awg, Bergshoeff:2017dqq, Bergshoeff:2014uea,
  Christensen:2013rfa, Christensen:2013lma, Hartong:2016nyx, Hartong:2014pma,
  Geracie:2015xfa, Geracie:2016dpu}, the latter has an advantage that it recasts
($d+1$)-dimensional Galilean-invariant physics in terms of ($d+2$)-dimensional
relativistic physics that we understand better. Unlike the NC framework, the
null background framework also explicitly manifests the Galilean (Milne) boost
symmetry. In this subsection, we follow~\cite{Banerjee:2015hra} to provide a
brief overview of the null background framework. The relation of these results
to the NC sources is reviewed later in \cref{sec:NC-formulation}.

\subsubsection{Null backgrounds}
\label{sec:null-backgrounds}

It is a well-understood fact that ($d+1$)-dimensional Galilean (Bargmann)
algebra can be embedded into a ($d+2$)-dimensional Poincar\'e algebra, seen as
the set of all generators that commute with a null momenta. To be more precise,
let us introduce an auxiliary direction $x^-$ in the Galilean theory and denote
the enlarged set of coordinates as $(x^\sM) = (x^-,t,x^i)$. In this language,
the Galilean transformations \eqref{eq:Gal-trans} can be understood as
$x^-$-independent Poincare transformations written in null coordinates
\begin{equation}
  x^\sM \to \Lambda^\sM{}_\sN\, x^\sN + a^\sM, \qquad
  \Lambda^\sM{}_\sN =
  \begin{pmatrix}
    1 & \half \vec v^2 & v_j \\
    0 & 1 & 0 \\
    0 & \Lambda^i{}_k v^k & \Lambda^i{}_j
  \end{pmatrix},
  \label{eq:null-trans}
\end{equation}
with $\Lambda^\sM{}_\sR \Lambda^\sN{}_\sS \eta_{\sM\sN} = \eta_{\sR\sS}$, where
$\eta_{\sM\sN}\df x^\sM \df x^\sN = -2\df x^-\df t + \delta_{ij}\df x^i \df
x^j$. The translations $a^-$ along the $x^-$-direction act trivially in the
theory and can be understood as global U(1) ``mass'' transformations associated
with the continuity equation. We can arrange various Galilean densities and
fluxes into a ($d+2$)-dimensional energy-momentum tensor
\begin{equation}
  T^{\sM\sN} =
  \begin{pmatrix}
    \times & \epsilon^t & \epsilon^j \\
    \epsilon^t &  \rho^t & \rho^j \\
    \epsilon^i & \rho^i & \tau^{ij}
  \end{pmatrix}.
  % T^{\sM}_{\,\,\,\,\sN} =
  % \begin{pmatrix}
  %   -\epsilon & \times & \epsilon^j \\
  %   -\rho &  -\epsilon & \rho^j \\
  %   -\rho^i & -\epsilon^i & t^{ij}
  % \end{pmatrix}.
  \label{eq:null-EMTensor}
\end{equation}
In terms of this, the conservation equations \eqref{eq:Gal-conservation} are
merely represented as the ($d+2$)-dimensional energy-momentum conservation
\begin{equation}\label{eq:null.conservation}
  \dow_\sM T^{\sM\sN} = 0.
\end{equation}
Note that the $\dow_-$ derivatives in this equation are trivially zero because
$T^{\sM\sN}$ does not have any $x^-$ dependence. Consequently,
``$T^{--} = \times$'' component in \cref{eq:null-EMTensor} represents an
irrelevant auxiliary entry that drops out of the dynamical equations. The
representation \eqref{eq:null-EMTensor} is particularly useful because the
transformation properties \eqref{eq:Gal-quantity-trans} drastically compactify
into the usual Poincar\'e transformation of the higher-dimensional
energy-momentum tensor
\begin{equation}
  T^{\sM\sN} \to \Lambda^\sM{}_\sR \Lambda^\sN{}_\sS\, T^{\sR\sS}.
\end{equation}

Recast in this language, we know precisely how to couple this theory to a curved
background. We promote $\eta_{\sM\sN}$ to an arbitrary background metric
$g_{\sM\sN}$. We also lift the auxiliary coordinate direction $\dow_-$ into a
vector field $V^\sM \dow_\sM$. As we don't want to probe the ``unphysical''
component $T^{--}$ of the energy-momentum tensor, we require
$V^\sM V^\sN g_{\sM\sN} = 0$. We further require the metric to be independent of
the auxiliary coordinate, i.e.  $\lie_V g_{\sM\sN} = 0$. The $x^-$-independent
Poincar\'e transformations get promoted to general $(d+2)$-dimensional
diffeomorphisms
\begin{equation}
  \label{eq:null.diffeo}
  x^\sM \to x'^\sM(x),
\end{equation}
which act on the background fields as usual
\begin{align}
  g_{\sM\sN}(x)
  &\to
    g'_{\sM\sN}(x')
    = \frac{\dow x^\sR}{\dow x'^\sM} \frac{\dow x^\sS}{\dow x'^\sN}
    g_{\sR\sS}(x), \nn\\
  V^\sM(x)
  &\to V'^\sM(x') = \frac{\dow x'^\sM}{\dow x^\sN} V^\sN(x_s).
    \label{eq:null.diffeo-background}
\end{align}
We can always partially fix this symmetry to set $V^\sM = \delta^\sM_-$
exactly. In this coordinate system, the residual symmetries $t\to t'(t,\vec x)$,
$x^i\to x'^i(t,\vec x)$ correspond to spacetime diffeomorphisms, while
$x^-\to x^- - \Lambda(x)$ corresponds to local U(1) mass gauge
transformations. The partial derivatives $\dow_\sM$ get promoted to the
covariant derivative $\nabla_\sM$ associated with $g_{\sM\sN}$ and the
Levi-Civita symbol gets promoted to the Levi-Civita tensor
$\epsilon_{-t123} = \sqrt{-g}$. Finally, in the presence of background sources,
the conservation equations \eqref{eq:null.conservation} modify to
\begin{equation}
  \nabla_\sM T^{\sM\sN} =  0.
  \label{eq:null-cons}
\end{equation}
These serve as the generalisation of \cref{eq:Gal-conservation} to include
background sources.

% We should also revisit the decomposition \eqref{eq:null-EMTensor} on curved
% space to
% \begin{equation}
%   T^{\sM\sN} =
%   \begin{pmatrix}
%     \times & \epsilon^\mu + T^{\mu\lambda} b_\lambda  \\
%     \epsilon &  T^{\mu\nu}
%   \end{pmatrix}, \qquad
%   % T^{\sM}_{\,\,\,\,\sN} =
%   % \begin{pmatrix}
%   %   -\epsilon & \times & \epsilon^j \\
%   %   -\rho &  -\epsilon & \rho^j \\
%   %   -\rho^i & -\epsilon^i & t^{ij}
%   % \end{pmatrix}.
%   \label{eq:null-EMTensor-curved}
% \end{equation}
% These definitions are chosen to ensure that they are invariant under mass gauge
% transformations.

Given a Galilean field theory described by some action $S$, the coupling structure
of $T^{\sM\sN}$ to $g_{\sM\sN}$ is as usual
\begin{equation}
  \delta S = \int \df^{d+2}x\, \sqrt{-g}\, \half T^{\sM\sN} \delta g_{\sM\sN}.
  \label{eq:null-action}
\end{equation}
\Cref{eq:null-cons} follows from here provided that $S$ respects
$(d+2)$-dimensional diffeomorphism invariance (that is same as
$(d+1)$-dimensional Galilean invariance). Note that the integrand above is
entirely independent of $x^-$, so the associated integral yields a trivial
volume factor and the actual integration is only performed over the physical
coordinate directions. With this in place, the retarded correlation functions of
conserved currents can be computed by taking variations with respect to the
sources~\cite{Kovtun:2012rj}. For instance, the retarded two-point functions are
given as
\begin{equation}
  G^{\text R}_{T^{\sM\sN}T^{\sR\sS}}(x,x')
  = 2\frac{\delta (\sqrt{-g}\, T^{\sM\sN}(x))}{\delta g_{\sR\sS}(x')}
  \bigg|_{\text{flat}}.
  \label{eq:null-retG}
\end{equation}
These formulas can be decomposed into components to arrive at the retarded
two-point functions of all the Galilean observables.

\subsubsection{Hydrodynamics on curved spacetime}
\label{sec:hydro-sources}

We would like to couple the hydrodynamic constitutive relations to background
sources. Following \cref{sec:null-backgrounds}, we first need to arrange the
Galilean fluid constitutive relations into a higher dimensional ``null fluid''
according to \cref{eq:null-EMTensor}. Using \cref{eq:internal-observables} and
\cref{eq:Gal-EC-non-cov}, it can be checked that the observables of Galilean
hydrodynamics and entropy current arrange themselves neatly
into~\cite{Banerjee:2015hra}
\begin{align}
  T^{\sM\sN}
  &= \rho\, u^\sM u^\sN + 2\varepsilon\, u^{(\sM}V^{\sN)}
    + 2 q^{(\sM} V^{\sN)} + t^{\sM\sN}, \nn\\
  S^\sM
  &= s\, u^\sM + \frac{1}{T} q^\sM + \Upsilon^\sM,
    \label{eq:null-fluid-EM}
\end{align}
with the comoving stress tensor $t^{\sM\sN}$, comoving heat flux $q^\sM$, and
the ``non-canonical'' part of the entropy current $\Upsilon^\sM$ are given by
\begin{align}
  t^{\sM\sN}
  &= p\, \Delta^{\sM\sN}
    - \eta\, \sigma^{\sM\sN}
    - \zeta\, \Delta^{\sM\sN}  \dow_\sR u^\sR, \nn\\
  q^\sM
  &= - \kappa\, \Delta^{\sM\sN} \dow_\sN T
    + \xi_\Omega \epsilon^{\sM\sN\sR\sS\sT} V_\sN u_\sR \dow_\sS u_\sT, \nn\\
  \Upsilon^\sM
  &= \lb 2K_0\mu + K_2T \rb \epsilon^{\sM\sN\sR\sS\sT} V_\sN u_\sR \dow_\sS u_\sT
  - K_0 \epsilon^{-\sM\sN\sR\sS} u_\sN \dow_\sR u_\sS.
\end{align}
Here $V^\sM \dow_\sM = \dow_-$ is the null Killing vector and
$u^\sM \dow_\sM = \half \vec u^2 \dow_- + \dow_t + u^i \dow_i$ is the null fluid
velocity, satisfying the normalisation conditions
\begin{equation}
  V^\sM V_\sM = u^\sM u_\sM = 0, \qquad V^\sM u_\sM = -1.
  \label{eq:null-norm}
\end{equation}
Also, $\Delta^{\sM\sN} = \eta^{\sM\sN} + 2 u^{(\sM}V^{\sN)}$ is a projector
transverse to $u^\sM$ and $V^\sM$. We have defined the covariantised fluid shear
tensor
$\sigma^{\sM\sN} = 2\Delta^{\sM\sR}\Delta^{\sN\sS} (\dow_{(\sR} u_{\sS)} - 1/d\,
P_{\sR\sS} \dow_\sT u^\sT)$. Note that $t^{\sM\sN}$ and $q^\sM$ satisfy
$t^{\sM\sN}V_{\sM} = t^{\sM\sN} u_\sN = q^\sM V_\sM = q^\sM u_\sM = 0$ and
$t^{\sM\sN} = t^{\sN\sM}$. The statement of the second law of thermodynamics
becomes $\dow_\sM S^\sM = \dow_t s^t + \dow_i s^i \geq 0$. We should note that
the second term in $\Upsilon^\sM$ is not boost-invariant. Nonetheless, its
divergence that contributes to entropy production is perfectly boost-invariant.

Coupling to curved spacetime is now straight-forward. Firstly, the normalisation
conditions on the null fluid velocity \eqref{eq:null-norm} lead us to promote
$u^\sM$ to
\begin{equation}
  u^\sM \dow_\sM
  = \half \lb \lb \frac{1 + u^i g_{i-}}{g_{t-}} \rb^2  g_{tt}
  - 2\frac{1 + u^i g_{i-}}{g_{t-}} u^i g_{it} + u^i u^j g_{ij} \rb \dow_-
  - \frac{1 + u^i g_{i-}}{g_{t-}} \dow_t
  + u^i \dow_i,
\end{equation}
while $\Delta^{\sM\sN} = g^{\sM\sN} + 2 u^{(\sM}V^{\sN)}$. As for the
hydrodynamic constitutive relations, we need to promote $t^{\sM\sN}$, $q^{\sM}$,
and $\Upsilon^\sM$ into covariant expressions accordingly and obtain
\begin{align}
  t^{\sM\sN}
  &= p\, \Delta^{\sM\sN}
  - \eta\, \sigma^{\sM\sN}
  - \zeta\, \Delta^{\sM\sN}\nabla_\sR u^\sR, \nn\\
  q^{\sM}
  &= - \kappa\, \Delta^{\sM\sN} \lb\nabla_\sN T - 2T \nabla_{[\sN} V_{\sR]} u^\sR \rb
    + \xi_\Omega \epsilon^{\sM\sN\sR\sS\sT} V_\sN u_\sR \nabla_\sS u_\sT
    + \xi_H \epsilon^{\sM\sN\sR\sS\sT} V_\sN u_\sR \nabla_\sS V_\sT, \nn\\
  \Upsilon^\sM
  &= \frac{a_2}{T} \epsilon^{\sM\sN\sR\sS\sT} V_\sN u_\sR \nabla_\sS u_\sT
    + \frac{a_1}{4T} \epsilon^{\sM\sN\sR\sS\sT} V_\sN u_\sR \nabla_\sS V_\sT
    - \frac{a_0}{2T} \epsilon^{-\sM\sN\sR\sS} u_\sN \nabla_\sR u_\sS,
    \label{eq:null-consti-cov}
\end{align}
where the shear tensor is now
$\sigma^{\sM\sN} = 2\Delta^{\sM\sR}\Delta^{\sN\sS} (\nabla_{(\sR} u_{\sS)} -
1/d\, P_{\sR\sS} \nabla_\sT u^\sT)$ and
\begin{gather}
  a_0 = 2K_0T, \qquad
  a_1 = 2K_1 T^3 + 2K_2 T^2\mu + 2K_0T\mu^2, \qquad
  a_2 = K_2 T^2 + 2K_0T\mu, \nn\\
  \xi_\Omega = a_2 - \frac{\varepsilon+p}{\rho} a_0, \qquad
  \xi_H = a_1 - \frac{\epsilon+p}{\rho} a_2,
  \label{eq:transconstraints}
\end{gather}
with arbitrary constants $K_0$, $K_1$, $K_2$. In writing these, we have
introduced a few new terms coupled to
$2\dow_{[\sM} V_{\sN]} = 2\partial_{[\sM} g_{\sN]-}$ that drop out on a flat
background. Their coupling is fixed by the second law up to the constant $K_1$;
see \cref{sec:second}. The respective dynamical equations are, of course, given
by substituting \cref{eq:null-fluid-EM} into \cref{eq:null-cons} with the
constitutive relations \eqref{eq:null-consti-cov}.

\subsection{Frame transformations and thermodynamic frame}
\label{sec:frame}

The issue of frame transformations in hydrodynamics is a subtle one. The key
idea is that while the definitions of fluid velocity, temperature, and chemical
potential are well posed in equilibrium on flat space, these become hazy
out-of-equilibrium or in the presence of background sources. For instance, back
in \cref{sec:conservation-eqs}, we defined fluid velocity $u^i$ as the flow of
mass $\rho^i/\rho^t$. However, one might alternatively associate $u^i$ with the
flow of entropy/heat defined as $s^i/s^t$ instead. While in equilibrium both of
these definitions agree, they are generically different as is clearly seen in
\cref{eq:Gal-EC-non-cov}. Similarly, we used the thermodynamic relations
\eqref{thermodynamics} to define temperature $T$ and chemical potential $\mu$ so
that the thermodynamic energy-density $\varepsilon$ and mass-density $\rho$ are
identified with the physical comoving energy-density
$\epsilon^t - \half \rho \vec u^2$ and mass-density $\rho^t$
respectively. However, in the process, entropy-density $s^t$ in
\cref{eq:Gal-EC-non-cov} is no longer the same as the thermodynamic entropy
density $s$. If one were to insist on identifying $s^t$ with $s$ by redefining
$T$ and $\mu$, one of the previous identifications will need to be spoiled. Of
course, any such choice is purely a matter of taste and convenience, and does
not affect the physical results in any way. Nonetheless, as it turns out,
choosing certain frames over others might have consequences for the well
posedness of the hydrodynamic differential equations, at least in the context of
relativistic hydrodynamics~\cite{Kovtun:2019hdm}.

Let us consider a general frame transformation of the hydrodynamic variables
compared to the definitions in the previous subsections 
\begin{equation}
  T \to T + \delta T, \qquad
  \mu \to \mu + \delta\mu, \qquad
  u^i \to u^i + \delta u^i,
  \label{eq:frame-trans-fields}
\end{equation}
where the shifts contain at least one derivative. The last of these induces a
transformation of the null fluid velocity $u^\sM \to u^\sM + \delta u^\sM$,
normalised as $V_\sM \delta u^\sM = u_\sM \delta u^\sM = 0$ up to leading order
in derivatives. The null fluid energy-momentum tensor and entropy current are
expressed in terms of the new variables, up to leading order in derivatives, as
\begin{align}
  T^{\sM\sN}
  &= \lb \rho + \delta\rho\rb u^\sM u^\sN
    + 2 j^{(\sM} u^{\sN)}
    + 2\lb \varepsilon + \delta\varepsilon\rb u^{(\sM}V^{\sN)}
    + 2 q^{(\sM} V^{\sN)}
    + t^{\sM\sN}, \nn\\
  S^\sM
  &= \lb s + \delta s\rb u^\sM - \frac{\mu}{T} j^\sM
    + \frac{1}{T} q^\sM + \Upsilon^\sM,
    \label{eq:null-consti-gen-frame}
\end{align}
where
\begin{gather}
  \delta\rho
  = \frac{\dow\rho}{\dow T}\delta T
  + \frac{\dow\rho}{\dow \mu}\delta \mu, \qquad
  \delta\varepsilon
  = \frac{\dow\varepsilon}{\dow T}\delta T
  + \frac{\dow\varepsilon}{\dow \mu}\delta \mu, \qquad
  \delta s
  = \frac{\dow s}{\dow T}\delta T
  + \frac{\dow s}{\dow \mu}\delta \mu, \nn\\
  j^\sM = \rho\, \delta u^\sM, \qquad
  q^\sM = q^\sM_{\text{mass}} + (\varepsilon + p) \delta u^\sM, \qquad
  t^{\sM\sN}
  = t^{\sM\sN}_{\text{mass}}
  + \lb s\, \delta T + \rho\, \delta \mu \rb \Delta^{\sM\sN}.
\end{gather}
Here $j^\sM$, $q^\sM$, and $t^{\sM\sN}$ should be understood as the ``comoving''
mass flux, energy flux, and stress tensor with respect to the redefined fluid
velocity. On the other hand, $q^\sM_{\text{mass}}$ and
$t^{\sM\sN}_{\text{mass}}$ are the comoving energy flux and stress tensor
defined with respect to the flow of mass from the previous subsections. Imposing
the frame-fixing condition
$T^{\sM\sN} V_\sN = - \rho\, u^\sM - \varepsilon\, V^\sM$, we recover our
original results, formally referred to as the ``mass frame''.

Another convenient choice of hydrodynamic frame from the perspective of the EFT
formalism are the so-called ``thermodynamic frames''. These are a class of
hydrodynamic frames defined abstractly as the following: when coupled to
time-independent background sources, i.e.  $\dow_t g_{\sM\sN} = 0$, the trivial
configuration given by
$\beta^\sM \equiv (u^\sM - \mu\, V^\sM)/(\kB T) = (\delta^\sM_t -
\mu_0V^\sM)/(\kB T_0)$ for constant $\mu_0$ and $T_0$ is a solution of the
equations of motion. Functionally, the constitutive relations in a thermodynamic
frame are those that satisfy the ``off-shell local second law of
thermodynamics'':
\begin{equation}
  \nabla_\sM S^\sM + \kB\beta_\sN \nabla_\sM T^{\sM\sN} \geq 0,
\end{equation}
without using the equations of motion. The effective action prescription
naturally generates conserved currents in a thermodynamic frame. For the present
case of interest, one such thermodynamic frame is given by a choice
$\delta T = \delta \mu = 0$ and
\begin{equation}
  \delta u^\sM
  = \frac{a_0}{\rho} \epsilon^{\sM\sN\sR\sS\sT} V_\sN u_\sR \nabla_\sS u_\sT
  + \frac{a_2}{2\rho} \epsilon^{\sM\sN\sR\sS\sT} V_\sN u_\sR H_{\sS\sT}.
\end{equation}
This leads to $\delta\rho = \delta\varepsilon = \delta s = 0$, along with
\begin{align}
  t^{\sM\sN}
  &= p\, \Delta^{\sM\sN}
  - \eta\, \sigma^{\sM\sN}
    - \zeta\, \Delta^{\sM\sN}\nabla_\sR u^\sR, \nn\\
  j^\sM
  &= a_0\, \epsilon^{\sM\sN\sR\sS\sT} V_\sN u_\sR \nabla_\sS u_\sT
  + a_2\, \epsilon^{\sM\sN\sR\sS\sT} V_\sN u_\sR \nabla_\sS V_\sT, \nn\\
  q^{\sM}
  &= - \kappa \Delta^{\sM\sN} \lb \nabla_\sN T
    + 2T u^\sR \nabla_{[\sM} V_{\sR]} \rb
    + a_2\, \epsilon^{\sM\sN\sR\sS\sT} V_\sN u_\sR \nabla_\sS u_\sT
    + a_1\, \epsilon^{\sM\sN\sR\sS\sT} V_\sN u_\sR \nabla_\sS V_\sT, \nn\\
  \Upsilon^\sM
  &= \frac{a_2}{T} \epsilon^{\sM\sN\sR\sS\sT} V_\sN u_\sR \nabla_\sS u_\sT
    + \frac{a_1}{4T} \epsilon^{\sM\sN\sR\sS\sT} V_\sN u_\sR H_{\sS\sT}
    - \frac{a_0}{2T} \epsilon^{-\sM\sN\sR\sS} u_\sN \nabla_\sR u_\sS.
    \label{eq:null-consti-cov-thermo}
\end{align}
We refer to this frame as the ``thermodynamic mass frame'', since the mass
density/flux does not receive any dissipative (entropy producing)
corrections. While this representation looks unarguably more involved than the
mass frame expressions, the simplicity lies in the fact that all the parity-odd
transport coefficients $a_0$, $a_1$, $a_2$ appearing here are simple polynomial
functions of $T$ and $\mu$ given in \cref{eq:transconstraints}, determined up to
three constants $K_0$, $K_1$, $K_2$. For reference, we note that in the absence
of sources, the associated constitutive relations are given as 
\begin{gather}
  \rho^t = \rho, \qquad
  \epsilon^t = \varepsilon + \half\rho\vec u^2
  + a_0\, \epsilon^{ijk}u_i \dow_j u_k, \qquad
  \pi^i = \rho^i = \rho\, u^i + a_0\, \epsilon^{ijk} \dow_j u_k, \nn\\
  \epsilon^i = \lb \varepsilon + p + \half\rho\vec u^2
  + a_0\, \epsilon^{ljk}u_l \dow_j u_k
  - \zeta\, \dow_k u^k
  \rb u^i
  - \eta\, \sigma^{ij} u_j 
  - \kappa\, \dow^i T
  + \lb a_2 + \half\vec u^2 a_0\rb \epsilon^{ijk} \dow_j u_k, \nn\\
  \tau^{ij} = \rho\, u^i u^j + p\,\delta^{ij}
  - \eta\, \sigma^{ij} -\zeta\,\delta^{ij}\dow_k u^k
  + 2a_0 u^{(i}\epsilon^{j)kl} \dow_k u_l.
  \label{eq:thermo-mass-frame-flat}
\end{gather}

\subsection{Generalities and adiabaticity equation}
\label{sec:GalAdiabaticity}

In the preceding subsections we have reviewed a generic framework of classical
Galilean hydrodynamics and discussed how to couple the hydrodynamic equations to
Galilean-covariant background sources. While we have focused our discussion to
one-derivative hydrodynamics, the machinery follows to higher-derivative orders
more or less in the same manner. Nonetheless, for our construction of effective
field theories, it will be beneficial draw out some essential all-order
features.

As we have seen, equations of $(d+1)$-dimensional classical Galilean
hydrodynamics can be represented in terms of $(d+2)$-dimensional conservation
equations \eqref{eq:null.conservation}. These are $d+2$ independent equations
that can be solved for $d+2$ independent hydrodynamic variables (in the grand
canonical ensemble): temperature $T$, chemical potential $\mu$, and null fluid
velocity $u^\sM$ normalised as \cref{eq:null-norm}. It is convenient to package
these into a single un-normalised vector field
\begin{equation}
  \beta^\sM = \frac{1}{\kB T} \lb u^\sM - \mu V^\sM \rb.
\end{equation}
The hydrodynamic constitutive relations are given as generic expressions for
$T^{\sM\sN}$ allowed by symmetries in terms of $\beta^\sM$, the background
metric $g_{\sM\sN}$, and the Killing vector $V^\sM$, arranged order-by-order in
a derivative expansion.  As such, $\beta^\sM$ are some arbitrary fields chosen
to describe the system and, like in any field theory, can admit arbitrary field
redefinitions known as the hydrodynamic frame redefinitions; see
\cref{sec:frame}. A commonplace hydrodynamic frame is the so-called
``mass-frame'' obtained by choosing $\beta^\sM$ such that
$T^{\sM\sN}V_\sN = -\rho\, u^\sM - \varepsilon\, V^\sM$, where $\rho$ and
$\varepsilon$ are thermodynamic mass and internal energy densities respectively
defined through the thermodynamic relations \cref{eq:GCthermo}. The
one-derivative constitutive relations in this frame are given by
\cref{eq:null-fluid-EM} along with \cref{eq:null-consti-cov}. However,
generically, it is convenient to not restrict to this frame and leave the
redefinition freedom unfixed for now.

The hydrodynamic constitutive relations are required to satisfy the second law
of thermodynamics, i.e. there must exist an entropy current $S^\sM$ and a
quadratic form $\Delta$ satisfying
\begin{equation}
  \nabla_\sM S^\sM = \kB\Delta \geq 0.
\end{equation}
For one-derivative hydrodynamics expressed in mass frame, this is given as
$S^\sM = s\,u^\sM + q^\sM/T + \Upsilon^\sM$ with the ``non-canonical'' entropy
current $\Upsilon^\sM$ provided in \cref{eq:null-consti-cov}. This requirement
is only imposed onshell, i.e. when the conservation equations are
satisfied. Nonetheless, it can be made into an offshell statement by adding
combinations of equations of motion
\begin{equation}
  \nabla_\sM S^\sM + \kB\beta_\sN \nabla_\sM T^{\sM\sN} = \kB\Delta \geq 0.
  \label{eq:offshell-second-law}
\end{equation}
In writing this equation, we have partially fixed the redefinition freedom in
$\beta^\sM$ by choosing it to be the multiplier for equations of motion required
to make the second law an offshell statement. Such frame choices are known as
the ``thermodynamic frames''. There is still some freedom left associated with
modifying the constitutive relations with combinations of equations of
motion. We know that conservation equations determine time-derivatives of
various hydrodynamic fields, so we can consistently fix the residual
redefinition freedom by eliminating
$u^\sM\lie_\beta g_{\sM\sN} = u^\sM \nabla_\sM \beta_\sN + u^\sM\nabla_\sN
\beta_\sM$ from the constitutive relations, where
$\lie_\beta g_{\sM\sN} = 2\nabla_{(\sM}\beta_{\sN)}$ is the Lie derivative of
$g_{\sM\sN}$ along $\beta^\sM$. This amounts to choosing the ``thermodynamic
mass frame''; the associated constitutive relations truncated to one-derivative
order are given in \cref{eq:null-consti-gen-frame} along with
\cref{eq:null-consti-cov-thermo}. The offshell second law
\eqref{eq:offshell-second-law} can be recast into a more useful form
\begin{equation}
  \nabla_\sM N^\sM = \half T^{\sM\sN} \lie_\beta g_{\sM\sN} + \Delta, \qquad
  \Delta \geq 0,
  \label{eq:adiabaticity}
\end{equation}
known as the \emph{adiabaticity equation}, where
$N^\sM = S^\sM/\kB + T^{\sM\sN}\beta_\sN$ is the free-energy
current,\footnote{Technically, the free energy current is $-\kB T N^\sM$, but we
  will use this terminology for simplicity.} up to terms proportional to $V^\sM$
that do not enter the equation. The constitutive relations allowed by the second
law of thermodynamics are the most generic solutions of \cref{eq:adiabaticity}
for some $N^\sM$ and $\Delta$. For instance, $N^\sM$ for one-derivative Galilean
fluid in the thermodynamic mass frame is given as
$N^\sM = p\, \beta^\sM + \Upsilon^\sM/\kB$. A general classification scheme for
Galilean constitutive relations based on the adiabaticity equation can be found
in~\cite{Jain:2018jxj}.

\begin{table}[t]
  \centering
  \begin{tabular}{cc|cc|c}
    \toprule
    && P & T & PT \\
    \midrule
    \multicolumn{2}{c|}{$X^0$, $t$, $\tau$} & $+$ & $-$ & $-$ \\
    \multicolumn{2}{c|}{$X^i$, $x^i$, $\sigma^i$} & $-$ & $+$ & $-$ \\
    $X^-$, $x^-$, $\sigma^-$ & $\varphi$ & $+$ & $-$ & $-$ \\
    \midrule

    \multicolumn{2}{c|}{$u^i$, $\beta^i$, $\bbbeta^i$} & $-$ & $-$ & $+$ \\
    \multicolumn{2}{c|}{$T$, $\beta^t$, $\bbbeta^\tau$} & $+$ & $+$ & $+$ \\
    $\mu$, $\beta^-$, $\bbbeta^-$ & $\mu$, $\Lambda_\beta$, $\Lambda_\bbbeta$
    & $+$ & $+$ & $+$ \\
    
    \midrule
    $T^{t-}$, $g_{t-}$ & $\epsilon^t$, $n_t$ & $+$ & $+$ & $+$ \\
    $T^{i-}$, $g_{i-}$ & $\epsilon^i$, $n_i$ & $-$ & $-$ & $+$ \\
    $T^{ij}$, $g_{ij}$ & $\tau^{ij}$, $h_{ij}$ & $+$ & $+$ & $+$ \\
    $T^{tt}$, $g_{tt}$ & $\rho^t$, $b_t$ & $+$ & $+$ & $+$ \\
    $T^{ti}$, $g_{ti}$ & $\pi^i = \rho^i$, $b_i$ & $-$ & $-$ & $+$ \\
    \bottomrule
  \end{tabular}
  \caption{Action of parity (P) and time-reversal (T) on various quantities in
    classical Galilean hydrodynamics and effective field theory. \SK double
    copies of various quantities in the effective theory, ``$1/2$'' or
    ``$r/a$'', behave the same as their unlabelled versions.\label{tab:CPT}}
\end{table}

Based on the physical system in mind, we can also impose other phenomenological
constraints on the hydrodynamic equations in addition to the second law of
thermodynamics and Galilean symmetries. For instance, it is usually the case
that the microscopic theory underlying the hydrodynamic system of interest has
some sort of discrete time reversal invariance, denoted by $\Theta$, like T or
more generally PT.\footnote{Since we have only included a mass current and no
  charge current in our discussion, we cannot talk about CPT.} The action of
$\Theta$ on various quantities is given in \cref{tab:CPT}. Full dissipative
hydrodynamics, of course, is not time reversal invariant because of entropy
production, but this leads to various non-trivial constraints on the
constitutive relations. For instance, constitutive relations in thermal
equilibrium must be $\Theta$-invariant; this can be imposed by requiring the
equilibrium partition function associated with hydrodynamics to be
$\Theta$-invariant~\cite{Banerjee:2015uta,Banerjee:2012iz,Jensen:2012jh}. Choosing
$\Theta=\text{T}$ does not impose any constraints on the constitutive relations,
while choosing $\Theta=\text{T}$ sets all the three constants $K_{0,1,2} = 0$,
getting rid of the parity-violating sector altogether. Another consequence of
the discrete symmetry are the Onsager's reciprocity relations, which require the
retarded two-point functions in \cref{eq:null-retG} to be symmetric
(see~\cite{Kovtun:2012rj}). This does not lead to any new constrains on
one-derivative Galilean hydrodynamics.

\vspace{1em}

This finishes our short course in Galilean hydrodynamics. We studied
one-derivative hydrodynamics in detail and derived the associated constitutive
relations using the second law of thermodynamics. We also discussed how to
couple the hydrodynamic equations to background sources, allowing us to compute
retarded response functions predicted by hydrodynamics, while more generally
serving as field theory sources coupled to conserved currents in the forthcoming
discussion of EFT for Galilean hydrodynamics. Finally, we discussed the offshell
second law of thermodynamics and the adiabaticity equation, which allows us to
organise the hydrodynamic constitutive relations allowed by the second law to
arbitrarily high orders in the derivative expansion. We are now ready to write
down an EFT that describes stochastic fluctuations on top of classical Galilean
hydrodynamics.

\section{Effective field theory for Galilean hydrodynamics}
\label{sec:GalSK}

In the previous section we reviewed a generic framework for classical Galilean
hydrodynamics, expressed in terms of a one-dimensional higher relativistic
``null fluid''~\cite{Banerjee:2015hra}. This framework can be utilised to write
down a Schwinger-Keldysh effective field theory for Galilean hydrodynamics,
following its relativistic cousin developed recently~\cite{Glorioso:2017lcn,
  Glorioso:2018wxw, Gao:2018bxz, Gao:2017bqf, Crossley:2015evo,
  Glorioso:2017fpd} (see~\cite{Glorioso:2018wxw} for a review). There are two
equivalent languages one can employ to describe the effective theory. One can
either look at the space of individual fluid elements, each equipped with a
comoving clock, and track how their spacetime position evolves as a function of
the comoving time. Alternatively, one can formulate the theory on the physical
spacetime and track the fluid elements occupying each spacetime point. We start
with the former fluid worldvolume formulation in \cref{sec:GalSK-FS}, where the
setup is fundamentally more natural, and subsequently move to the latter
physical spacetime formulation in \cref{sec:GalSK-PS}, which is computationally
more useful as a field theory. Irrespective of the language employed, the most
interesting part of the framework are a set of SK constraints in
\cref{sec:SKaction-null}, representing that the theory describes
out-of-equilibrium fluctuations of a quantum system in a state of thermal
equilibrium. Notably, the theory must satisfy a discrete KMS symmetry, which
imposes the (non-linear) fluctuation-dissipation theorem on the hydrodynamic
correlation functions and is responsible for the emergent second law of
thermodynamics in the classical limit. In \cref{sec:classicalLimit-null}, we
construct the most generic effective action describing stochastic fluctuations
in Galilean hydrodynamics, in a limit where all the quantum fluctuations have
been suppressed, and discuss the emergence of the classical constitutive
relations and second law. Later in \cref{sec:1derGal}, we shall apply these
ideas explicitly to one-derivative Galilean hydrodynamics.

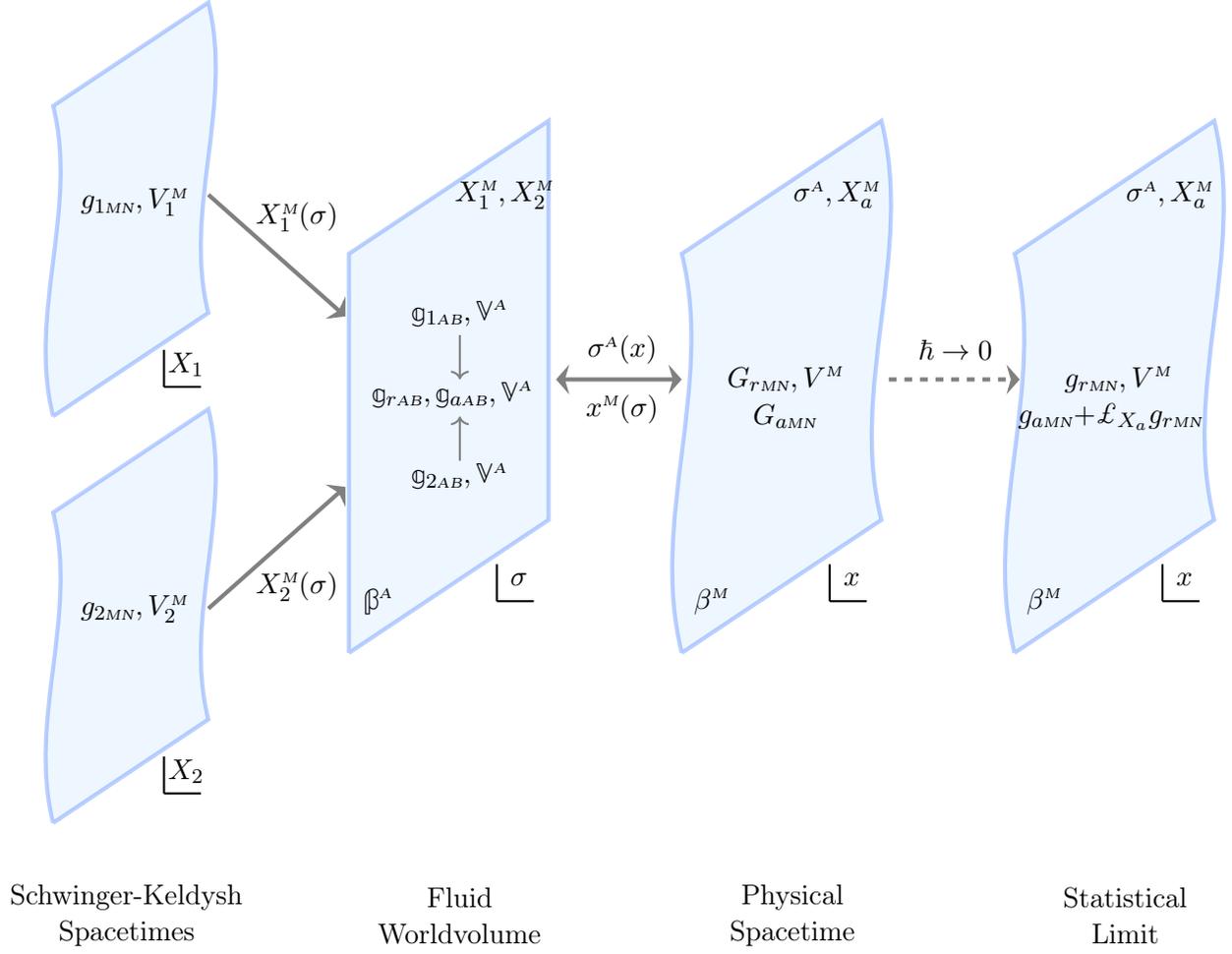
\begin{figure}[t]
  \centering
  \hspace*{-1em}
  \begin{tikzpicture}
    \begin{scope}[shift={(0.5,0)}]
      \draw[ultra thick,lightblueborder,fill=lightblue,opacity=0.3,scale=0.7]
      (0,0) -- (3,2) .. controls (2.5,4) and (3.5,6) .. (3,8) -- (0,6)
      .. controls (0.5,4) and (-0.5,2) .. (0,0);
    
      \begin{scope}[shift={(1.5,0.4)}]
        \draw[thick] (0,0) -- (0,0.5); \draw[thick] (0,0) -- (0.5,0); \node at
        (0.3,0.3) {$X_1$};
      \end{scope}
      
      \begin{scope}[shift={(1.1,2.4)}]
        \node at (0,0.5) {$g_{1\sM\sN},V_1^\sM$};
      \end{scope}
    \end{scope}

  \begin{scope}[shift={(0.5,-5.5)}]
      \draw[ultra thick,lightblueborder,fill=lightblue,opacity=0.3,scale=0.7]
      (0,0) -- (3,2) .. controls (2.5,4) and (3.5,6) .. (3,8) -- (0,6)
      .. controls (0.5,4) and (-0.5,2) .. (0,0);
    
    \begin{scope}[shift={(1.5,0.4)}]
      \draw[thick] (0,0) -- (0,0.5); \draw[thick] (0,0) -- (0.5,0); \node at
      (0.3,0.3) {$X_2$};
    \end{scope}

    \begin{scope}[shift={(1.1,2.4)}]
      \node at (0,0.5) {$g_{2\sM\sN},V_2^\sM$};
    \end{scope}
  \end{scope}

  \node at (1.5,-6.5) {\SK};
  \node at (1.5,-7) {Spacetimes}; 

  \draw[ultra thick,gray,right] (2.6,3) -- (4.4,1.4);
  \node at (3.8,2.7) {$X_1^\sM(\sigma)$};

  \draw[ultra thick,gray,right] (2.6,-2.6) -- (4.4,-1);
  \node at (3.8,-2.3) {$X_2^\sM(\sigma)$};

  \begin{scope}[shift={(4.5,-3.2)}]
    \draw[ultra thick,lightblueborder,fill=lightblue,opacity=0.3,scale=0.9]
    (0,0) -- (3,2) -- (3,8) -- (0,6) -- (0,0);

    \node at (2.1,6.2) {$X_1^\sM, X_2^\sM$}; 
    
    \begin{scope}[shift={(2,0.7)}]
      \draw[thick] (0,0) -- (0,0.5); \draw[thick] (0,0) -- (0.5,0); \node at
      (0.3,0.3) {$\sigma$};
    \end{scope}

    \begin{scope}[shift={(1.5,4.1)}]
      \node at (0,0.5) {$\bbg_{1\sA\sB},\bbV^\sA$};
    \end{scope}

    \begin{scope}[shift={(1.5,1.9)}]
      \node at (0,0.5) {$\bbg_{2\sA\sB},\bbV^\sA$};
    \end{scope}

    \draw[thick,gray,->] (1.5,4.3) -- (1.5,3.7);
    \draw[thick,gray,->] (1.5,2.6) -- (1.5,3.2);

    \begin{scope}[shift={(1.4,2.9)}]
      \node at (0,0.6) {$\bbg_{r\sA\sB},\bbg_{a\sA\sB},\bbV^\sA$};
    \end{scope}

    \node at (0.4,0.7) {$\bbbeta^\sA$};
  \end{scope}

  \node at (6,-6.5) {Fluid};
  \node at (6,-7) {Worldvolume}; 
  
  \draw[ultra thick,gray,leftright] (7.3,0.5) -- (9,0.5);
  \node at (8.2,0.9) {$\sigma^\sA(x)$};
  \node at (8.2,0.1) {$x^\sM(\sigma)$};

  \begin{scope}[shift={(9,-3.2)}]
    \draw[ultra thick,lightblueborder,fill=lightblue,opacity=0.3,scale=0.9]
    (0,0) -- (3,2) .. controls (2.5,4) and (3.5,6) .. (3,8) -- (0,6)
    .. controls (0.5,4) and (-0.5,2) .. (0,0);

    \node at (2.1,6.2) {$\sigma^\sA, X_a^\sM$}; 
    
    \begin{scope}[shift={(2,0.7)}]
      \draw[thick] (0,0) -- (0,0.5); \draw[thick] (0,0) -- (0.5,0); \node at
      (0.3,0.3) {$x$};
    \end{scope}

    \begin{scope}[shift={(1.4,3.2)}]
      \node at (0,0.5) {$G_{r\sM\sN},V^\sM$};
      \node at (0,0) {$G_{a\sM\sN}$};
    \end{scope}

    \node at (0.4,0.7) {$\beta^\sM$};
  \end{scope}

  \node at (10.5,-6.5) {Physical};
  \node at (10.5,-7) {Spacetime}; 
  
  \draw[ultra thick,gray,dashed,right] (11.8,0.5) -- (13.5,0.5);
  \node at (12.7,0.9) {$\hbar\to0$};

  \begin{scope}[shift={(13.5,-3.2)}]
    \draw[ultra thick,lightblueborder,fill=lightblue,opacity=0.3,scale=0.9]
    (0,0) -- (3,2) .. controls (2.5,4) and (3.5,6) .. (3,8) -- (0,6)
    .. controls (0.5,4) and (-0.5,2) .. (0,0);

    \node at (2.1,6.2) {$\sigma^\sA, X_a^\sM$}; 
    
    \begin{scope}[shift={(2,0.7)}]
      \draw[thick] (0,0) -- (0,0.5);
      \draw[thick] (0,0) -- (0.5,0);
      \node at (0.3,0.3) {$x$};
    \end{scope}

    \begin{scope}[shift={(1.4,3.2)}]
      \node at (0,0.5) {$g_{r\sM\sN},V^\sM$};
      \node at (-.1,0) {$g_{a\sM\sN} {+} \lie_{X_a}g_{r\sM\sN}$};
    \end{scope}

    \node at (0.4,0.7) {$\beta^\sM$};
  \end{scope}

  \node at (15,-6.5) {Statistical};
  \node at (15,-7) {Limit};  
\end{tikzpicture}
\caption{Schematic representation of the effective field theory framework for
  Galilean hydrodynamics in null background formulation.\label{fig:null-schematic}}
\end{figure}

\subsection{Fluid worldvolume formulation} 
\label{sec:GalSK-FS}

The Schwinger-Keldysh effective field theory for hydrodynamics is setup as a
sigma model. We start with a ($d+2$)-dimensional ``fluid worldvolume'' with
coordinates $\sigma^\sA$. Building upon our discussion in
\cref{sec:null-backgrounds}, this spacetime is required to carry a vector field
$\bbV^\sA(\sigma)$ characterising the auxiliary ``null-direction'' needed for
describing a $(d+1)$-dimensional Galilean fluid. The fluid worldvolume also
carries a preferred ``thermal time vector'' $\bbbeta^\sA(\sigma)$ that defines
the reference frame associated with the global thermal state. We will require
$\bbbeta^\sA$ to not depend on the auxiliary null direction, i.e.
$\lie_\bbV \bbbeta^\sA = \bbV^\sB\dow_\sB\bbbeta^\sA - \bbbeta^\sB\dow_\sB
\bbV^\sA = 0$. We can always choose a basis
$(\sigma^\sA) = (\sigma^-,\tau,\sigma^i)$ on the fluid worldvolume such that
$\bbV^\sA = \delta^\sA_-$ and
$\bbbeta^\sA = \beta_0(\delta^\sA_\tau - \mu_0\delta^\sA_-)$, where
$\beta_0 = (k_{\text B}T_0)^{-1}$ is the inverse temperature and $\mu_0$ the
mass chemical potential of the global thermal state. In this basis, the
coordinate $\tau$ can be understood as the time coordinate associated with the
thermodynamic rest frame, $\sigma^i$ as a set of internal labels associated with
each ``fluid element'', while the coordinate $\sigma^-$ as the global U(1)
``mass/particle number'' phase of each element. However, for the majority of our
discussion below, we shall leave $\bbV^\sA$ and $\bbbeta^\sA$ explicit.

The fluid worldvolume carries the dynamical fields of the theory:
$X^\sM_s(\sigma)$ with $s=1,2$. These fields should be understood as the SK
double copies of null spacetime coordinates of a given fluid element; see
\cref{fig:null-schematic}. Decomposing the coordinate fields as
$X^\sM_{1,2} = X^\sM_r \pm \hbar/2\, X^\sM_a$, the average combination
$X^\sM_r(\sigma)$ is understood as the true physical spacetime coordinates of
the fluid elements, while $X^\sM_a(\sigma)$ as the associated stochastic
noise. The effective theory needs to satisfy global Galilean symmetries
\eqref{eq:null-trans} independently on the two SK spacetimes. To probe the
associated Noether currents, we introduce a pair of null background sources on
the SK spacetimes $g_{s\sM\sN}(X_s)$ and null vector fields $V^\sM_s(X_s)$ such
that $V_s^\sM V_s^\sN g_{s\sM\sN} = 0$ and $\lie_{V_s}g_{s\sM\sN} = 0$. The full
system is now required to be invariant under $(d+2)$-dimensional diffeomorphisms
\begin{subequations}
  \begin{gather}
    X^\sM_s(\sigma) \to X'^\sM_s(X_s(\sigma)),
  \end{gather}
  which act on the background metric and null vectors as
  \begin{align}
    g_{s\sM\sN}(X_s)
    &\to
    g'_{s\sM\sN}(X'_s)
    = \frac{\dow X^\sR_s}{\dow X'^\sM_s} \frac{\dow X^\sS_s}{\dow X'^\sN_s}
    g_{s\sR\sS}(X_s), \nn\\
    V^\sM(X_s)
    &\to V'^\sM(X'_s) = \frac{\dow X'^\sM_s}{\dow X^\sN_s} V^\sN(X_s).
  \end{align}%
  \label{eq:SK-phys-diffeo}%
\end{subequations}
We remind the reader that due to the presence of a null isometry, these are
merely the $(d+2)$-dimensional representation of the $(d+1)$-dimensional local
Galilean transformations. Since the choice of SK null vectors is entirely in our
hands, we pick them to have the same functional form on the two SK spacetimes,
i.e.  $V_1^\sM(X) = V_2^\sM(X)$ when evaluated on the same numerical coordinates
$X^\sM$; this shall be useful later. To be able to describe $(d+1)$-dimensional
Galilean physics, the dynamical fields $X^\sM_s$ should only be independent
functions of $\tau$ and $\sigma^i$, and not of the auxiliary coordinate
$\sigma^-$. This can be fixed by requiring that the pushforward of SK null
vectors $V_s^\sM(X_s)$ onto the physical spacetime is the fixed worldvolume
vector $\bbV^\sM(\sigma)$, i.e.
\begin{gather}
  \bbV^\sA(\sigma) \dow_\sA X_s^\sM(\sigma) = V_s^\sM(X_s(\sigma)).
  \label{eq:Vconstraint-fluidST}
\end{gather}
We can also pullback the background metric over to the fluid worldvolume to
define invariants under SK spacetime diffeomorphisms
\begin{align}
  \bbg_{s\sA\sB}(\sigma)
  &= g_{s\sM\sN}(X_s(\sigma))\,
    \dow_\sA X^\sM_s(\sigma) \dow_\sB X^\sN_s(\sigma).
    \label{eq:invariants-null}
\end{align}
All the dependence on the dynamical and background fields in the effective
theory enter via these invariants. Note that
$\bbV^\sA \bbV^\sB \bbg_{s\sA\sB} = 0$ and $\lie_\bbV \bbg_{s\sA\sB} = 0$,
similar to the null background sources on the SK spacetimes. In fact, we can
choose the average combination
$\bbg_{r\sA\sB} = (\bbg_{1\sA\sB}+\bbg_{2\sA\sB})/2$ as a metric on the fluid
worldvolume.

The choice of coordinates $\sigma^\sA$ on the fluid worldvolume is entirely
arbitrary and can, therefore, be arbitrarily redefined without changing any
physics. This leads to the invariance of the theory under local worldvolume
diffeomorphisms
\begin{subequations}
  \begin{gather}
    \sigma^\sA \to \sigma'^\sA(\sigma).
  \end{gather}
  These transformations act naturally on various fluid worldvolume objects
  \begin{gather}
    \bbg_{s\sA\sB}(\sigma) \to \bbg'_{s\sA\sB}(\sigma') =
    \frac{\dow\sigma^\sC}{\dow \sigma'^\sA}
    \frac{\dow \sigma^\sD}{\dow \sigma'^\sB} \bbg_{s\sC\sD}(\sigma), \nn\\
    \bbV^\sA(\sigma) \to \bbV'^\sA(\sigma') =
    \frac{\dow\sigma'^\sA}{\dow\sigma^\sB}\bbV^\sB(\sigma), \qquad
    \bbbeta^\sA(\sigma) \to \bbbeta'^\sA(\sigma') =
    \frac{\dow\sigma'^\sA}{\dow\sigma^\sB}\bbbeta^\sB(\sigma).
  \end{gather}%
  \label{eq:fluid-spacetime-symm-null}%
\end{subequations}
As mentioned previously, we can partially fix this local symmetry to explicitly
set $\bbV^\sA = \delta^\sA_-$ and
$\bbbeta^\sA = \beta_0(\delta^\sA_\tau-\mu_0\delta^\sA_-)$. The residual
symmetry transformations, in this case, are the arbitrary spatial relabelling of
the fluid elements $\sigma^i\to\sigma'^i(\vec\sigma)$, spatial redefinitions of
the local time coordinate $\tau \to \tau + f(\vec\sigma)$, and that of the
auxiliary coordinate (read U(1) mass/particle number phase)
$\sigma^-\to\sigma^- + \lambda(\vec\sigma)$. Physically, we can understand the
time-independence of these transformations as the requirement that labelling
scheme and phases chosen at one point in time cannot be arbitrarily changed as
the fluid evolves.  If we were interested in a state with spontaneously broken
symmetries, like superfluids or crystals, the system could contain additional
preferred coordinate directions allowing further fixing of the symmetries
\eqref{eq:fluid-spacetime-symm-null}.

The effective action $S$ for Galilean hydrodynamics is the most generic
functional made out of the constituent fields $g_{s\sM\sN}$, $X^\sM_s$, and
their derivatives, respecting the SK spacetime symmetries
\eqref{eq:SK-phys-diffeo} and the fluid worldvolume symmetries
\eqref{eq:fluid-spacetime-symm-null}. Using the invariants
\eqref{eq:invariants-null}, the effective action can be written in terms of a
Lagrangian density
\begin{equation}
  S[\bbg_1,\bbg_2;\bbV,\bbbeta] = \int \df^{d+2}\sigma\sqrt{-\bbg_r}\,
  \mathcal{L}[\bbg_1,\bbg_2;\bbV,\bbbeta].
  \label{eq:SKaction-null}
\end{equation}
The Lagrangian $\mathcal{L}$ is a scalar on the fluid worldvolume constructed by
proper contraction of $\sA,\sB,\ldots$ indices. Notice that the integrand in
\cref{eq:SKaction-null} is entirely independent of the fiducial $\sigma^-$
direction, therefore the integration over this coordinate merely spits out a
constant volume factor. We can define the SK double copies of the null fluid
energy-momentum tensor associated with the action \eqref{eq:SKaction-null} by
performing variations with respect to the sources
\begin{align}
  T^{\sM\sN}_1(X_1)
  = \frac{2\hbar}{\sqrt{-g_1(X_1)}}
    \frac{\delta S}{\delta g_{1\sM\sN}(X_1)}, \qquad
  T^{\sM\sN}_2(X_2)
  = \frac{-2\hbar}{\sqrt{-g_2(X_2)}}
  \frac{\delta S}{\delta g_{2\sM\sN}(X_2)}.
  \label{eq:doubleEMTensor-null}
\end{align}
The minus sign in the second expression stems from the second copy of the
energy-momentum tensor being inserted on the time-reversed part of the SK
contour. We have taken $S$ to be unitless, leading to the additional factors of
$\hbar$ in these formulae; this will facilitate taking a small $\hbar$ limit of
the effective action later. The classical equations of motion of $X^\sM_s$ imply
the respective conservation equations for the energy-momentum tensors
\begin{equation}
  \nabla^s_\sM T^{\sM\sN}_s = 0.
\end{equation}
Here $\nabla^s_\sM$ are covariant derivatives associated with
$g_{s\sM\sN}$. Hence we see that the conservation equations, used as equations
of motion in classical hydrodynamics, follow from a variational principle in the
effective field theory. We can use the dictionary \eqref{eq:null-EMTensor} to
read out the respective Galilean fluid observables.

\subsection{Physical spacetime formulation}
\label{sec:GalSK-PS}

The fluid worldvolume formalism of the effective field theory discussed above is
quite elegant as it makes most of the underlying structure manifest. However,
for most practical purposes and to make contact with classical hydrodynamics, it
is useful to translate to a physical spacetime formulation instead. To this end,
let us choose the single-copy physical spacetime coordinates as the average
fields $x^\sM \equiv X_r^\sM(\sigma)$. We can define the inverse maps
$\sigma^\sA = \sigma^\sA(x)$ via $X_r^\sM(\sigma(x)) = x^\sM$ and accordingly
$X^\sM_a(x) = X_a^\sM(\sigma(x))$. With this rewriting, the set of dynamical
fields in the theory are the fluid element labels $\sigma^\sA(x)$ and the
stochastic noise fields $X^\sM_a(x)$ written as a function of physical spacetime
coordinates.

Rewriting the invariants in \cref{eq:invariants-null} in the ``$r/a$'' basis as
$\bbg_{1,2\,\sA\sB} = \bbg_{r\sA\sB} \pm \hbar/2\, \bbg_{a\sA\sB}$, we can
define their pullback onto the physical spacetime as
\begin{align}
  G_{r\sM\sN}(x)
  &= \bbg_{r\sA\sB}(\sigma(x)) \dow_\sM\sigma^\sA(x) \dow_\sN\sigma^\sB(x)
    = g_{r\sM\sN}(x) + \mathcal{O}(\hbar), \nn\\
  G_{a\sM\sN}(x)
  &= \bbg_{a\sA\sB}(\sigma(x)) \dow_\sM\sigma^\sA(x) \dow_\sN\sigma^\sB(x)
    = g_{a\sM\sN}(x) + \lie_{X_a} g_{r\sM\sN}(x) + \mathcal{O}(\hbar).
    \label{eq:semiclassical-expansion-null}
\end{align}
Here $\lie_{X_a}$ denotes a Lie derivative along $X^\sM_a$. In the second
equalities above, we have expanded the respective quantities up to leading order
in $\hbar$, referred to as the ``statistical limit'', with
$g_{1,2\,\sM\sN}(x) = g_{r\sM\sN}(x) \pm \hbar/2\, g_{a\sM\sN}(x)$ evaluated on
the physical spacetime.\footnote{Generally, the procedure of adding/subtracting
  the two sets of background fields is not useful as they transform under
  independent symmetries. However, as we see explicitly, in the statistical
  ($\hbar\to 0$) limit such an operation is well defined with the average and
  difference combinations being related to the pullback of the worldvolume
  invariants.\label{foot:add-background}} We can identify the average
combination $g_{r\sM\sN}$ with the single-copy background metric $g_{\sM\sN}$
introduced in \cref{sec:source-coupling-null}. We can choose $G_{r\sM\sN}$ as
the metric on the physical spacetime. Note that $G_{r\sM\sN}$ and $G_{a\sM\sN}$
are \emph{not} purely background fields; they are generically given by a
complicated combination of background and dynamical fields. Nonetheless, in the
statistical limit, the average combination $G_{r\sM\sN}$ reduces to the average
background metric $g_{r\sM\sN}$. Similarly, using \cref{eq:Vconstraint-fluidST},
we can deduce the pushforward of the null isometry $\bbV^\sA$ as
\begin{align}
  V^\sM(x)
  = \bbV^\sA(\sigma(x)) \frac{\dow X_r^\sM(\sigma(x))}{\dow\sigma^\sA(x)}
  = \half \Big( V_1^\sM(x) + V_2^\sM(x) \Big) + \mathcal{O}(\hbar).
  \label{eq:SK-null-vector-null}
\end{align}
The choice $V^\sM_1(X) = V^\sM_2(X)$ results in the condition
$\lie_V X_a^\sM = 0$. It is easy to see that
$V^\sM V^\sN G_{r,a\sM\sN} = \lie_V G_{r,a\sM\sN} = 0$, providing the physical
spacetime with the null background structure. 

On the other hand, pushforward of the thermal time vector $\bbbeta^\sA$ onto the
physical spacetime results in the ``hydrodynamic fields''
\begin{equation}
  \beta^\sM(x)
  = \bbbeta^\sA(\sigma(x)) \dow_\sA X_r^\sM(\sigma(x)).
  \label{eq:SK-hydro-fields-null}
\end{equation}
The condition $\lie_\bbV \bbbeta^\sA = 0$ leads to $\lie_V \beta^\sM = 0$.  The
conventional hydrodynamic fields: null fluid velocity $u^\sM(x)$ (normalised as
$u^\sM u^\sN G_{r\sM\sN} = 0$, $V^\sM u^\sN G_{r\sM\sN} = -1$), temperature
$T(x)$, and mass chemical potential $\mu(x)$ can be defined in terms of
$\beta^\sM$ as
\begin{gather}
  \kB T(x) = \frac{1}{\beta(x)}, \qquad
  u^\sM(x) = \frac{\beta^\sM(x)}{\beta(x)}
  + \mu(x) V^\sM,
  \qquad
  \mu(x) = \half \frac{\beta^\sR(x)\beta^\sS(x) G_{r\sR\sS}(x)}{\beta(x)^2},
  \label{eq:SK-conventional-hydro-fields-null}
\end{gather}
where $\beta(x) = -\beta^\sM(x)V^{\sN}(x)G_{r\sM\sN}(x)$. Note that in the
statistical limit, $G_{r\sM\sN}$, $G_{a\sM\sN}$, and $V^\sM$ are entirely
independent of the non-stochastic dynamical fields $\sigma^\sA(x)$ and all
dependence thereof comes only via the hydrodynamic fields. This justifies the
validity of the choice of degrees of freedom in classical hydrodynamics.

Working in the physical spacetime formulation, the fluid worldvolume symmetries
\eqref{eq:fluid-spacetime-symm-null} are explicitly realised. As a payoff,
however, we need to implement the ``average part''\footnote{Explicitly in terms of
  \cref{eq:SK-phys-diffeo}, one finds
  \begin{align*}
    x'^\sM(x)
    &= \half \lb
      X'^\sM_1(x+\hbar/2\,X_a(x)) + X'^\sM_2(x- \hbar/2\,X_a(x)) \rb
      = \half  \lb X'^\sM_1(x) + X'^\sM_2(x) \rb + \mathcal{O}(\hbar).
  \end{align*}
  In the statistical limit, these turn into the average combinations of the \SK
  spacetime transformations.  } of the \SK spacetime symmetries
\eqref{eq:SK-phys-diffeo}, pulled down to the physical spacetime through the
coordinate maps $x^\sM = X_r^\sM(\sigma)$, leading to physical spacetime
diffeomorphisms
\begin{subequations}
  \begin{equation}
    x^\sM \to x'^\sM(x).
  \end{equation}
  This acts on various physical spacetime fields in the theory as expected
  \begin{gather}
    G_{r,a\sM\sN}(x) \to G'_{r,a\sM\sN}(x') = \frac{\dow x^\sR}{\dow x'^\sM}
    \frac{\dow x^\sS}{\dow x'^\sN} G_{r,a\sR\sS}(x), \nn\\
    V^\sM(x) \to V'^\sM(x')
    = \frac{\dow x'^\sM}{\dow x^\sN} V^\sN(x), \qquad
    \beta^\sM(x) \to \beta'^\sM(x')
    = \frac{\dow x'^\sM}{\dow x^\sN} \beta^\sN(x).
  \end{gather}
  \label{eq:PS-symm-null}%
\end{subequations}%
Note that, unlike \cref{eq:SK-phys-diffeo}, we only have a single copy of
physical spacetime diffeomorphisms here.

The effective action \eqref{eq:SKaction-null} can also be translated into a
physical spacetime representation. One can treat
$\sigma^\sA \to x^\sM = X_r^\sM(\sigma)$ as a diffeomorphism and represent the
coordinate transformed effective action \eqref{eq:SKaction-null} as
\begin{equation}
  S[G_{r},G_{a};\beta,V] = \int \df^{d+2}x\sqrt{-G_r}\,
  \mathcal{L}[G_{r},G_{a};\beta,V].
  \label{eq:S-FS-null}
\end{equation}
The Lagrangian $\mathcal{L}$ is a scalar with appropriate contraction of
$\sM,\sN,\ldots$ indices among the constituent fields. Given the effective
action, we can define the ``$r/a$'' basis of null background conserved
energy-momentum tensors according to
\begin{align}
  T^{\sM\sN}_r(x)
  &= \frac{2}{\sqrt{-G_r(x)}}
    \frac{\delta S}{\delta g_{a\sM\sN}(x)}
    = \frac{1/2}{\sqrt{-G_r(x)}}
    \lb \sqrt{-g_1(x)}\, T_1^{\sM\sN}(x) + \sqrt{-g_2(x)}\, T_2^{\sM\sN}(x) \rb, \nn\\
  T^{\sM\sN}_a(x)
  &= \frac{2}{\sqrt{-G_r(x)}}
    \frac{\delta S}{\delta g_{r\sM\sN}(x)}
    = \frac{1/\hbar}{\sqrt{-G_r(x)}}
    \lb \sqrt{-g_1(x)}\, T_1^{\sM\sN}(x) - \sqrt{-g_2(x)}\, T_2^{\sM\sN}(x) \rb.
  \label{eq:singleEMTensor-null}
\end{align}
The second equalities here follow from the expressions in
\cref{eq:doubleEMTensor-null}. The average combination $T^{\sM\sN}_r$ is
understood as the physical null background energy-momentum tensor, while the
difference combination $T^{\sM\sN}_a$ is understood as the associated stochastic
noise. While both of these operators are conserved on flat spacetime, they are
not individually conserved in the presence of background sources. Nonetheless,
in the statistical limit, the physical average combination satisfies the
classical conservation equation \eqref{eq:null-cons}. The classical
energy-momentum tensor $T^{\sM\sN}$ from \cref{sec:ClassicalGalilean} is
obtained from $T^{\sM\sN}_r$ by going to the statistical ($\hbar\to0$) limit and
switching off all the stochastic ``$a$'' type fields $g_{a\sM\sN}$ and
$X_a^\sM$.

% \begin{align}
%   \nabla_\sM T^{\sM\sN}_r = \mathcal{O}(\hbar), \qquad
%   \nabla_\sM T^{\sM\sN}_a =
%   - T^{\sM\sR}_r g_r^{\sN\sS} \lb \nabla_\sM g_{a\sR\sS}
%   - \half\nabla_\sS g_{a\sM\sR} \rb
%   + \mathcal{O}(\hbar),
%   \label{eq:2copy-classical-EOM}
% \end{align}
% making the average combination conserved. 

\subsection{Schwinger-Keldysh generating functional}
\label{sec:SKaction-null}

Let us momentarily return to our original starting point of the EFT on the fluid
worldvolume. We argued that Galilean hydrodynamics can be described by an
effective action \eqref{eq:SKaction-null}, written as a functional of two copies
of background fields $g_{s\sM\sN}(X_s)$ and two copies of dynamical fields
$X_s^\sM(\sigma)$, consistent with certain spacetime symmetries. The generating
functional for the theory, which allows us to compute out-of-equilibrium thermal
correlators of various hydrodynamic operators, is obtained by the path integral
\begin{equation}
  \exp W[g_{1},g_{2};V_1,V_2]
  = \int \mathcal{D} X_1 \mathcal{D} X_2\,
  \exp\lb iS[\bbg_{1},\bbg_{2};\bbbeta,\bbV]\rb.
  \label{eq:SK-Z-null-FS}
\end{equation}
We have absorbed the conventional weight factor of $1/\hbar$ in the exponential
within the definition of $S$ making it unitless. This will allow the path
integral to have a well-defined statistical ($\hbar\to0$) limit. In the physical
spacetime representation, we can equivalently write\footnote{In principle, one
  needs to be careful about the Jacobian of the field redefinition. However, as
  the transformation from $X_{1,2}^\mu$ to $X_a^\mu$, $\sigma^\sA$ is purely
  algebraic, i.e. does not include any derivatives, the ghosts obtained by
  exponentiating the Jacobian do not propagate and can be
  ignored~\cite{Arzt:1993gz}.}
\begin{equation}
  \exp W[g_{r},g_{a};V]
  = \int \mathcal{D} \sigma \mathcal{D} X_a\,
  \exp\lb iS[G_{r},G_{a};\beta,V]\rb.
  \label{eq:SK-Z-null}
\end{equation}
This generating functional can be used to compute out-of-equilibrium thermal
correlators of conserved Galilean densities and fluxes in the ``$r/a$'' basis
through
\begin{align}
  G_{T_\alpha^{\sM\sN}\ldots}(x,\ldots)
  = i^{n_a} \lb \frac{-i\delta}{\delta g_{\bar\alpha\sM\sN}(x)} \ldots \rb
  W[g_r,g_a].
  \label{eq:corr-funs}
\end{align}
where $\alpha=r,a$ and $\bar\alpha = a,r$ is its conjugate, while $n_a$ is the
number of $a$ type fields in the correlator on the left. Following general SK
machinery (see e.g.~\cite{Wang:1998wg}), the correlators of the type
``$ra\ldots a$'' are interpreted as the fully retarded correlation functions,
``$a\ldots ar$'' as the fully advanced, ``$rr\ldots r$'' as the symmetric
correlation functions, and various other time-ordering schemes in between. For
instance, the retarded, advanced, and symmetric two point functions can be
computed via
\begin{align}
  G^{\text{R}}_{T^{\sM\sN}T^{\sR\sS}}(x,x')
  &= G_{T^{\sM\sN}_r T^{\sR\sS}_a}(x,x')
  = \frac{-i\delta^2 W}{\delta g_{a\sM\sN}(x)\delta g_{r\sR\sS}(x')}, \nn\\
  G^{\text{A}}_{T^{\sM\sN}T^{\sR\sS}}(x,x')
  &= G_{T^{\sM\sN}_a T^{\sR\sS}_r}(x,x')
    = \frac{-i\delta^2 W}{\delta g_{r\sM\sN}(x)\delta g_{a\sR\sS}(x')}, \nn\\
  G^{\text{S}}_{T^{\sM\sN}T^{\sR\sS}}(x,x')
  &= G_{T^{\sM\sN}_r T^{\sR\sS}_r}(x,x')
    = \frac{-i\delta^2 W}{\delta g_{a\sM\sN}(x)\delta g_{a\sR\sS}(x')}.
    \label{eq:null-retG-SK}
\end{align}
The retarded correlator can be compared to the expression \eqref{eq:null-retG}
from classical hydrodynamics.

We have strategically avoided mentioning the ``$aa$'' correlator in
\cref{eq:null-retG-SK}; in general, all the correlators of the type
``$aa\ldots a$'' must be zero for $W$ describing a bona fide out-of-equilibrium
thermal field theory. Similarly, the retarded and advanced correlators, in
momentum space, must be complex conjugates of each other (in momentum space). On
the other hand, the retarded and symmetric correlators must be related by the
fluctuation-dissipation theorem. There are similar constraints for higher-point
functions as well; see~\cite{Wang:1998wg}. These are compactly represented in
terms of the generating functional as
\begin{gather}
  W^*[g_1,g_2;V_1,V_2] = W[g_2,g_1;V_2,V_1], \qquad
  W[g,g;V,V] = 0, \qquad
  \mathrm{Re}\, W[g_1,g_2;V_1,V_2] \leq 0, \nn\\
  W[g_1,g_2;V_1,V_2] = W[\tilde g_1,\tilde g_2;\tilde V_1,\tilde V_2].
  \label{eq:SKW1}
\end{gather}
The last of these is known as the \emph{KMS condition}, with KMS conjugation of
the background fields defined as
\begin{gather}
  \tilde g_{1\sM\sN}(x) = \Theta g_{1\sM\sN}(x), \qquad
  \tilde g_{2\sM\sN}(x) = \Theta g_{2\sM\sN}(t+i\hbar\beta_0, \vec x), \nn\\
  \tilde V_1^\sM(x) = \Theta V_1^\sM(x), \qquad
  \tilde V_2^\sM(x) = \Theta V_2^\sM(t + i\hbar\beta_0, \vec x).
\end{gather}
Here $\beta_0 = (\kB T_0)^{-1}$ is the constant inverse temperature of the
global thermal state and $\Theta$ represents a discrete symmetry involving a
time-flip that the system might enjoy, for example time-reversal T or spacetime
parity PT; see~\cref{tab:CPT}.

The conditions \eqref{eq:SKW1} can be implemented in the field theory by
requiring the effective action to obey the
constraints~\cite{Crossley:2015evo}\footnote{See~\cref{foot:noghosts}.}
\begin{subequations}
  \begin{gather}
    S^*[\bbg_1,\bbg_2;\bbbeta,\bbV] = -S[\bbg_2,\bbg_1;\bbbeta,\bbV], \\
    S[\bbg,\bbg;\bbbeta,\bbV] = 0,  \\
    \mathrm{Im}\, S[\bbg_1,\bbg_2;\bbbeta,\bbV] \geq 0, \\
    S[\bbg_1,\bbg_2;\bbbeta,\bbV]
    = S[\tilde\bbg_1,\tilde\bbg_2;\tilde\bbbeta,\tilde\bbV],
    \label{eq:SK-KMS-FS}
  \end{gather}
  \label{eq:SK-cons-FS}%
\end{subequations}%
where the KMS conjugation of the respective fields is defined according to
\begin{gather}
  \tilde X_1^\sM(\sigma) = \Theta X_1^\sM(\sigma), \qquad
  \tilde X_2^\sM(\sigma) = \Theta X_2^\sM(\sigma + i\hbar\,\Theta\bbbeta(\sigma))
  - i\hbar\beta_0 \delta^\sM_{t}, \nn\\
  \tilde\bbV^\sA(\sigma) = \Theta\bbV^\sA(\sigma), \qquad
  \tilde\bbbeta^\sA(\sigma) = \Theta\bbbeta^\sA(\sigma),
  \label{eq:KMS-trans-FS}
\end{gather}
leading to
\begin{gather}
  \tilde\bbg_{1\sA\sB}(\sigma) = \Theta\bbg_{1\sA\sB}(\sigma), \nn\\
  \tilde\bbg_{2\sA\sB}(\sigma)
  = \dow_\sA \Big(\sigma^\sC+i\hbar\,\Theta\bbbeta^\sC(\sigma)\Big)
  \dow_\sB \Big(\sigma^\sD+i\hbar\,\Theta\bbbeta^\sD(\sigma)\Big)\, 
  \Theta\bbg_{2\sC\sD}(\sigma + i\hbar\,\Theta\bbbeta(\sigma)).
\end{gather}
The argument $(\sigma + i\hbar\,\Theta\bbbeta(\sigma))$ should be understood as
a vector, i.e. $\sigma^\sA + i\hbar\,\Theta\bbbeta^\sA(\sigma)$.  The condition
$\lie_\bbV\bbbeta^\sA = 0$ ensures that \cref{eq:Vconstraint-fluidST} remains
satisfied by the conjugated quantities. The KMS conjugation in the ``1'' sector
is merely a $\Theta$-conjugation, while in the ``2'' sector it is given by a
$\Theta$-conjugation followed by a diffeomorphism along
$i\hbar\,\Theta\beta^\sM(\sigma)$. The KMS transformation in the ``2'' sector is
highly non-trivial and is quite hard to implement in its full
generality. Fortunately, the transformation becomes better-behaved in the small
$\hbar$ limit
\begin{equation}
  \tilde\bbg_{1\sA\sB}(\sigma) = \Theta\bbg_{1\sA\sB}(\sigma), \qquad
  \tilde\bbg_{2\sA\sB}(\sigma) = \Theta\bbg_{2\sA\sB}(\sigma)
  - i\hbar\,\Theta\lie_\bbbeta\bbg_{2\sA\sB}(\sigma),
\end{equation}
which is much easier to implement in the effective theory than the full quantum
version. See appendix A of~\cite{Glorioso:2016gsa} for more discussion.

The constraints \eqref{eq:SK-cons-FS} can also be stated in the physical
spacetime language, i.e.
\begin{subequations}
  \begin{gather}
    S^*[G_r,G_a;\beta,V] = -S[G_r,-G_a;\beta,V], \label{eq:SK-conj-null} \\
    S[G_r,G_a=0;\beta,V] = 0, \label{eq:SK-lin-null} \\
    \mathrm{Im}\, S[G_r,G_a;\beta,V] \geq 0, \label{eq:SK-pos-null} \\
    S[G_r,G_a;\beta,V] = S[\tilde G_r,\tilde G_a;\tilde\beta,\tilde
    V], \label{eq:SK-KMS-null}
  \end{gather}
  \label{eq:SK-constraints-null}%
\end{subequations}
where the KMS conjugation of various fields can be derived using
\cref{eq:KMS-trans-FS}. However, the full quantum expressions are quite
complicated as the physical coordinates $x^\sM = X_r^\sM(\sigma)$ non-trivially
mix the ``1'' and ``2'' type spacetime fields. Nonetheless, in the statistical
limit, we can derive
\begin{gather}
  \tilde\sigma^\sA(x) = \Theta\sigma^\sA(x) + \mathcal{O}(\hbar), \qquad
  \tilde X^\sM_a(x) = \Theta X^\sM_a(x)
  - i\Theta\beta^\sM(x) + i\beta_0\delta^\sM_t + \mathcal{O}(\hbar),
  \nn\\
  \tilde g_{r\sM\sN}(x) = \Theta g_{r\sM\sN}(x) + \mathcal{O}(\hbar), \qquad
  \tilde g_{a\sM\sN}(x) = \Theta g_{a\sM\sN}(x)
  + i\beta_0\Theta\dow_t g_{r\sM\sN}(x)
  + \mathcal{O}(\hbar), \nn\\
  \tilde G_{r\sM\sN}(x) = \Theta G_{r\sM\sN}(x) + \mathcal{O}(\hbar), \qquad
  \tilde G_{a\sM\sN}(x) = \Theta G_{a\sM\sN}(x)
  + i\Theta\lie_\beta G_{r\sM\sN}(x) + \mathcal{O}(\hbar), \nn\\
  \tilde V^\sM(x) \to \Theta V^\sM(x) + \mathcal{O}(\hbar), \qquad
  \tilde\beta^\sM(x) \to \Theta\beta^\sM(x) + \mathcal{O}(\hbar).
  \label{eq:KMS-conjugation-G}
\end{gather}
Recall that in small $\hbar$ limit,
$G_{r\sM\sN}(x) = g_{r\sM\sN}(x) + \mathcal{O}(\hbar)$. Note that the Lie
derivative operator $\lie_\beta$ is odd under $\Theta$. Together, these
requirements constrain the most generic form of an EFT for Galilean
hydrodynamics. At the classical level, these constraints conspire to ensure that
the Onsager's relations and local second law of thermodynamics are satisfied
order-by-order in the derivative expansion~\cite{Glorioso:2016gsa}; see the next
subsection.

\subsection{Explicit effective action and emergent second law}
\label{sec:classicalLimit-null}

We would like to write down the most generic effective action describing
Galilean hydrodynamics guided by the structure outlined above. Let us work in
the physical spacetime formulation, wherein the effective action must respect
the physical spacetime symmetries \eqref{eq:PS-symm-null} and SK constraints
\eqref{eq:SK-constraints-null}; note that the fluid worldvolume symmetries are
manifestly preserved in the physical spacetime representation. The discussion in
the fluid worldvolume language is analogous. As mentioned in
\cref{sec:SKaction-null}, we do not have the tools to implement the full
non-local KMS symmetry in the quantum regime, so we are forced to work in the
statistical ($\hbar\to0$) limit. In this limit, all the quantum fluctuations
freeze out and we are only left with statistical/stochastic fluctuations. Even
so, the structure of the simplified effective theory is extremely rich and is
suitable to describe the effect of stochastic interactions in hydrodynamic
correlation functions. One could, in principle, reintroduce quantum effects
perturbatively in $\hbar$, which we shall not explore in the present work.

The most generic Lagrangian consistent with \cref{eq:SK-lin-null} can be
arranged in a power series in $G_{a}$ starting from the linear term, with
factors of $i$ chosen for consistency with \cref{eq:SK-conj-null}. We have
\begin{equation}
  \mathcal{L}
  = \sum_{m=1}^\infty \frac{(-i)^{m+1}}{2^m}
  \mathcal{F}_m(\underbrace{G_a,G_a,\ldots}_{\times m})
  + \mathcal{O}(\hbar).
  \label{eq:generalL_1}
\end{equation}
Here $\cF_m(\circ,\cdots)$ are a set of totally symmetric real multi-linear
operators made out of $G_{r\sM\sN}$, $V^\sM$ and $\beta^\sM$, allowing $m$
number of arguments. The underbrace notation is meant to denote a repeated set
of arguments. We can perform a change of basis on the operators given by
\begin{align}
  \cF_{m}(\underbrace{\circ,\ldots}_{\times m})
  = \sum_{n=m}^{2m+1} \frac{c_{mn}}{2^{n-m}}
  \cD_{n}(\underbrace{\circ,\ldots}_{\times m},
  \underbrace{\lie_\beta G_r,\ldots}_{\times n-m}), \quad
  c_{mn} =
  \begin{cases}
    (-)^{n/2+1} {\textstyle\binom{n/2}{n-m}} & n~\text{even} \\
    (-)^{(n+1)/2} \frac{m+1}{n+1} {\textstyle\binom{(n+1)/2}{n-m}} & n~\text{odd}
  \end{cases},
\end{align}
where $\cD_m(\circ,\cdots)$, again, are symmetric real multi-linear maps similar
to $\cF_m(\circ,\cdots)$. Manipulating the double summations, one finds that
this operation recasts the Lagrangian into
\begin{align}
  \mathcal{L}
  &= \half\cD_1(G_a)
    + i \sum_{n=1}^\infty
    \frac{1}{2^{2n}}
    \cD_{2n}(\underbrace{G_a,\ldots}_{\times n},
    \underbrace{G_a {+} i \lie_\beta G_r,
    \ldots}_{\times n}) \nn\\
  &\qquad
    + \sum_{n=1}^\infty
    \frac{1}{2^{2n+1}}
    \cD_{2n+1}(G_a{+}{\textstyle\frac{i}{2}}\lie_\beta G_r,
    \underbrace{G_a,\ldots}_{\times n},
    \underbrace{G_a {+} i \lie_\beta G_r,\ldots}_{\times n})
    + \mathcal{O}(\hbar).
    \label{eq:L-G-null}%
\end{align}
This comprises the most general effective action for Galilean hydrodynamics in
the statistical limit. The utility of this form is that under the KMS
conjugation \eqref{eq:KMS-conjugation-G}, the arguments of the operators under
the summation map to each other up to a $\Theta$-conjugation, i.e. the $n$
instances of $G_a \to \Theta G_a + i\Theta\lie_\beta G_r$ and
$G_a {+} i\lie_\beta G_r \to \Theta G_a$ map to each other, while
$G_a {+} \frac{i}{2}\lie_\beta G_r \to \Theta G_a {+} \frac{i}{2}\Theta
\lie_\beta G_r$ maps to itself. Only the $\mathcal{D}_1(G_a)$ term outside the
summation poses an exception to this general rule and leads to our first
constraint
\begin{subequations}
  \begin{equation}
    \half\mathcal{D}_1(\lie_\beta G_r) = \nabla_\sM \mathcal{N}^\sM_0
    \quad \text{for some vector~} \mathcal{N}^\sM_0,
    \label{eq:D1-cond-null}
  \end{equation}
  This ensures that the term does
  not generate a $G_a$-independent piece under KMS transformation (up to a total
  derivative), in accordance with \cref{eq:SK-lin-null}. Since the operators
  $\mathcal{D}_m$ are made out of only ``$r$'' type fields, KMS merely acts on
  these as a $\Theta$-conjugation. Given that the arguments already map to each
  other up to a $\Theta$-conjugation under KMS, in order to respect
  \cref{eq:SK-KMS-null} we just need to ensure that $\mathcal{D}_m$ has the
  right sign under $\Theta$. We can express this requirement as
  \begin{equation}
    \mathcal{D}_m(G_a,G_a,\ldots)
    \quad \text{is $\Theta$-even}~\forall m.
    \label{eq:Dn-conjugation-null}
  \end{equation}
  Note that $\lie_\beta$ is a $\Theta$-odd operation, therefore swapping any of
  the $G_a$'s in the arguments above with $i\lie_\beta G_r$ flips the
  $\Theta$-parity of the resultant term. As a consequence, the Lagrangian
  \eqref{eq:L-G-null} as a whole does not have a definite sign under
  $\Theta$. This should be expected due to the presence of dissipation. This
  only leaves us with the inequality constraint in \cref{eq:SK-pos-null},
  leading to
  \begin{equation*}
    \frac{(-1)^{m+1}}{2^{2m}} \cF_{2m}(\underbrace{G_a,\ldots}_{\times 2m})
    =
    (-1)^{m+1}
    \sum_{n=2m}^{4m+1} \frac{c_{2m,n}}{2^{n}}
    \cD_{n}(\underbrace{G_a,\ldots}_{\times 2m},
  \underbrace{\lie_\beta G_r,\ldots}_{\times n-2m})
    \geq 0,
  \end{equation*}
  for arbitrary field configurations. In practise, this condition can be
  implemented order-by-order in $G_a$ and derivative
  expansion~\cite{Bhattacharyya:2014bha, Bhattacharyya:2013lha}, leading to a
  simple condition
  \begin{equation}
    \cD_{2}(G,G)\big|_{\text{leading order}} \geq 0
    \quad \text{for arbitrary~} G_{\sM\sN},
    \label{eq:D2-cond-null}
  \end{equation}
  \label{eq:KMSConstraints-null}%
\end{subequations}
If the leading derivative
piece in $\mathcal{D}_2$ happens to be zero for a system (which amounts to zero
viscosities and conductivity), the requirement can, in principle, shift to
higher order in derivatives or $G_a$, leading to different equality/inequality
constraints; see~\cite{Jensen:2018hse} for an analogue of this for relativistic
fluids.

The constraints \eqref{eq:KMSConstraints-null} are the field theory realisation
of the second law of thermodynamics and Onsager's relations. To see this, let us
perform integration by parts to define
\begin{equation}
  \cD_m(G,\underbrace{\circ,\ldots}_{\times m-1})
  = G_{\sM\sN} \cT^{\sM\sN}_m(\underbrace{\circ,\ldots}_{\times m-1})
  + \nabla_\sM \cN^\sM_m(G;\underbrace{\circ,\ldots}_{\times m-1}),
  \label{eq:byparts-null}
\end{equation}
for arbitrary symmetric tensor $G_{\sM\sN}$. This equation essentially says that
if $\mathcal{D}_m$ acts on its argument $G_{\sM\sN}$ as a differential operator,
we can remove the derivatives by adding a total derivative term. The classical
constitutive relations $T^{\sM\sN}$ are defined as the average quantities
$T^{\sM\sN}_r$ evaluated on a configuration with zero ``a'' type fields
$X^\sM_a = g_{a\sM\sN} = 0$. Using the variational formulae
\eqref{eq:singleEMTensor-null} and the definitions \eqref{eq:byparts-null}, we
can obtain
\begin{align}
  T^{\sM\sN}
  &= \cT^{\sM\sN}_1 - \half\cT^{\sM\sN}_{2}(\lie_\beta G_r) 
    - \frac18 \cT^{\sM\sN}_{3}(\lie_\beta G_r,\lie_\beta G_r).
    \label{eq:classical-consti-null}
\end{align}
Note that the operators $\mathcal{D}_n$ for $n>3$ do not contribute to the
classical constitutive relations. Using \eqref{eq:byparts-null} again, one can
derive an identity satisfied by these constitutive relations
\begin{equation}
  \half T^{\sM\sN} \lie_\beta g_{\sM\sN}
  = \nabla_\sM N^\sM - \Delta, \qquad
  \Delta \geq 0,
  \label{eq:SK-adiabaticity-null}
\end{equation}
where
\begin{align}
  N^\sM
  &= \mathcal{N}^\sM_0
    + \half\mathcal{N}^\sM_1(\lie_\beta G_r)
    - \frac14 \mathcal{N}^\sM_2(\lie_\beta G_r,\lie_\beta G_r)
    - \frac{1}{16} \mathcal{N}^\sM_3(\lie_\beta G_r,\lie_\beta G_r,\lie_\beta G_r),
  \nn\\
  \Delta
  &= \frac14 \cD_{2}(\lie_\beta G_r,\lie_\beta G_r) 
    + \frac{1}{16} \cD_{3}(\lie_\beta G_r,\lie_\beta G_r,\lie_\beta G_r).
    \label{eq:NDelta-classical-null}
\end{align}
In obtaining this, we have used the condition \eqref{eq:D1-cond-null} and the
identification of $G_{r\sM\sN}$ with the single-copy background metric
$g_{\sM\sN}$ in the statistical limit. The positivity of $\Delta$, within a
derivative expansion, follows from the condition \eqref{eq:D2-cond-null}. The
condition \eqref{eq:SK-adiabaticity-null} is precisely the adiabaticity
condition \eqref{eq:adiabaticity} of Galilean hydrodynamics. Consequently, the
entropy current defined as $S^\sM = \kB N^\sM - T^{\sM\sN}u_\sN/T$ follows the
second law of thermodynamics, $\nabla_\sM S^\sM = \kB\Delta \geq 0$, onshell.
We have, therefore, arrived at a derivation of the second law of thermodynamics
within the hydrodynamic field theoretic framework, originally due
to~\cite{Crossley:2015evo} for relativistic fluids.

Finally, the condition \eqref{eq:Dn-conjugation-null}, along with
\cref{eq:byparts-null} and \cref{eq:classical-consti-null}, implies that the
contribution from $\mathcal{D}_1$ and $\mathcal{D}_3$ to the classical
constitutive relations must be $\Theta$-preserving (same as expected for the
respective operators in \cref{tab:CPT}), while that from $\mathcal{D}_2$ must be
$\Theta$-violating (opposite to expected for respective operators). Noting that
$\mathcal{D}_3$ only contributes to the constitutive relations at two-derivative
order and higher, these conditions imply, for instance, that one-derivative
dissipative transport (produces entropy) must be $\Theta$-violating, while
one-derivative adiabatic transport (does not produce entropy) must be
$\Theta$-preserving. This should be physically expected for a dissipative
system. In linear hydrodynamics, these requirements implement Onsager's
reciprocity relations.

\subsection{Field redefinitions and frame transformations}
\label{sec:field-redef}

The dynamical fields in the effective field theory can, in general, be subjected
to arbitrary field redefinitions. However, having chosen the fields to realise a
particular representation of the KMS transformations in
\cref{eq:KMS-conjugation-G} fixes a large part of this freedom. For
concreteness, let us work in the physical spacetime representation; we also
restrict to the statistical limit for simplicity. We consider a general
transformation of the ``$a$'' type fields
\begin{subequations}
  \begin{equation}
    X^\sM_a \to X^\sM_a + if^\sM_a,
  \end{equation}
  where $f^\sM_a$ can be arbitrary symmetry-respecting functions of the building
  blocks $G_{r,a\sM\sN}$, $V^\sM$, and $\beta^\sM$. Due to the condition
  \eqref{eq:SK-lin-null}, these quantities must be at least linear in
  $G_{a\sM\sN}$, and due to \cref{eq:SK-conj-null}, each occurrence of
  $G_{a\sM\sN}$ must be accompanied with a factor of $i$. Finally, to respect
  the KMS transformations \eqref{eq:KMS-conjugation-G}, the hydrodynamic fields
  $\beta^\sM$ must transform as
  \begin{equation}
    \beta^\sM \to \beta^\sM + (f_a^\sM - \Theta \tilde f_a^\sM)
    + \mathcal{O}(\hbar).
    \label{eq:redef-beta-null}
  \end{equation}
  \label{eq:SK-redef-null}%
\end{subequations}%
Interestingly, we find that the redefinitions of $\beta^\sM$ are entirely
constrained by the redefinitions of $X^\sM_a$. This, in turn, constrains the
redefinitions in the remaining dynamical fields $\sigma^\sA$ through the
definitions \eqref{eq:SK-hydro-fields-null}. Note that the KMS ``tilde''
conjugation of $G_{r\sM\sN}$, $V^{\sM}$, and $\beta^\sM$ in the statistical
limit in \cref{eq:KMS-conjugation-G} is same as a $\Theta$-conjugation, while
that of $G_{a\sM\sN}$ is same as a $\Theta$-conjugation up to terms involving
$\lie_\beta G_{r\sM\sN}$. Hence, KMS conjugation of $f^\sM_a$ followed by
$\Theta$-conjugation yields back the original quantities up to terms involving
$\lie_\beta G_{r\sM\sN}$. It follows that the allowed shifts of $\beta^\sM$ in
\cref{eq:redef-beta-null} are purely non-hydrostatic, i.e they involve at least
one instance of $\lie_\beta G_{r\sM\sN}$. In fact, since the constitutive
relations following from the most generic KMS-respecting effective Lagrangian
\eqref{eq:L-G-null} always satisfy the adiabaticity equation
\eqref{eq:SK-adiabaticity-null}, these allowed redefinitions merely correspond
to the residual field redefinitions among thermodynamic frames; see
\cref{sec:frame}.

During our discussion of classical Galilean hydrodynamics in
\cref{sec:GalAdiabaticity}, we fixed the residual redefinition freedom among
thermodynamic frames by eliminating any dependence on
$u^\sM \delta_\scB g_{\sM\sN}$ from the constitutive relations using the
classical equations of motion. We can make an equivalent statement in the field
theory as well. Note that any infinitesimal redefinitions of $X_a^\sM$ change
the Lagrangian by terms involving equations of motion. Hence, we can entirely
fix the redefinitions \eqref{eq:SK-redef-null} by choosing to skip terms
involving $u^\sM \delta_\scB G_{r\sM\sN}$ while constructing the effective
Lagrangian \eqref{eq:L-G-null}, which can always be eliminated using equations
of motion. This is the analogue of working in the thermodynamic mass frame from
classical hydrodynamics. Note that $G_{r\sM\sN}$ are identified with the average
sources $g_{\sM\sN}$ in the statistical limit.

\vspace{1em}

This concludes our formal discussion of the \SK effective field theory framework
for Galilean hydrodynamics. We outlined a set of effective fields and rules that
must be respected by an effective action describing Galilean hydrodynamics. We
then proceeded to apply these rules in small $\hbar$ limit to construct the most
generic effective theory governing stochastic fluctuations in a Galilean
fluid. We also illustrated how the classical local second law of thermodynamics
and Onsager's relations emerge within the field theory.

\section{Effective action for one-derivative Galilean fluids}
\label{sec:1derGal}

Our discussion of the EFT framework for Galilean hydrodynamics so far has been
quite formal. As a concrete realisation of these ideas, we now write down the
explicit effective action describing one-derivative Galilean fluids from
\cref{sec:ClassicalGalilean}, using the generic machinery from
\cref{sec:classicalLimit-null}. We then proceed to linearise this effective
action around a fluid configuration at rest and isolate an interacting
perturbative field theory describing stochastic fluctuations.

\subsection{Non-linear effective action}
\label{sec:non-linear-1der}

Truncation of the EFT at a given derivative order requires us to pick a
derivative counting scheme for various background and dynamical fields, based on
the physical system under consideration. A natural choice is to treat the
fluctuations of the hydrodynamic fields $\beta^\sM$ (or equivalently of the
conjugate conserved densities $T_r^{t\sM}$) and noise fields $X^\sM_a$ at the
same order, taken to be $\mathcal{O}(\dow^0)$ for reference. It follows that
$G_{a\sM\sN}$ and $\lie_\beta G_{r\sM\sN}$ should be treated as
$\mathcal{O}(\dow^1)$. For consistency, we must choose the background fields
$g_{r\sM\sN}$ as $\mathcal{O}(\dow^0)$ and $g_{a\sM\sN}$ as
$\mathcal{O}(\dow^1)$. This counting scheme ensures that the KMS constraints in
the statistical limit do not mix between derivative orders. Since the physical
energy-momentum tensor is given by a variational derivative of the action with
respect to $g_{a\sM\sN}$, the ``derivative order'' of the hydrodynamic
constitutive relations, or simply that of hydrodynamics, is given by one less
than the order of the Lagrangian. It follows that the $\cD_m$ operators in
\cref{eq:L-G-null} start contributing to the constitutive relations at
$\mathcal{O}(\dow^{m-1})$ in the derivative expansion.

Therefore, one-derivative hydrodynamics is entirely characterised by the most
generic expressions for the operators $\cD_1(\circ)$ and $\cD_2(\circ,\circ)$
written in terms of the metric $G_{r\sM\sN} = g_{r\sM\sN}$, null isometry
$V^\sM$, and the thermal vector $\beta^\sM$, satisfying the constraints
\eqref{eq:KMSConstraints-null}. The operator $\mathcal{D}_1$ needs to involve
all the allowed terms up to one derivative order, while $\mathcal{D}_2$ only
needs to involve zero derivative terms. Let us start with
$\mathcal{D}_1$. Focusing on $d=3$ spatial dimensions for parity-violating
terms, while generic $d$ for parity-even terms, we propose that it takes the
form
\begin{subequations}
  \begin{align}
    \cD_1(G)
  &= \bigg[
    \rho\, u^\sM u^\sN
    + 2\varepsilon\, u^{(\sM} V^{\sN)} + p\, \Delta^{\sM\sN} \nn\\
  &\qquad
    + 2 u^{(\sM} \lb
    a_0 \epsilon^{\sN)\sP\sR\sS\sT} V_\sP u_\sR \dow_\sS u_\sT
    + a_2 \epsilon^{\sN)\sP\sR\sS\sT} V_\sP u_\sR \dow_\sS V_{\sT}
    \rb \nn\\
  &\qquad
    + 2 V^{(\sM} \lb
    a_2 \epsilon^{\sN)\sP\sR\sS\sT} V_\sP u_\sR \dow_\sS u_\sT
    + a_1 \epsilon^{\sN)\sP\sR\sS\sT} V_\sP u_\sR \dow_\sS V_{\sT}
    \rb
    \bigg] G_{\sM\sN},
  \end{align}
\end{subequations}
for arbitrary symmetric tensor $G_{\sM\sN}$. Various quantities appearing here
are the same as introduced in \cref{sec:ClassicalGalilean}. The fluid velocity
$u^\sM$, temperature $T$, and mass chemical potential $\mu$ are defined in terms
of the field theoretic structures in
\cref{eq:SK-conventional-hydro-fields-null}, while
$\Delta^{\sM\sN} = G_r^{\sM\sN} + 2 u^{(\sM}V^{\sN)}$. The mass density
$\rho(T,\mu)$ and internal energy density $\varepsilon(T,\mu)$ are fixed in
terms of the pressure $p(T,\mu)$ using the thermodynamic relations
\eqref{eq:GCthermo}, while the parity-violating coefficients $a_0(T,\mu)$,
$a_1(T,\mu)$, and $a_2(T,\mu)$ are fixed in terms of three constants given in
\cref{eq:transconstraints}. With these in place, $\mathcal{D}_1$ is the most
generic operator truncated at one-derivative order that satisfies the condition
\eqref{eq:D1-cond-null}, with $\mathcal{N}_0^\sM = p\, \beta^\sM + \Upsilon^\sM$
where $\Upsilon^\sM$ is given in \cref{eq:null-consti-cov}. To implement the
condition \eqref{eq:Dn-conjugation-null} we need to pick a $\Theta$
operator. Choosing $\Theta$ to be just the time-reversal operator T does not
lead to any constraints. If we were to choose $\Theta$ to be PT, instead, the
parity-violating terms above will no longer be allowed. Moving on, for the
$\mathcal{D}_2$ operator we find
\begin{align}
  \cD_2(G,G')
  &= \kB T \bigg[
    2\eta\, \Delta^{\sM(\sR} \Delta^{\sS)\sN} 
    + \lb \zeta - {\textstyle\frac2d}\eta \rb \Delta^{\sM\sN}\Delta^{\sR\sS}
    + 4T\kappa\, V^{(\sM}\Delta^{\sN)(\sR}V^{\sS)}
    \bigg] G_{\sM\sN}G'_{\sM\sN},
    \label{eq:1der-G}
\end{align}
for generic tensors $G_{\sM\sN}$ and $G'_{\sM\sN}$. Note that $\mathcal{D}_2$ is
symmetric under the exchange of its arguments. We have chosen not to write down
any terms along the vector $u^\sM$, because the respective contribution will
couple to $u^\sM \delta_\scB G_{r\sM\sN}$ in the effective action which we have
chosen to eliminate using the classical equations of motion; see
\cref{sec:field-redef}. All the terms in $\mathcal{D}_2$ respect
\eqref{eq:Dn-conjugation-null} for any choice of $\Theta$. On the other hand,
the dissipative transport coefficients: shear viscosity $\eta(T,\mu)$, bulk
viscosity $\zeta(T,\mu)$, and thermal conductivity $\kappa(T,\mu)$ are
constrained to be non-negative due to the condition \eqref{eq:D2-cond-null}.

Plugging the operators \eqref{eq:1der-G} into \cref{eq:L-G-null}, we can
explicitly work out the effective action for one-derivative Galilean
hydrodynamics in the statistical limit
\begin{align}
  \mathcal{L}
  &= \frac12 \lb \rho\, u^\sM u^\sN
    + 2\varepsilon\, u^{(\sM} V^{\sN)} + p\, \Delta^{\sM\sN} \rb
    \lb g_{a\sM\sN} + \lie_{X_a} g_{r\sM\sN} \rb \nn\\
  &\qquad
    + \lb a_0 \epsilon^{\sN\sP\sR\sS\sT} V_\sP u_\sR \dow_\sS u_\sT
    + a_2 \epsilon^{\sN\sP\sR\sS\sT} V_\sP u_\sR \dow_\sS V_{\sT}
    \rb
    \lb u^\sM g_{a\sM\sN} + u^\sM \lie_{X_a} g_{r\sM\sN} \rb \nn\\
  &\qquad
    +  \lb a_2 \epsilon^{\sN\sP\sR\sS\sT} V_\sP u_\sR \dow_\sS u_\sT
    + a_1 \epsilon^{\sN\sP\sR\sS\sT} V_\sP u_\sR \dow_\sS V_{\sT}
    \rb
    \lb V^\sM g_{a\sM\sN} + \lie_{X_a} V_{\sN} \rb \nn\\
  &\qquad
    + \frac{i\kB T}{4}\lb 2\eta \Delta^{\sM(\sR} \Delta^{\sS)\sN}
    + \lb \zeta - {\textstyle\frac{2}{d}}\eta \rb \Delta^{\sM\sN}\Delta^{\sR\sS} \rb
    \lb g_{a\sM\sN} + \lie_{X_a}g_{r\sM\sN} \rb
    \lb g_{a\sR\sS} + \lie_{(X_a+i\beta)}g_{r\sR\sS} \rb \nn\\
  &\qquad
    + i\kB T^2\kappa\, \Delta^{\sM\sR} \lb g_{a\sM\sN}V^\sN + \lie_{X_a} V_\sM \rb
    \lb g_{a\sR\sS}V^\sS + \lie_{(X_a+i\beta)}V_\sR \rb, \nn\\
  &= \half T^{\sM\sN} \lb g_{a\sM\sN} + \lie_{X_a} g_{r\sM\sN} \rb \nn\\
  &\qquad
    + \frac{i\kB T}{4}\lb 2\eta \Delta^{\sM(\sR} \Delta^{\sS)\sN}
    + \lb \zeta - {\textstyle\frac{2}{d}}\eta \rb \Delta^{\sM\sN}\Delta^{\sR\sS} \rb
    \lb g_{a\sM\sN} + \lie_{X_a}g_{r\sM\sN} \rb
    \lb g_{a\sR\sS} + \lie_{X_a}g_{r\sR\sS} \rb \nn\\
  &\qquad
    + i\kB T^2\kappa\, \Delta^{\sM\sR} \lb g_{a\sM\sN}V^\sN + \lie_{X_a} V_\sM \rb
    \lb g_{a\sR\sS}V^\sS + \lie_{X_a}V_\sR \rb,
    \label{eq:1der-action} 
\end{align}
where we have expanded the definitions of $G_{r\sM\sN}$ and $G_{a\sM\sN}$ in the
statistical limit using \cref{eq:semiclassical-expansion-null}. Here
$\lie_{X_a}$ denotes a Lie derivative along $X_a^\sM$ and $\lie_{(X_a+i\beta)}$
along $X^\sM_a + i\beta^\sM$. In the second equality, we have introduced the
classical constitutive relations $T^{\sM\sN}$ in the thermodynamic mass frame
given in \cref{eq:null-consti-gen-frame,eq:null-consti-cov-thermo}. Varying the
action with respect to $g_{a\sM\sN}$, in a configuration with
$X_a^\sM = g_{a\sM\sN} = 0$, leads to the classical constitutive relations in
the thermodynamic mass frame, given in
\cref{eq:null-consti-gen-frame,eq:null-consti-cov-thermo}, while extremising
with respect to $X^\sM_a$ leads to the associated classical conservation
equations \eqref{eq:null-cons}. The respective expressions in the Newton-Cartan
language are presented in \cref{sec:NC-formulation}.

Turning off the background fields, i.e. setting $g_{r\sM\sN} = \eta_{\sM\sN}$,
$g_{a\sM\sN} = 0$, choosing the coordinates $(x^\sM) = (x^-,t,x^i)$ such that
$V^\sM = \delta^\sM_-$, and identifying $\varphi_a = - X_a^-$, the effective
action \eqref{eq:1der-action} results in 
\begin{align}
  \mathcal{L}
  &= \rho \lb \dow_t + u^i \dow_i \rb \varphi_a
    - \lb \varepsilon + {\textstyle\half} \rho\, \vec u^2
    + a_0 \epsilon^{ijk} u_i \dow_j u_k \rb
    \lb \dow_t + u^i \dow_i \rb X^t_{a} \nn\\
  &\qquad
    + \lb \rho\, u^i + a_0\, \epsilon^{ijk} \dow_j u_k \rb
    \lb \dow_t + u^k \dow_k \rb X_{ai}
    + p \lb \dow_i X_{a}^i - u^{i} \dow_i X^t_{a} \rb \nn\\
  &\qquad
    + a_0\, \epsilon^{ijk} \dow_j u_k
    \lb \dow_i \varphi_{a}
    + u^l \dow_i X_{al} \rb
    - \lb a_2 + a_0\half\vec u^2\rb \epsilon^{ijk}\dow_j u_k\, \dow_i X^t_a \nn\\
  &\qquad
    + 2i\kB T \eta \lb \dow_{(i}X_{aj)} - u_{(i} \dow_{j)}X^t_a \rb
    \lb 
    \dow^{(i}X_{a}^{j)} - u^{(i} \dow^{j)}X^t_a + i/T\, \dow^{(i} u^{j)} \rb \nn\\
  &\qquad
    + i\kB T \lb \zeta - {\textstyle\frac{2}{d}}\eta\rb
    \lb \dow_{i}X_{a}^i - u^i \dow_{i}X^t_a \rb
    \lb 
    \dow_k X_{a}^k - u^k \dow_k X^t_a + i/T\, \dow_k u^k \rb \nn\\
  &\qquad
    + i \kB \kappa\, \dow_i X^t_a \lb T^2 \dow^i X^t_a - i \dow^i T \rb \nn\\
  &= \rho^t \dow_t \varphi_a
    + \pi^i \dow_i \varphi_a
    - \epsilon^t \dow_t  X^t_{a}
    - \epsilon^i \dow_i  X^t_{a}
    + \pi^i \dow_t X_{ai} 
    + \tau^{ij} \dow_i X_{aj} \nn\\
  &\qquad
    + 2i\kB T \eta \Big( \dow_{(i}X_{aj)} - u_{(i} \dow_{j)}X^t_a \Big)
    \Big( \dow^{(i}X_{a}^{j)} - u^{(i} \dow^{j)}X^t_a \Big)
    + i\kB T \lb \zeta - {\textstyle\frac{2}{d}}\eta\rb
    \lb \dow_k X_{a}^k - u^k \dow_k X^t_a \rb^2 \nn\\
  &\qquad
    + i \kB T^2 \kappa\, \dow_i X^t_a \dow^i X^t_a.
    \label{eq:1der-action-noback}
\end{align}
Again, in the second step, we have substituted the thermodynamic mass frame
constitutive relations from \cref{eq:thermo-mass-frame-flat}. All the
parity-violating terms get absorbed within the constitutive relations.
Non-trivial contributions only arise due to the dissipative terms. This action
is merely the generalisation of the effective action
\eqref{eq:1der-action-flat-intro} presented in the introduction, to include
parity-violating effects. Note that the effective action
\eqref{eq:1der-action-noback} is applicable at the full non-linear level; in the
next subsection we inspect this in a linearised limit.

\subsection{Linearised stochastic fluctuations}
\label{sec:linear-Gal}

\subsubsection{Perturbative expansion in fluctuations}

Let us consider a Galilean fluid at rest coupled to a flat background. In the
EFT terms, the fluid is described by the effective action
\eqref{eq:1der-action-noback}, with the equilibrium state
\begin{equation}
  \tau = t, \qquad
  \sigma^i = x^i, \qquad
  \varphi_r = 0, \qquad
  X^t_a = X^i_a = \varphi_a = 0.
\end{equation}
The hydrodynamic fields in this state are given using
\cref{eq:SK-hydro-fields-null} as $\beta^\sM = \delta^\sM_\sA\bbbeta^\sA$. We
can choose the thermal reference vector to be
$\bbbeta^\sA = \beta_0(\delta^\sM_t- \mu_0\delta^\sM_-)$, which implies that
$T = T_0$, $\mu = \mu_0$, and $u^i = 0$ in equilibrium. We would like to expand
the effective action perturbatively in fluctuations around the equilibrium
state. In the ``$a$'' sector we can work with the fields $X^t_a$, $X^i_a$,
$\varphi_a$ directly, while in the ``$r$'' sector it is instead convenient to
work with the fluctuations in the physical conserved densities:
$\delta\rho = \rho^t - \rho_0$, $\delta\epsilon = \epsilon^t - \varepsilon_0$,
$\pi^i$ given in \cref{eq:thermo-mass-frame-flat}. These are related to
fluctuations of hydrodynamic variables $\delta T = T - T_0$,
$\delta\mu = \mu - \mu_0$, and $u^i$ through
\begin{align}
  u^i
  &= \frac{1}{\rho_0}\pi^i
    - \frac{2K_0T_0}{\rho^2_0} \epsilon^{ijk} \dow_j\pi_k \nn\\
  &\quad
    - \frac{1}{\rho^2_0}\pi^i\delta\rho
    - \frac{2K_0T_0}{\rho_0^3} \epsilon^{ijk} \pi_j \dow_k \delta\rho
    - \lb\frac{\dow}{\dow\rho}\frac{2K_0T}{\rho^2}\rb_0
    \epsilon^{ijk} \dow_j\pi_k \delta\rho
    - \lb\frac{2K_0}{\rho^2} \frac{\dow T}{\dow\varepsilon}\rb_0
    \epsilon^{ijk} \dow_j\pi_k \delta \epsilon
    + \ldots, \nn\\
  \rho(T,\mu)
  &= \rho_0 + \delta\rho, \qquad
    \varepsilon(T,\mu) = \varepsilon_0 + \delta\epsilon
    - \frac{1}{2\rho_0} \vec\pi^2 + \ldots.
    \label{eq:hydro-expansion}
\end{align}
These expressions are valid until quadratic order in fields and leading order in
derivatives. The subscript ``$0$'' on various coefficients represents evaluation
on the equilibrium configuration $T=T_0$, $\mu=\mu_0$. The thermodynamic
derivatives have been performed in the microcanonical ensemble controlled by
($\varepsilon$, $\rho$). On the other hand, the hydrodynamic fields are related
to $\delta\sigma^\tau = \sigma^\tau - t$, $\delta\sigma^i = \sigma^i - x^i$, and
$\varphi_r$ through \cref{eq:SK-conventional-hydro-fields-null} leading to
\begin{gather}
  \delta T = T_0 \dow_t\delta\sigma^\tau
  - T_0 \dow_t\delta\sigma^k \dow_k\delta\sigma^\tau + \ldots, \qquad
  u^i = - \dow_t\delta\sigma^i + \dow_t\delta\sigma^k \dow_k\delta\sigma^i
  \ldots,
  \nn\\
  \delta\mu = 
  \dow_t\lb \varphi_r + \mu_0\delta\sigma^\tau \rb
  - \dow_t\delta\sigma^k\dow_k \lb \varphi_r + \mu_0\delta\sigma^\tau \rb
  + \ldots.
  \label{eq:Tdef-linear}
\end{gather}

We can substitute \cref{eq:hydro-expansion} into the constitutive relations for
$\epsilon^i$ and $\tau^{ij}$ in \cref{eq:thermo-mass-frame-flat} and truncate
the expressions to leading order in derivatives and quadratic order in
fluctuations to find\footnote{Tip: it is actually easier to start with the mass
  frame constitutive relations, where $u^i$ is just $\pi^i/\rho^t$.}
\begin{align}
  \epsilon^i
  &= \frac{\varepsilon_0 + p_0}{\rho_0} \pi^i
    - \lb \kappa\frac{\dow T}{\dow\rho} \rb_0 \dow^i \rho
    - \lb \kappa\frac{\dow T}{\dow\varepsilon} \rb_0 \dow^i \varepsilon
    + \lb\frac{\xi_\Omega}{\rho}\rb_0 \epsilon^{ijk} \dow_j \pi_k \nn\\
  &\quad
    + \frac{1}{\rho_0}
    \lb 1 + \frac{\dow p}{\dow\varepsilon} \rb_0 \pi^i\delta\epsilon
    + \frac{1}{\rho_0} \lb \frac{\dow p}{\dow \rho}
    - \frac{\epsilon + p}{\rho} \rb_0 \pi^i\delta\rho \nn\\
  &\quad
    - \lB \frac{\dow}{\dow\rho}\lb \kappa\frac{\dow T}{\dow\rho} \rb \rB_0
    \delta\rho \dow^i \rho
    - \lB \frac{\dow}{\dow\varepsilon}\lb\kappa\frac{\dow T}{\dow\rho}\rb\rB_0
    \delta\varepsilon \dow^i \rho  \nn\\
  &\quad
    - \lB \frac{\dow}{\dow\rho}\lb\kappa\frac{\dow T}{\dow\varepsilon}\rb\rB_0
    \delta\rho\dow^i \delta\varepsilon
    - \lB \frac{\dow}{\dow\varepsilon}\lb\kappa
    \frac{\dow T}{\dow\varepsilon}\rb\rB_0
    \delta\varepsilon\dow^i \delta\varepsilon
    \nn\\
  &\quad
    + \half \lb \frac{\kappa}{\rho}
    \frac{\dow T}{\dow\varepsilon} \rb_0 \dow^i\vec\pi^2
    - \frac{\eta_0}{\rho^2_0}
    \lb \half \dow^i \vec\pi^2 + \pi^k \dow_k \pi^i \rb
    - \frac{\zeta_0 - \frac2d \eta_0}{\rho_0^2} \pi^i \dow_k \pi^k \nn\\
  &\quad
    + \lb \frac{\dow}{\dow\rho} \frac{\xi_\Omega}{\rho} \rb_0
    \epsilon^{ijk} \dow_j \pi_k \delta\rho
    + \lb \frac{\dow}{\dow\varepsilon} \frac{\xi_\Omega}{\rho}\rb_0
    \epsilon^{ijk} \dow_j \pi_k \delta \epsilon
    + \bfrac{\xi_\Omega}{\rho^2}_0 \epsilon^{ijk}\pi_j \dow_k \delta\rho
    + \ldots, \nn\\
  \tau^{ij}
  &=
    p_0 \delta^{ij}
    + \lb \frac{\dow p}{\dow\rho}\rb_0\delta^{ij} \delta\rho
    + \lb \frac{\dow p}{\dow\varepsilon}\rb_0\delta^{ij} \delta\epsilon
    - \frac{\eta_0}{\rho_0} 2\dow^{(i}\pi^{j)}
    - \frac{\zeta_0 - \frac2d \eta_0}{\rho_0} \delta^{ij}\dow_k\pi^k \nn\\
  &\quad
    + \frac{1}{\rho_0} \pi^i \pi^j
    - \frac{1}{2\rho_0} \lb \frac{\dow p}{\dow\varepsilon}\rb_0\delta^{ij}
    \vec\pi^2
    + \half \lb \frac{\dow^2 p}{\dow\rho^2}\rb_0\delta^{ij} \delta\rho^2
    + \lb \frac{\dow^2 p}{\dow\rho\delta\varepsilon}\rb_0\delta^{ij}
    \delta\rho\delta\epsilon
    + \half \lb \frac{\dow^2 p}{\dow\varepsilon^2}\rb_0\delta^{ij} \delta\epsilon^2
    \nn\\
  &\qquad
    - \lb \frac{\dow}{\dow\rho} \frac{\eta}{\rho} \rb_0 2\dow^{(i}\pi^{j)}\delta\rho
    - \lb \frac{\dow}{\dow\varepsilon} \frac{\eta}{\rho} \rb_0
    2\dow^{(i}\pi^{j)}\delta\epsilon
    + \frac{\eta_0}{\rho^2_0} 2\pi^{(i}\dow^{j)} \delta\rho
    \nn\\
  &\quad
    - \lb\frac{\dow}{\dow\rho}\frac{\zeta - \frac2d \eta}{\rho}\rb_0
    \,\delta^{ij}\dow_k\pi^k \delta\rho
    - \lb\frac{\dow}{\dow\varepsilon}\frac{\zeta - \frac2d \eta}{\rho}\rb_0
    \,\delta^{ij}\dow_k\pi^k \delta\epsilon
    + \frac{\zeta_0- \frac2d \eta_0}{\rho_0^2} \,\delta^{ij} \pi^k \dow_k
    \delta\rho
    + \ldots.
    \label{eq:consti-expansion}
\end{align}
The parity-odd transport coefficient $\xi_\Omega$ has been defined in
\cref{eq:transconstraints}. If required, this expansion can be extended to
higher orders in fluctuations.

\subsubsection{Free theory}

We can substitute the expansion of constitutive relations from
\cref{eq:consti-expansion} into \cref{eq:1der-action-noback} and work out the
hydrodynamic effective action order-by-order in fluctuations. Note that the
effective action is always one order higher in derivatives and fluctuations
compared to the constitutive relations. For example, truncated to quadratic
order in fields, we find the ``free'' Lagrangian 
\begin{align}
  \mathcal{L}_{\text{free}}
  &= - \varphi_a \lb \dow_t \delta\rho + \dow_i\pi^i \rb
    + X^t_{a} \lb \dow_t \delta \epsilon
    - \kappa \frac{\dow T}{\dow \varepsilon} \dow^i\dow_i \delta \epsilon
    - \kappa \frac{\dow T}{\dow\rho} \dow^i \dow_i \delta\rho
    \rb 
    + \frac{\varepsilon+p}{\rho} X^t_a \dow_i \pi^i
    \nn\\
  &\qquad
    - X^i_{a} \lb
    \dow_t\pi_i
    + \frac{\dow p}{\dow\rho} \dow_{i} \delta\rho
    + \frac{\dow p}{\dow \varepsilon} \dow_{i} \delta \epsilon
    - \frac{\eta}{\rho}\dow^k\dow_k \pi_i
    - \frac{\zeta + {\textstyle\frac{d-2}{d}}\eta}{\rho}
    \dow_j \dow_i \pi^i
    \rb \nn\\
  &\qquad
    + i\kB T^2\kappa\, \dow^i X^t_a\dow_i X^t_a
    + i\kB T\eta\, \dow_{i}X_{aj} \dow^{i}X^{j}_a
    + i\kB T\lb \zeta + {\textstyle\frac{d-2}{d}}\eta\rb (\dow_k X^k_a)^2,
    \label{eq:free-action}
\end{align}
where we have ignored certain total derivative terms. We have also dropped the
subscript ``$0$'' from the coefficients for clarity. We can write this out in
momentum space
\begin{align}
  \mathcal{L}
  &=
  \begin{pmatrix}
    \varphi_a(-p) \\ - X^t_a(-p) \\ X^j_a(-p)
  \end{pmatrix}^\rmT
  \begin{pmatrix}
    i\omega & 0
    & - ik_i \\
    - \kappa\, \dow T/\dow\rho\, k^2
    & i\omega - \kappa\, \dow T/\dow\varepsilon\, k^2
    & - (\varepsilon+p)/\rho\, ik_i \\
    - \dow p/\dow\rho\, ik_j
    & - \dow p/\dow\varepsilon\, ik_j
    & i\omega \delta_{ij} - \eta/\rho\, \delta_{ij} k^2
    - (\zeta - \frac2d \eta)/\rho\, k_i k_j
  \end{pmatrix}
  \begin{pmatrix}
    \delta\rho(p) \\ \delta\epsilon(p) \\ \pi^i(p)
  \end{pmatrix} \nn\\
  &\qquad
    + \frac{i}{2}
    \begin{pmatrix}
      \varphi_a(-p) \\ - X^t_a(-p) \\ X^j_a(-p)
  \end{pmatrix}^\rmT
  \begin{pmatrix}
    0 & 0 & 0 \\
    0 & 2\kB T^2\kappa k^2 & 0 \\
    0 & 0
    & 2\kB T\eta \delta_{ij} k^2
    + 2\kB T(\zeta - \frac2d \eta) k_i k_j
  \end{pmatrix}
  \begin{pmatrix}
    \varphi_a(p) \\ - X^t_a(p) \\ X^i_a(p)
  \end{pmatrix} \nn\\
  &= \varphi^I_{a}(-p) K_{I}^{~J}(p) \mathcal{O}_J(p)
    + \frac{i}{2} \varphi^I_{a}(-p) G_{IJ}\varphi^J_{a}(p).
\end{align}
Here $p=(\omega, k)$ in the arguments collectively denotes frequency and
momentum. In the second step, we have collectively denoted the fluctuations in
conserved operators by $\mathcal{O}_I = (\delta\rho, \delta\epsilon, \pi_i)$ and
the auxiliary fields by $\varphi^I_{a} = (\varphi_a,-X^t_a,X^i_a)$. From here one
can read out the momentum-space free propagators
\begin{gather}
  \langle \mathcal{O}_I(p)\varphi^J_{a}(-p) \rangle_0
  = i(K^{-1})^{~J}_{I}(p), \qquad
  \langle\varphi^I_{a}(p)\mathcal{O}_J(-p) \rangle_0
  = i(K^{-1*})^{~I}_{J}(p), \nn\\
  \langle \mathcal{O}_I(p)\mathcal{O}_{J}(-p) \rangle_0
  = (K^{-1}G K^{-\rmT *})_{IJ}(p), \qquad
  \langle\varphi^I_{a}(p)\varphi^J_{a}(-p) \rangle_0 = 0.
\end{gather}

The poles of the propagators are controlled by $\det K$. As expected, we find a
total of $(d+2)$ modes: a pair of sound mode, a longitudinal charge diffusion
mode, and $d-1$ copies of transverse shear diffusion mode. In small momentum
limit, they are given by
\begin{equation}
  \omega = \pm v_s k - \frac{i}{2}\Gamma_s k^2, \qquad
  \omega = -iD_\parallel k^2, \qquad
  \omega = -iD_\perp k^2,
\end{equation}
where the speed of sound, attenuation constant, and diffusion constants are
respectively given
as\footnote{$\frac{\dow T}{\dow\rho} + \frac{\varepsilon+p}{\rho} \frac{\dow
    T}{\dow\varepsilon} = \frac{T}{\rho} \frac{\dow p}{\dow\varepsilon}$.}
\begin{gather}
  v_s^2 = \frac{\dow p}{\dow\rho}
  + \frac{\varepsilon+p}{\rho} \frac{\dow p}{\dow \varepsilon}, \qquad
  \Gamma_s = \frac{\zeta + 2\frac{d-1}{d}\eta}{\rho}
  + \frac{T\kappa}{v_s^2\rho} \lb\frac{\dow p}{\dow\varepsilon}\rb^2, \nn\\
  D_\parallel = \frac{\kappa}{v_s^2}
  \lb \frac{\dow p}{\dow\rho} \frac{\dow T}{\dow\varepsilon}
  - \frac{\dow p}{\dow\varepsilon} \frac{\dow T}{\dow \rho}
  \rb \qquad
  D_\perp = \frac{\eta}{\rho}.
\end{gather}

One can see that the propagators $\langle \mathcal{O}_I\varphi^J_{a} \rangle_0$
are purely retarded, while $\langle\varphi^I_{a}\mathcal{O}_{J} \rangle_0$ are
purely advanced. Coupling the theory to background sources, one can indeed
verify that they are related to retarded and advanced propagators of the
hydrodynamic densities $\mathcal{O}_I$ respectively, while
$\langle \mathcal{O}_I\mathcal{O}_J \rangle$ are the symmetric propagators. To
wit
\begin{gather}
  G^{\rmR,0}_{\mathcal{O}_I\mathcal{O}_J}(p)
  = -i\omega (K^{-1})^{~K}_{I}(p)\, \chi_{KJ}, \qquad
  G^{\rmA,0}_{\mathcal{O}_I\mathcal{O}_J}(p)
  = i\omega (K^{-1*})^{~K}_{J}(p)\, \chi_{KI}, \nn\\
  G^{\rmS,0}_{\mathcal{O}_I\mathcal{O}_J}(p)
  = (K^{-1}G K^{-\rmT *})_{IJ}(p),
\end{gather}
up to contact terms. Here $\chi_{IJ}$ is the susceptibility matrix
\begin{equation}
  \chi_{IJ} =
  \begin{pmatrix}
    \dow\rho/\dow\mu
    & \dow\varepsilon/\dow\mu & 0 \\
    \dow\varepsilon/\dow\mu 
    & T\, \dow\epsilon/\dow T + \mu\, \dow\epsilon/\dow\mu & 0 \\
    0 & 0 & \rho
  \end{pmatrix}
  =
  \begin{pmatrix}
    T\dow (\mu/T)/\dow\rho
    & 1/T\, \dow T/\dow\rho  & 0 \\
    1/T\, \dow T/\dow\rho & 1/T\, \dow T/\dow\varepsilon & 0 \\
    0 & 0 & 1/\rho
  \end{pmatrix}^{-1}.
\end{equation}
It can be checked that the retarded function satisfies the Onsager's relations:
$G^{\rmR,0}_{\mathcal{O}_I\mathcal{O}_J} =
G^{\rmR,0}_{\mathcal{O}_J\mathcal{O}_I}$. In addition, the three correlations
functions are related by
\begin{gather}
  G^{\rmA,0}_{\mathcal{O}_I\mathcal{O}_J}
  = (G^{\rmR,0}_{\mathcal{O}_I\mathcal{O}_J})^*, \qquad
  G^{\rmS,0}_{\mathcal{O}_I\mathcal{O}_J}
  = \frac{2\kB T}{\omega} \mathrm{Im}\, G^{\rmR,0}_{\mathcal{O}_I\mathcal{O}_J}.
\end{gather}
The second of these is the well-known two-point fluctuation-dissipation theorem.

\subsubsection{Interactions}

To account for interactions between hydrodynamic and stochastic degrees of
freedom, we need to expand the effective action to higher order in
fluctuations. Substituting \cref{eq:consti-expansion} into
\cref{eq:1der-action-noback}, we find leading order three-point interaction
terms coming from ideal hydrodynamics
\begin{equation}
  \mathcal{L}^{\text{1der}}_{\text{int}}
  = \frac{1}{\rho} \pi^i \pi^j \dow_i X_{aj}
    - \lb \alpha_{\rho} \delta\rho + \alpha_{\epsilon} \delta \epsilon \rb
    \pi^i \dow_i X^t_a
    + \lb\half\beta_{\rho} \delta\rho^2
    + \half\beta_{\epsilon} \delta\epsilon^2
    + \beta_{\rho\epsilon} \delta\epsilon\delta\rho
    + \half\beta_\pi \vec\pi^2
    \rb \dow_{i}X^i_{a},
    \label{eq:1der-3pt}
\end{equation}
where we have defined the couplings 
\begin{gather}
  \alpha_{\rho}
  = \frac1\rho\lb \frac{\dow p}{\dow\rho} - \frac{\varepsilon+p}{\rho} \rb, \qquad
  \alpha_{\epsilon}
  = \frac1\rho\lb 1 + \frac{\dow p}{\dow \varepsilon} \rb, \nn\\
  \beta_\rho = \frac{\dow^2 p}{\dow\rho^2}, \qquad
  \beta_{\epsilon} = \frac{\dow^2 p}{\dow\varepsilon^2}, \qquad
  \beta_{\rho\epsilon} = \frac{\dow^2 p}{\dow\rho\dow\varepsilon}, \qquad
  \beta_\pi = -\frac{1}{\rho}\frac{\dow p}{\dow\varepsilon}.
\end{gather}
Similarly, we can work out two-derivative three-point interactions due to
one-derivative corrections to the constitutive relations. We have unitary
interactions involving two hydrodynamic and one stochastic fields like above
\begin{align}
  \mathcal{L}_{\text{int}}^{\text{2der,rra}}
    &= 
    \frac{\eta}{\rho^2} \pi^i \dow_i \pi^j \dow_j X^t_a
    - \frac{\eta}{\rho^2}
      \delta\rho\,\pi_i \dow^k \dow_k X_{a}^i
      - \lb \theta^\eta_\rho \delta\rho + \theta^\eta_\epsilon \delta\epsilon \rb
      \lb \dow_{j} \pi_{i} \dow^{i}X^j_{a} + \dow_i\pi_j \dow^i X_{a}^j \rb
      \nn\\
    &\qquad
      + \frac{\zeta - {\textstyle\frac{2}{d}}\eta}{\rho^2}
      \dow_k \pi^k\, \pi^{i}\dow_iX^t_a
      - \frac{\zeta + \frac{d-2}{d}\eta}{\rho^2}
      \delta\rho\, \pi^{i} \dow_{i} \dow_j X_{a}^j
      - \lb \theta^\zeta_\rho \delta\rho + \theta^\zeta_\epsilon \delta\epsilon \rb
    \dow_k \pi^k \dow_{i}X^i_{a}
 \nn\\
    &\qquad
      - \lb
      \half \lambda_\rho \delta\rho^2
      + \half \lambda_\epsilon \delta\epsilon^2 
      + \lambda_{\rho\epsilon} \delta\rho\,\delta\epsilon
      + \half \lambda_\pi \vec\pi^2
      \rb \dow^k \dow_k X^t_a
      - \lambda'_{\rho\epsilon} \lb \delta\rho\, \dow^i \delta\epsilon
      - \delta\epsilon\, \dow^i \delta\rho\rb \pi^{i}\dow_iX^t_a \nn\\
    &\qquad
      - \lb \theta^\xi_\rho \delta\rho
      + \theta^\xi_\epsilon \delta \epsilon \rb
      \epsilon^{ijk} \dow_j \pi_k \dow_i X^t_a,
      \label{eq:2der-3pt-rra}
\end{align}
with couplings 
\begin{gather}
  \theta^\eta_\rho
  = \frac1\rho\frac{\dow\eta}{\dow\rho}, \qquad
  \theta^\eta_\epsilon
  = \frac1\rho\frac{\dow\eta}{\dow\varepsilon}, \qquad
  \theta^\zeta_\rho
  = \frac1\rho\frac{\dow(\zeta - \frac{2}{d}\eta)}{\dow\rho}, \qquad
  \theta^\zeta_\epsilon
  = \frac1\rho\frac{\dow(\zeta - \frac{2}{d}\eta)}{\dow\varepsilon}, \nn\\
    \lambda_\rho
  = \frac{\dow}{\dow\rho}
  \lb \kappa\, \frac{\dow T}{\dow\rho} \rb, \quad
  \lambda_\epsilon
  = \frac{\dow}{\dow\varepsilon}
  \lb \kappa\, \frac{\dow T}{\dow\varepsilon} \rb, \quad
  \lambda_{\rho\epsilon}
  = \half\lB \frac{\dow}{\dow\varepsilon}
  \lb \kappa\, \frac{\dow T}{\dow\rho} \rb
  + \frac{\dow}{\dow\rho}
  \lb \kappa\, \frac{\dow T}{\dow\varepsilon} \rb \rB, \nn\\
  \lambda'_{\rho\epsilon}
  = \half\lb \frac{\dow\kappa}{\dow\varepsilon} \frac{\dow T}{\dow\rho}
  - \frac{\dow\kappa}{\dow\rho} \frac{\dow T}{\dow\varepsilon} \rb,
  \qquad
  \lambda_\pi = \frac{1}{\rho} \lb \frac{\eta}{\rho}
  - \kappa\, \frac{\dow T}{\dow\varepsilon} \rb, \qquad
  \theta^\xi_\rho
  = \frac1\rho\frac{\dow\xi_\Omega}{\dow\rho}, \qquad
  \theta^\xi_\epsilon
  = \frac1\rho\frac{\dow\xi_\Omega}{\dow\varepsilon}.
\end{gather}
Secondly, we have non-unitary terms involving one hydrodynamic and two
stochastic fields
\begin{align}
  \mathcal{L}_{\text{int}}^{\text{2der,raa}}
  &= i\kB T^2 \lb \frac{\tilde\lambda_\rho}{\dow T/\dow\rho} \delta\rho
    + \frac{\tilde\lambda_\epsilon}{\dow T/\dow\varepsilon}
    \delta\epsilon \rb \dow^i X^t_a\dow_i X^t_a
     \nn\\
  &\qquad
    + i\kB T\rho \lb \tilde\theta^\eta_\rho\delta\rho
    + \tilde\theta^\eta_\epsilon\delta\epsilon \rb
    \lb \dow_{i}X_{aj} \dow^{i}X^{j}_a + \dow_{i}X_{aj} \dow^{j}X^{i}_a \rb
    - \frac{2iT\kB\eta}{\rho}
    \lb \pi^i\dow^jX^t_a + \pi^j\dow^iX^t_a \rb \dow_{i}X_{aj}
    \nn\\
  &\qquad
    + i\kB T\rho \lb \tilde\theta^\zeta_\rho\delta\rho
    + \tilde\theta^\zeta_\epsilon \delta\epsilon \rb
  (\dow_k X^k_a)^2
  - \frac{2i\kB T(\zeta - {\textstyle\frac{2}{d}}\eta)}{\rho}
    \pi^{k}\dow_{k}X^t_a \dow_{i}X^i_{a},
    \label{eq:2der-3pt-raa}
\end{align}
with coefficients
\begin{gather}
  \tilde\theta^\eta_\rho = \frac{1}{T\rho} \frac{\dow(T\eta)}{\dow\rho}, \qquad
  \tilde\theta^\eta_\epsilon = \frac{1}{T\rho}
  \frac{\dow(T\eta)}{\dow\varepsilon}, \qquad
  \tilde\theta^\zeta_\rho = \frac{1}{T\rho}
  \frac{\dow(T\zeta - \frac{2}{d}T\eta)}{\dow\rho}, \qquad
  \tilde\theta^\zeta_\epsilon = \frac{1}{T\rho}
  \frac{\dow(T\zeta - \frac{2}{d}T\eta)}{\dow\varepsilon}, \nn\\
  \tilde\lambda_\rho
  = \frac{1}{T^2} \frac{\dow T}{\dow\rho}
  \frac{\dow(T^2\kappa)}{\dow\rho}, \qquad
  \tilde\lambda_\epsilon
  = \frac{1}{T^2} \frac{\dow T}{\dow\varepsilon}
  \frac{\dow(T^2\kappa)}{\dow\varepsilon}.
\end{gather}
Note that not all the couplings appearing above are independent. In total, there
can only be 15 couplings at this order in the parity even sector: pressure $p$,
two thermodynamic first derivatives of $p$, and three second derivatives, three
dissipative coefficients, and their 2 thermodynamic derivatives each. In the
parity violating sector in $d=3$, there are just constants $K_0$ and $K_2$;
$K_1$ does not contribute in the absence of background fields. Most couplings
appearing above are related due to the underlying Galilean symmetry of the
effective theory. However, there are certain relations between unitary and
non-unitary parts of the interaction Lagrangian; these can be understood as a
realisation of three-point fluctuation dissipation theorem. If required, we can
work out higher point interactions in a similar manner.

Few comments are in order regarding the validity of the perturbative expansion
itself, as we do not seem to have a small parameter controlling the strength of
interactions. Since we are interested in a low-energy effective theory, we can
treat momentum itself as the small parameter for the purposes of loop
counting. In hydrodynamics, due to the presence of the sound mode, we typically
take the frequency to scale as $\omega\sim k$ and $T_0 \sim 1$. Since the
Lagrangian must scale as $\mathcal{L} \sim \omega k^d \sim k^{d+1}$,
\cref{eq:free-action} implies that all the fields $\delta\rho$,
$\delta\epsilon$, $\pi_i$, $\varphi_a$, $X_a^t$, $X_a^i$ scale as $k^{d/2}$. As
a consequence, all the leading derivative three-point couplings in
\cref{eq:1der-3pt} are irrelevant by $k^{-d/2}$ in the RG sense, while the
subleading couplings in \cref{eq:2der-3pt-rra,eq:2der-3pt-raa} are irrelevant by
$k^{-d/2-1}$. Similarly, all higher derivative and non-linear couplings are
further suppressed within the effective theory. Care should be taken in lower
number of dimensions; in $d=2$ spatial dimensions the interaction terms in
\cref{eq:1der-3pt} are as important as the dissipative corrections in the free
Lagrangian \eqref{eq:free-action}, while in $d=1$ they become more
important. The perturbative expansion is still valid in these dimensions, but
the results are qualitatively different; see~\cite{Chen-Lin:2018kfl} for an
example in the simpler case of energy diffusion model.

\subsection{Incompressible and diffusion limits}

As is evident from our discussion above, the full interacting theory of
stochastic hydrodynamics is quite rich. For practical applications, therefore,
it is often convenient to work in certain simplifying limits. Most popular of
these is the incompressible limit where one ignores the density fluctuations
$\delta\rho$. Inspecting the full non-linear effective action
\eqref{eq:1der-action-noback}, we can infer that the field $\varphi_a$ does not
appear in the interactions at all and merely becomes a Lagrange multiplier for
freezing the longitudinal momentum modes $\dow_i\pi^i = 0$. Correspondingly, we
also choose $\dow_iX^i_a = 0$. The resulting non-linear effective action is
given as
\begin{align}
  \mathcal{L}^{\text{incom}}
  &=
    - \epsilon^t \dow_t  X^t_{a}
    - \epsilon^i \dow_i  X^t_{a}
    + \pi_i \dow_t X_{ai}
    + \tau^{ij} \dow_i X_{aj} \nn\\
  &\qquad
    + 2i\kB T \eta \Big( \dow_{(i}X_{aj)} - u_{(i} \dow_{j)}X^t_a \Big)
    \Big( \dow^{(i}X_{a}^{j)} - u^{(i} \dow^{j)}X^t_a \Big)
    + i \kB T^2 \kappa\, \dow_i X^t_a \dow^i X^t_a.
\end{align}
We can apply these constraints to the linearised effective action in
\cref{sec:linear-Gal} and obtain
\begin{align}
  \mathcal{L}^{\text{incom}}_{\text{free}}
  &= X^t_{a} \lb \dow_t - \kappa \frac{\dow T}{\dow \varepsilon} \dow^i\dow_i
    \rb \delta \epsilon
    + i\kB T^2\kappa\, \dow^i X^t_a\dow_i X^t_a
    - X^i_{a} \lb \dow_t - \frac{\eta}{\rho}\dow^k\dow_k \rb \pi_i
    + i\kB T\eta\, \dow_{i}X_{aj} \dow^{i}X^{j}_a, \nn\\
  \mathcal{L}^{\text{incom}}_{\text{int}}
  &= \frac{1}{\rho} \pi^i \pi^j \dow_i X_{aj}
    - \alpha_{\epsilon} \delta \epsilon\,\pi^i \dow_i X^t_a
    + \frac{\eta}{\rho^2} \pi^i \dow_i \pi^j \dow_j X^t_a
    - \theta^\eta_\epsilon \delta\epsilon
      \lb \dow_{j} \pi_{i} \dow^{i}X^j_{a} + \dow_i\pi_j \dow^i X_{a}^j \rb \nn\\
    &\qquad
      - \half \lb \lambda_\epsilon \delta\epsilon^2
      + \lambda_\pi \vec\pi^2 \rb \dow^k \dow_k X^t_a
      - \theta^\xi_\epsilon\, \epsilon^{ijk} \dow_j \pi_k \dow_i X^t_a 
      + \frac{i\kB T^2 \tilde\lambda_\epsilon}{\dow T/\dow\varepsilon}
      \delta\epsilon\, \dow^i X^t_a\dow_i X^t_a
      \nn\\
  &\qquad
    + i\kB T\rho\,\tilde\theta^\eta_\epsilon\delta\epsilon
    \lb \dow_{i}X_{aj} \dow^{i}X^{j}_a + \dow_{i}X_{aj} \dow^{j}X^{i}_a \rb
    - \frac{2iT\kB\eta}{\rho}
    \lb \pi^i\dow^jX^t_a + \pi^j\dow^iX^t_a \rb \dow_{i}X_{aj}.
\end{align}
At the free level, the theory decouples into transverse and longitudinal
modes. There is no sound. We still keep the transverse diffusion mode at
$D_\perp = \eta/\rho$, but the longitudinal diffusion mode now shifts to
$D_\parallel = \kappa(\dow T/\dow\varepsilon)$. Since there is no sound, the
counting scheme slightly changes. Frequency scaling is now controlled by the
diffusion modes as $\omega\sim k^2$. The fields still scale as $k^{d/2}$, but
the leading derivative interactions terms are only irrelevant by $k^{-d/2+1}$
and the sub-leading interactions by $k^{-d/2}$. In $d=2$, this causes the
leading interaction couplings (first two terms in
$\mathcal{L}^{\text{incom}}_{\text{int}}$) to become marginal. In $d=1$, these
couplings are actually relevant and the perturbation theory breaks down.

If we were to ignore the momentum modes altogether by setting
$\pi^i = X^i_a = 0$, we arrive at the theory of energy diffusion derived
in~\cite{Chen-Lin:2018kfl}, truncated to three-point couplings\footnote{Our
  $X^t_a$ field is related to the $\varphi_a$ field used
  in~\cite{Chen-Lin:2018kfl} as $X^t_a = - \varphi_a$.}
\begin{align}
  \mathcal{L}^{\text{diff}}_{\text{free}}
  &= X^t_{a} \lb \dow_t - \kappa \frac{\dow T}{\dow \varepsilon} \dow^i\dow_i
    \rb \delta \epsilon
    + i\kB T^2\kappa\, \dow^i X^t_a\dow_i X^t_a, \nn\\
  \mathcal{L}^{\text{diff}}_{\text{int}}
  &= - \half \lambda_\epsilon \delta\epsilon^2 \dow^k \dow_k X^t_a
      + \frac{i\kB T^2 \tilde\lambda_\epsilon}{\dow T/\dow\varepsilon}
      \delta\epsilon\, \dow^i X^t_a\dow_i X^t_a.
\end{align}
This theory has also been studied in detail in~\cite{Kovtun:2012rj}. We are
only left with the longitudinal diffusion mode at
$D_\parallel = \kappa(\dow T/\dow\varepsilon)$. This simplified theory also gets
rid of the leading-derivative order interaction terms that were problematic in
lower number of dimensions. However, it does serve as a toy model to probe the
structure of stochastic field theories.

\section{Coupling to Newton-Cartan sources}
\label{sec:NC-formulation}

In the previous section, we presented an EFT for $(d+1)$-dimensional Galilean
hydrodynamics in terms of $(d+2)$-dimensional uncharged relativistic null
fluids. From an implementation viewpoint, this language is immensely helpful as
our understanding of relativistic symmetries is much more mature than the
Galilean ones. However, it does make the underlying $(d+1)$-dimensional physics
more obscure. To remedy this situation, we now present a translation of our
results to a manifestly $(d+1)$-dimensional language by ``gauge-fixing'' the
auxiliary null direction. We first introduce the NC backgrounds in
\cref{sec:null-redn}, seen as a reduction of null backgrounds, along with their
coupling to the constitutive relations of classical Galilean hydrodynamics,
mirroring \cref{sec:ClassicalGalilean}. Later in \cref{sec:GalSK-NC}, we
translate the EFT framework for Galilean hydrodynamics to the NC language
following our discussion from \cref{sec:GalSK}. At the end of the day, once the
effective action describing our system of interest has been obtained, the path
one takes to arrive at the same is immaterial.

\subsection{Classical hydrodynamics with Newton-Cartan sources}
\label{sec:null-redn}

Let us rewind back to \cref{sec:source-coupling-null}, where we exploited the
embedding of $(d+1)$-dimensional Galilean transformations into
$(d+2)$-dimensional Poincare algebra to introduce a set of background sources
$g_{\sM\sN}$ coupled to the Galilean conserved currents. To this end, we
introduced an auxiliary null direction $V^\sM$ on the background.  To recast
these results into a manifestly $(d+1)$-dimensional language, we just need to
``undo'' this embedding by choosing a coordinate system $(x^\sM) = (x^-,x^\mu)$
and partially fix the $(d+2)$-dimensional background diffeomorphisms to set
$V^\sM = \delta^\sM_-$~\cite{Banerjee:2015hra}. This procedure is formally known
as null reduction and yields the so-called Newton-Cartan (NC) formulation of
Galilean field theories. The price we pay is that the background now contains a
non-invertible ``spatial'' metric, a torsional spacetime connection, and the
Galilean boost symmetry is no longer manifest, making it slightly more technical
to implement on its own. In the following, we shall derive the NC results via
null reduction. An independent review of the coupling of Galilean field theories
to NC sources can be found in~\cite{Jensen:2014aia}.

\subsubsection{Newton-Cartan structure}

With the choice of coordinate decomposition mentioned above, a generic null
background metric $g_{\sM\sN}$, satisfying $V^\sM V^\sN g_{\sM\sN} = g_{--} = 0$
and $\lie_V g_{\sM\sN} = \dow_- g_{\sM\sN} = 0$, can be decomposed
as\footnote{We use the notations of~\cite{Jensen:2014aia}. In the work
  of~\cite{Bergshoeff:2014uea, Christensen:2013rfa, Christensen:2013lma}, the
  clock-form is equivalently represented by $\tau_\mu = - n_\mu$. In the work of
  ~\cite{Banerjee:2017ouw, Banerjee:2016qxf, Banerjee:2015hra, Banerjee:2015uta,
    Jain:2018jxj}, the mass gauge field has been represented by
  $b_\mu = A_\mu$.}
\begin{equation}
  g_{\sM\sN}(x) \df x^\sM \df x^\sN
  = -2\lb \df x^- - A_\nu(x) \df x^\nu \rb n_\mu(x) \df x^\mu 
  + h_{\mu\nu}(x) \df x^\mu \df x^\nu,
  \label{eq:metric-decomposition}
\end{equation}
In this section, the arguments ``$(x)$'' denote the dependence only on the
coordinates $x^\mu$ and not on $x^-$. Here $n_\mu$ is called the
\emph{clock-form} or the temporal metric, $h_{\mu\nu}$ the \emph{spatial
  metric}, and $A_\mu$ is the \emph{mass gauge field}. They make up the complete
set of Newton-Cartan sources. One can check that, in total, these have $(d+1)$
more components than $g_{\sM\sN}$. To fix this redundancy, we require that the
spatial metric is degenerate, i.e. given a ``frame-velocity vector'' $v^\mu$
normalised as $v^\mu n_\mu = 1$, we have $v^\mu h_{\mu\nu} = 0$. This still
leaves us with arbitrary redefinitions of $v^\mu$ that act as
\begin{equation}
  v^\mu \to v^\mu + \psi^\mu, \qquad
  h_{\mu\nu} \to h_{\mu\nu} - 2 n_{(\mu} \psi_{\nu)} + n_\mu n_\nu \psi^2, \qquad
  A_\mu \to A_\mu + \psi_{\mu} - \half n_\mu \psi^2,
  \label{eq:MilneBoost}
\end{equation}
where $\psi^\mu$ is an arbitrary vector satisfying $\psi^\mu n_\mu = 0$; we have
denoted $\psi_\mu = h_{\mu\nu}\psi^\nu$ and
$\psi^2 = \psi^\mu \psi^\nu h_{\mu\nu}$. This is known as the Milne-boost
symmetry. Much of the complication in the NC construction stems from trying to
make this symmetry manifest in EFTs. Luckily for us, we get this for free
through the null background construction. We can similarly decompose the inverse
null background metric as
\begin{equation}
  g^{\sM\sN}(x) \dow_\sM \otimes \dow_\sN
  = - 2 v^\mu(x) \lb \dow_\mu + A_\mu(x) \dow_- \rb \otimes \dow_- 
  + h^{\mu\nu}(x) \lb \dow_\mu + A_\mu(x) \dow_- \rb
  \otimes \lb \dow_\nu + A_\nu(x) \dow_- \rb.
  \label{eq:inverse-decomposition}
\end{equation}
We have introduced the ``inverse spatial metric'' $h^{\mu\nu}$, defined via
$h^{\mu\nu} h_{\nu\rho} + v^\mu n_\rho = \delta^\mu_\rho$ and
$n_\mu h^{\mu\nu} = 0$, which is invariant under Milne boosts. Note that
$h^{\mu\nu}$ is \emph{not} the inverse of $h_{\mu\nu}$. We will denote
$h^{\mu}_{~\nu} = h^{\mu\lambda}h_{\lambda\nu}$. The NC volume element is
defined as $\epsilon^{\mu\nu\rho\sigma} = -\epsilon^{-\mu\nu\rho\sigma}$, so
that $\epsilon^{t123} = 1/\sqrt{\gamma}$ with
$\gamma = \det(h_{\mu\nu}+n_\mu n_\nu)$.

The residual transformations from the $(d+2)$-dimensional null background
diffeomorphisms \eqref{eq:null.diffeo} are merely the $(d+1)$-dimensional
diffeomorphisms and U(1) mass gauge transformations
\begin{subequations}
  \begin{equation}
    x^\mu \to x'^\mu(x), \qquad
    x^- \to x^- - \Lambda(x).
  \end{equation}
  These act on the background fields as expected
  \begin{gather}
    n_\mu(x)
    \to n'_\mu(x') = \frac{\dow x^\nu}{\dow x'^\mu} n_\nu(x), \qquad
    h_{\mu\nu}(x)
    \to h'_{\mu\nu}(x')
      = \frac{\dow x^\rho}{\dow x'^\mu}\frac{\dow x^\sigma}{\dow x'^\nu}
      h_{\rho\sigma}(x), \nn\\
    A_\mu(x)
    \to A'_\mu(x') = \frac{\dow x^\nu}{\dow x'^\mu} \lb A_\nu(x)
      + \dow_\nu\Lambda(x) \rb,
  \end{gather}
  along with
  \begin{gather}
    v^\mu(x)
    \to v'^\mu(x') = \frac{\dow x'^\mu}{\dow x^\nu} v^\nu(x), \qquad
    h^{\mu\nu}(x)
    \to h'^{\mu\nu}(x')
      = \frac{\dow x'^\mu}{\dow x^\rho}\frac{\dow x'^\nu}{\dow x^\sigma}
      h^{\rho\sigma}(x).
  \end{gather}%
  \label{eq:NC.diffeo}%
\end{subequations}
These transformation properties follow directly from
\cref{eq:null.diffeo-background}.

The NC connection $\tilde\Gamma^\lambda_{\mu\nu}$ is defined such that the
associated covariant derivative operator $\tilde\nabla_\mu$ satisfies
$\tilde\nabla_\mu n_\nu = \tilde\nabla_\mu h^{\nu\rho} = 0$. It is not possible
to construct such a connection that is simultaneously invariant under mass gauge
transformations and Milne boosts as well. Sacrificing the Milne boosts, we can
write down one such connection\footnote{It is known that this connection is not
  unique; see~\cite{Jensen:2014aia} for a detailed discussion.}
\begin{align}
  \tilde\Gamma^\lambda_{\mu\nu}
  = v^{\lambda} \dow_\mu n_{\nu}
    + \half  h^{\lambda\rho}
  \lb \dow_\mu h_{\nu\rho} + \dow_\nu h_{\mu\rho} - \dow_\rho h_{\mu\nu} \rb
  + n_{(\mu} F_{\nu)\rho} h^{\rho\lambda}.
  \label{eq:NC-connection}
\end{align}
Here $F_{\mu\nu} = 2\dow_{[\mu}A_{\nu]}$ is the mass gauge field
strength. Notice that the NC connection is torsional:
$2\tilde\Gamma^\lambda_{[\mu\nu]} = v^\lambda F^n_{\mu\nu}$, where
$F^n_{\mu\nu} = 2\dow_{[\mu}n_{\nu]}$ is the temporal torsion. We can also
generically introduce spatial and ``mass'' torsion, but it is not imposed upon
us by the framework (see~\cite{Jain:2018jxj}). It is easy to see that the
connection is not left invariant by Milne boosts \eqref{eq:MilneBoost}
\begin{equation}
  \tilde\Gamma^\lambda_{\mu\nu}
  \to \tilde\Gamma^\lambda_{\mu\nu}
  + \half \psi^{\lambda} F^n_{\mu\nu}
  - \psi_{(\mu} F^n_{\nu)\rho} h^{\rho\lambda}
  + \half \psi^2 n_{(\mu} F^n_{\nu)\rho} h^{\rho\lambda}.
\end{equation}
 The null
background Levi-Civita connection $\Gamma^\sR_{\sM\sN}$ can be decomposed in
terms of $\tilde\Gamma^\lambda_{\mu\nu}$ as
\begin{align}
  \Gamma^-_{\sM\sN} \df x^\sM \df x^\sN
  &= \lb h_{\lambda(\nu}\tilde\nabla_{\mu)} v^\lambda
  - \tilde\nabla_{(\mu} A_{\nu)} \rb \df x^\mu \df x^\nu
  - \lb \df x^- - A_{\nu} \df x^\nu \rb
  \lb v^\rho - h^{\rho\lambda}A_\lambda \rb F^n_{\rho\mu} \df x^\mu, \nn\\
  \Gamma^\lambda_{\sM\sN} \df x^\sM \df x^\sN
  &= \tilde\Gamma^\lambda_{(\mu\nu)} \df x^\mu \df x^\nu
  + \lb \df x^- - A_{\nu}\df x^\nu \rb h^{\lambda\rho} F^n_{\rho\mu} \df x^\mu.
\end{align}
Note that
$h_{\nu\lambda}\tilde\nabla_\mu v^\lambda = \half F_{\mu\nu} - v^\rho \lb
n_{(\mu} F_{\nu)\rho} + \dow_{(\mu} h_{\nu)\rho} - \half \dow_\rho h_{\mu\nu}
\rb$ and
$\tilde\nabla_\lambda h_{\mu\nu} = -2n_{(\mu}h_{\nu)\rho} \tilde\nabla_\lambda
v^\rho $. Also that
$\Gamma^\sM_{\sM\nu} = \tilde\Gamma^\mu_{\mu\nu} + F^n_{\nu\mu} v^\mu =
\frac{1}{\sqrt{\gamma}} \dow_\nu \sqrt{\gamma}$. These identities are helpful in
the explicit derivations below.

To revert back to a flat background, we just need to set $n_\mu = \delta_\mu^t$,
$v^\mu = \delta^\mu_t$, $A_\mu = 0$,
$h_{\mu\nu} = \delta_{\mu i} \delta^{i}_\nu$, and
$h^{\mu\nu} = \delta^{\mu i} \delta_{i}^\nu$. This also sets the connection
$\tilde\Gamma^\lambda_{\mu\nu} = 0$, along with $F_{\mu\nu} = F^n_{\mu\nu} = 0$.

\subsubsection{Coupling to Galilean field theories}

We can reduce the null conservation equations \eqref{eq:null.conservation} to
obtain the covariant version of the Galilean conservation equations in the NC
language. To this end, we introduce the covariant mass current $\rho^\mu$,
energy current $\epsilon^\mu$, and spatial stress tensor $\tau^{\mu\nu}$ in
terms of the higher dimensional energy-momentum tensor $T^{\sM\sN}$ as
\begin{equation}
  \rho^\mu = T^{\mu\nu} n_\nu, \qquad
  \epsilon^\mu = T^{\mu-} - T^{\mu\rho} A_\rho, \qquad
  \tau^{\mu\nu} = h^\mu_{~\rho} h^\nu_{~\sigma} T^{\rho\sigma}.
\end{equation}
Note that $\tau^{\mu\nu} = \tau^{\nu\mu}$ and $\tau^{\mu\nu} n_\nu = 0$. It is
also convenient to define the momentum density $\pi^\mu = h^\mu_{~\nu} \rho^\nu$
that is equal to the mass flux. These definitions are chosen so that they are
invariant under mass gauge transformations, however, $\epsilon^\mu$ and
$\tau^{\mu\nu}$ are consequently not boost-invariant. This is what one would
physically expect, because the notion of energy and momentum depends on the
choice of the Galilean observer. The conservation equations
\eqref{eq:null.conservation} then become
\begin{align}
  % \text{Energy-momentum conservation:}\qquad
  % &\lb \tilde\nabla_\mu + n_{\mu\lambda} v^\lambda \rb t^\mu_{~\nu}
  % = n_{\nu\mu} t^\mu_{~\rho} v^{\rho}
  % - h_{\nu\lambda} \rho^\mu \tilde\nabla_\mu v^\lambda, \nn\\
  \text{Energy conservation:}\qquad
  &\lb \tilde\nabla_\mu + F^n_{\mu\lambda} v^\lambda \rb \epsilon^\mu
  = v^\nu F^n_{\nu\mu} \epsilon^\mu
    - \lb v^\mu \pi^\lambda + \tau^{\mu\lambda} \rb h_{\lambda\nu} \tilde\nabla_\mu v^\nu, \nn\\
  \text{Momentum conservation:}\qquad
    &\lb \tilde\nabla_\mu + F^n_{\mu\lambda} v^\lambda \rb
      \lb v^\mu \pi^\nu + \tau^{\mu\nu} \rb
  = - h^{\nu\lambda} F^n_{\lambda\mu} \epsilon^\mu
    - \rho^\mu \tilde\nabla_\mu v^\nu, \nn\\
  \text{Continuity equation:}\qquad
  &\lb \tilde\nabla_\mu + F^n_{\mu\lambda} v^\lambda \rb \rho^\mu
    = 0.
    \label{eq:NC.Conservation}
\end{align} 
These are the covariant version of the Galilean conservation equations
\eqref{eq:Gal-conservation}. Aside from the torsional contributions coupled to
$F^n_{\mu\nu}$, the energy and momentum conservation is sourced by the
``pseudo-power'' and ``pseudo-force'' contributions coupled to the derivatives
of frame velocity.

Given some action $S$ describing a Galilean field theory, the coupling between
various currents and sources takes the form following from
\cref{eq:null-action}\footnote{These can also be stated in the form advocated
  in~\cite{Jensen:2014aia}. Note that
  $v^\mu \delta h_{\mu\nu} = - \delta h_{\mu\nu} \delta v^\mu$ and
  $\delta h_{\mu\nu} = - h_{\mu\rho} h_{\nu\sigma} \delta h^{\rho\sigma}$, which
  recasts the integrand into
  $\rho^\mu \delta A_\mu - \epsilon^\mu \delta n_\mu - \pi_{\mu} \delta v^\mu -
  \half \tau_{\mu\nu} \delta h^{\mu\nu}$, where $\pi_\mu = h_{\mu\nu}\pi^\nu$
  and $\tau_{\mu\nu} = h_{\mu\rho} h_{\nu\sigma}\tau^{\rho\sigma}$.}
\begin{equation}
  \delta S = \int \df^{d+1}x\, \sqrt{\gamma}\, \lb
  \rho^\mu \delta A_\mu
  - \epsilon^\mu \delta n_\mu
  + \lb v^\mu \pi^{\nu} + \half \tau^{\mu\nu} \rb \delta h_{\mu\nu} \rb.
  \label{eq:action-variation-NC}
\end{equation}
Conservation equations \eqref{eq:NC.Conservation} follow from requiring the
action to be invariant under the Galilean transformations
\eqref{eq:NC.diffeo}. On the other hand, invariance under Milne boosts
\eqref{eq:MilneBoost} leads to the identity between momentum density and mass
flux $\pi^{\mu} = h^\mu_{~\nu}\rho^\nu$. The retarded two-point functions of
various quantities follow from the reduction of \cref{eq:null-retG}. Up to
contact terms, we find
\begin{gather}
  G^{\text R}_{\rho^{\mu} \rho^{\nu}}
  = \frac{1}{\sqrt{\gamma}}
  \frac{\delta (\sqrt{\gamma}\, \rho^{\mu})}{\delta A_\nu}, \qquad
  G^{\text R}_{\epsilon^{\mu} \epsilon^{\nu}}
  = - \frac{1}{\sqrt{\gamma}}
  \frac{\delta (\sqrt{\gamma}\, \epsilon^\mu)}{\delta n_\nu}, \qquad
  G^{\text R}_{\tau^{\mu\nu}\tau^{\rho\sigma}} 
  = h^\mu_{~\lambda} h^\nu_{~\tau} h^\rho_{~\alpha} h^\sigma_{~\beta} \frac{2}{\sqrt{\gamma}}
  \frac{\delta (\sqrt{\gamma}\,\tau^{\lambda\tau})}{\delta h_{\alpha\beta}},
  \nn\\
  G^{\text R}_{\rho^{\mu} \epsilon^{\nu}}
  % - (\rho v^{\mu} v^\nu + 2v^{(\mu} p^{\nu)} + t^{\mu\nu})
  = - \frac{1}{\sqrt{\gamma}}
  \frac{\delta (\sqrt{\gamma}\, \rho^{\mu})}{\delta n_\nu}, \qquad
  G^{\text R}_{\epsilon^{\mu} \rho^{\nu}}
  = \frac{1}{\sqrt{\gamma}}
  \frac{\delta (\sqrt{\gamma}\, \epsilon^\mu)}{\delta A_\nu}, \nn\\
  G^{\text R}_{\tau^{\mu\nu}\rho^{\lambda}}
  = h_{~\rho}^\mu h_{~\sigma}^\nu \frac{2}{\sqrt{\gamma}}
  \frac{\delta (\sqrt{\gamma}\, \rho^{\lambda})} {\delta h_{\rho\sigma}}, \qquad
  G^{\text R}_{\rho^{\lambda}\tau^{\mu\nu}}
  = h^\mu_{~\rho} h^\nu_{~\sigma} \frac{1}{\sqrt{\gamma}}
  \frac{\delta (\sqrt{\gamma}\, \tau^{\rho\sigma})}{\delta A_\lambda}, \nn\\
  G^{\text R}_{\tau^{\mu\nu} \epsilon^\lambda}
  = h_{~\rho}^\mu h_{~\sigma}^\nu \frac{2}{\sqrt{\gamma}}
  \frac{\delta (\sqrt{\gamma}\, \epsilon^{\lambda})}{\delta h_{\rho\sigma}},
  \qquad
  G^{\text R}_{\epsilon^\lambda \tau^{\mu\nu}}
  = - h^\mu_{~\rho} h^\nu_{~\sigma} \frac{1}{\sqrt{\gamma}}
  \frac{\delta (\sqrt{\gamma}\,\tau^{\rho\sigma})}{\delta n_\lambda}.
  \label{eq:NC-retG}
\end{gather}
The factors of $h^\mu_{~\nu}$ project out the appropriate spatial components for
the spatial stress tensor $\tau^{\mu\nu}$ on the left hand side.

Let us say that we are provided with a ``lab frame'' characterised by the time
coordinate $t$. We can use this to fix the Milne boost symmetry
\eqref{eq:MilneBoost} exactly by choosing the Galilean frame velocity to be
$v^\mu \propto \delta^\mu_t$. The background sources then become
\begin{gather}
  A_\mu \df x^\mu = A_t \df t + A_i \df x^i, \qquad
  n_\mu \df x^\mu = n_t \df t + n_i \df x^i, \qquad
  h_{\mu\nu} \df x^\mu \df x^\nu = h_{ij} \df x^i \df x^j, \nn\\
  v^\mu \dow_\mu = \frac{1}{n_t}\dow_t, \qquad
  h^{\mu\nu} \dow_\mu \otimes \dow_\nu
  = \frac{h^{ij} n_i n_j}{n_t^2} \dow_t\otimes\dow_t
  - 2 \frac{h^{ij} n_j}{n_t} \dow_t\otimes\dow_i
  + h^{ij} \dow_i\otimes\dow_j, \nn\\
  h^\mu_{~\nu} \df x^\nu \dow_\mu
  = - \frac{n_i}{n_t} \df x^i \dow_t + \delta^j_i \df x^i \dow_j, \qquad
  \gamma = n_t^2 h.
  \label{eq:non-cov-sources}
\end{gather}
Here $h^{ij}$ is the inverse of $h_{ij}$ and $h = \det h_{ij}$. The action
variation \eqref{eq:action-variation-NC} takes a more natural form
\begin{equation}
  \delta S = \int \df^{d+1}x\, n_t \sqrt{h}\, \lb
  \rho^t \delta A_t
  + \rho^i \delta A_i
  - \epsilon^t \delta n_t
  - \epsilon^i \delta n_i
  + \half \tau^{ij} \delta h_{ij} \rb,
\end{equation}
with $h_{ij}$ sourcing the stress tensor. The response functions follow
accordingly from \cref{eq:NC-retG}.

\subsubsection{Hydrodynamics on curved spacetime}

We can use the generic discussion above to reduce the constitutive relations of
Galilean hydrodynamics to the NC language. We decompose the null fluid velocity
as $u^\sM \dow_\sM = u^- \dow_- + u^\mu \dow_\mu$. The normalisation conditions
\eqref{eq:null-norm} imply that $u^\mu n_\mu = 1$ and
$u^- = \half \vec u^2 + u^\mu A_\mu$, where $\vec u^2 = u^\mu u^\nu
h_{\mu\nu}$. Here $u^\mu$ is understood as the covariant velocity of the
Galilean fluid. We can further define the spatial part of the fluid velocity as
$\vec u^\mu = h^\mu_{~\nu} u^\nu = u^\mu - v^\mu$ with
$\vec u_\mu = h_{\mu\nu} u^\nu$. The null fluid constitutive relations and
entropy current from \eqref{eq:null-fluid-EM} can be reduced to obtain the
respective NC versions\footnote{Here $\epsilon^\mu$ and $\tau^{\mu\nu}$ are
  defined in an arbitrary Galilean frame of reference characterised by
  $v^\mu$. In hydrodynamics, we can explicitly fix the Milne boost symmetry by
  choosing $v^\mu = u^\mu$ or $\vec u^\mu = 0$. This is equivalent to working in
  the local rest frame of the fluid. We can always return to an arbitrary frame
  by performing an inverse Milne boost with $\psi^\mu = - \vec u^\mu$. However,
  with this choice, the flat background limit is slightly non-trivial
  \begin{gather}
    n_\mu \df x^\mu|_{\text{flat}} = \df t, \qquad h_{\mu\nu} \df x^\mu \df
    x^\nu|_{\text{flat}} = \vec u^2 \df t^2 - 2u_i \df x^i \df t +
    \delta_{ij}\df x^i \df x^j, \qquad
    h^{\mu\nu}\dow_\mu\otimes\dow_\nu = \delta^{ij}\dow_i \otimes\dow_j,  \nn\\
    A_\mu \df x^\mu|_{\text{flat}} = - \half \vec u^2 \df t + u_i \df x^i,
    \qquad \tilde\Gamma^\lambda_{\mu\nu}|_{\text{flat}} = 0.
  \end{gather}
  Since the energy and momentum defined in fluid's rest frame are not
  individually conserved, we have chosen to instead work in an arbitrary lab
  frame of reference in the main text.\label{foot:FluidRestFrame}}
\begin{subequations}
  \begin{gather}
    \rho^\mu = \rho\, u^\mu, \qquad
    \epsilon^\mu = \lb \varepsilon + \half\rho\,
    \vec u^2 \rb u^{\mu} + q^\mu + t^{\mu\nu}\vec u_\nu, \qquad
    \tau^{\mu\nu} =
    \lb \rho\, \vec u^\mu \vec u^\nu + t^{\mu\nu} \rb, \nn\\
    s^\mu = s\, u^\mu + \frac{1}{T} q^\mu + \Upsilon^\mu.
    \label{eq:NC-currents}
  \end{gather}
  The comoving heat flux $q^\mu$ and stress tensor $t^{\mu\nu}$ satisfy
  $q^\mu n_\mu = t^{\mu\nu} n_\nu = 0$ and $t^{\mu\nu} = t^{\nu\mu}$, while the
  non-canonical entropy current $\Upsilon^\mu$ is un-normalised. Up to
  one-derivative order, these are given by the reduction of
  \cref{eq:null-consti-cov} leading to
  \begin{align}
    t^{\mu\nu}
    &= p\, h^{\mu\nu}
      - \eta\, \sigma^{\mu\nu}
      - \zeta\, h^{\mu\nu}
      \lb \tilde\nabla_\lambda + F^n_{\lambda\rho} v^\rho \rb u^\lambda, \nn\\
    q^{\mu}
    &= - \kappa\, h^{\mu\nu} \lb \dow_\nu T + T F^n_{\nu\rho} u^\rho \rb
      - \half\lambda_\Omega \epsilon^{\mu\nu\rho\sigma} n_\nu \Omega_{\rho\sigma}
      + \half\lambda_H \epsilon^{\mu\nu\rho\sigma} n_\nu F^n_{\rho\sigma}, \nn\\
    \Upsilon^\mu
    &= - \frac{a_2}{2T} \epsilon^{\mu\nu\rho\sigma} n_\nu \Omega_{\rho\sigma}
      + \frac{a_1}{4T} \epsilon^{\mu\nu\rho\sigma} n_\nu F^n_{\rho\sigma} \nn\\
    &\qquad
      + \frac{a_0}{2T} \epsilon^{\mu\nu\rho\sigma}
      \lb \vec u_\nu + A_\nu - \half\vec u^2 n_\nu \rb
      \dow_\rho \lb \vec u_\sigma + A_\sigma - \half\vec u^2 n_\sigma \rb,
      \label{eq:NC-consti}
  \end{align}
\end{subequations}
where the fluid shear and vorticity tensors are given as
\begin{align}
  \sigma^{\mu\nu}
  &= 2 \lb h^{\rho(\mu} h^{\nu)}_{~\lambda}
  - \frac{1}{d} h^{\mu\nu} h^\rho_{~\lambda} \rb
  \lb \tilde\nabla_\rho u^\lambda + u^\lambda F^n_{\rho\sigma} u^\sigma \rb, \nn\\
  \Omega_{\mu\nu}
  &= h^\rho_{~\mu} h^\sigma_{~\nu} \lb
  2\dow_{[\rho}\vec u_{\sigma]}
  + F_{\rho\sigma} - \half \vec u^2 F^n_{\rho\sigma} \rb.
\end{align}
All the coefficients appearing here are functions of $T$ and $\mu$. The
thermodynamic coefficients $p$, $\varepsilon$, $\rho$, and $s$ satisfy the
relations \eqref{eq:GCthermo}, the dissipative transport coefficients $\eta$,
$\zeta$, and $\kappa$ are required to be positive semi-definite, while the
adiabatic transport coefficients $\lambda_\Omega$, $\lambda_H$, $a_0$, $a_1$,
and $a_2$ are given in terms of three arbitrary constants $K_0$, $K_1$, $K_2$
according to \cref{eq:transconstraints}. The hydrodynamic equations of motion in
the Newton-Cartan language are given by substituting the constitutive relations
into \cref{eq:NC.Conservation}. These constitutive relations satisfy the second
law of thermodynamics
\begin{equation}
  \lb \tilde\nabla_\mu + n_{\mu\lambda} v^\lambda \rb s^\mu \geq 0,
\end{equation}
on the solutions of the equations of motion.

As discussed in \cref{sec:frame}, the constitutive relations specified above are
written in the ``mass frame''. One can arbitrarily depart from this choice of
frame, however, by redefining $u^\mu$, $T$, and $\mu$ according to
\cref{eq:frame-trans-fields}. For instance, the thermodynamic mass frame is
arrived at by keeping the definitions of $T$ and $\mu$ intact, but redefining
the fluid velocity according to
\begin{equation}
  u^\mu \to u^\mu
  - \frac{a_0}{\rho} \epsilon^{\mu\nu\rho\sigma} n_\nu \Omega_{\rho\sigma}
  + \frac{a_2}{2\rho} \epsilon^{\mu\nu\rho\sigma} n_\nu F^n_{\rho\sigma}.
\end{equation}
The transformed set of constitutive relations are given by the reduction of
\cref{eq:null-consti-gen-frame,eq:null-consti-cov-thermo}
\begin{subequations}
  \begin{gather}
    \rho^\mu = \rho\, u^\mu + j^\mu, \qquad \epsilon^\mu = \lb \varepsilon +
    \half\rho\, \vec u^2 + j^\nu \vec u_\nu \rb u^{\mu}
    + q^\mu + t^{\mu\nu}\vec u_\nu + \half \vec u^2 j^\mu, \nn\\
    \tau^{\mu\nu} = \rho\, \vec u^\mu \vec u^\nu + t^{\mu\nu} + 2\vec u^{(\mu}
    j^{\nu)}, \qquad s^\mu = s\, u^\mu - \frac{\mu}{T} j^\mu + \frac{1}{T} q^\mu
    + \Upsilon^\mu,
  \end{gather}
  where
  \begin{align}
    t^{\mu\nu}
    &= p\, h^{\mu\nu}
      - \eta\, \sigma^{\mu\nu}
      - \zeta\, h^{\mu\nu}
      \lb \tilde\nabla_\lambda + F^n_{\lambda\rho} v^\rho \rb u^\lambda, \nn\\
    j^{\mu}
    &= - \frac{a_0}{2} \epsilon^{\mu\nu\rho\sigma} n_\nu \Omega_{\rho\sigma}
      + \frac{a_2}{2} \epsilon^{\mu\nu\rho\sigma} n_\nu F^n_{\rho\sigma}, \nn\\
    q^{\mu}
    &= - \kappa\, h^{\mu\nu} \lb \dow_\nu T + T F^n_{\nu\rho} u^\rho \rb
      - \frac{a_2}{2} \epsilon^{\mu\nu\rho\sigma} n_\nu \Omega_{\rho\sigma}
      + \frac{a_1}{2} \epsilon^{\mu\nu\rho\sigma} n_\nu F^n_{\rho\sigma}, \nn\\
    \Upsilon^\mu
    &= - \frac{a_2}{2T} \epsilon^{\mu\nu\rho\sigma} n_\nu \Omega_{\rho\sigma}
      + \frac{a_1}{4T} \epsilon^{\mu\nu\rho\sigma} n_\nu F^n_{\rho\sigma} \nn\\
    &\qquad
      + \frac{a_0}{2T} \epsilon^{\mu\nu\rho\sigma}
      \lb \vec u_\nu + A_\nu - \half\vec u^2 n_\nu \rb
      \dow_\rho \lb \vec u_\sigma + A_\sigma - \half\vec u^2 n_\sigma \rb.
  \end{align}
  \label{eq:NC-consti-thermo}%
\end{subequations}
In this frame, the constitutive relations can be checked to satisfy the
adiabaticity equation obtained by reducing \cref{eq:adiabaticity}
\begin{equation}
  \lb \tilde\nabla_\mu + n_{\mu\lambda} v^\lambda \rb N^\mu
  = \rho^\mu \delta_\scB A_\mu
  - \epsilon^{\mu} \delta_\scB n_\mu
  + \lb v^\mu p^\nu + \half \tau^{\mu\nu} \rb \delta_\scB h_{\mu\nu}
  + \Delta, \qquad
  \Delta \geq 0,
  \label{eq:adiabaticity-NC} 
\end{equation}
with the free energy current
\begin{equation}
  \kB T N^\mu = Ts^\mu
  + \rho^\mu(\mu-{\textstyle\half} \vec u^2)
  - \epsilon^{\mu}
  + (v^\mu \pi^\nu + \tau^{\mu\nu}) \vec u_\nu
  = p\,u^\mu + T \Upsilon^\mu.
\end{equation}
The operator $\delta_\scB$ above combines a Lie derivative along
$\beta^\mu = u^\mu/(\kB T)$, denoted by $\lie_\beta$, and a gauge shift along
$\Lambda_\beta = - \beta^- = (\mu - \half u^2 - A_\mu u^\mu)/(\kB T)$. Explicitly
\begin{gather}
  \delta_\scB n_\mu = \lie_\beta n_\mu
  = - \lb \frac{1}{\kB T^2} \dow_\mu T + F^n_{\mu\nu} \beta^\nu \rb, \qquad
  \delta_\scB h_{\mu\nu} = \lie_\beta h_{\mu\nu}
  = 2\lb h_{\lambda(\mu}\tilde{\nabla}_{\nu)}\beta^\lambda
  - n_{(\mu}h_{\nu)\rho} \beta^\lambda \tilde\nabla_\lambda v^\rho \rb, \nn\\
  \delta_\scB A_\mu
  = \lie_\beta A_\mu + \dow_\mu \Lambda_\beta
  = \dow_\mu\bfrac{\mu - \half \vec u^2}{\kB T}
  - F_{\mu\nu}\beta^\nu.
\end{gather}

The discussion of discrete symmetries is precisely the same as
\cref{sec:GalAdiabaticity}. Requiring the underlying microscopic theory to be
invariant under $\Theta=\text{T}$ symmetry does not lead to any new constraints
on the constitutive relations, while $\Theta=\text{PT}$ switches off the entire
parity-violating sector. See \cref{tab:CPT} for the action of discrete
symmetries on various fields. These constraints follow from requiring the
associated equilibrium partition function to respect
$\Theta$~\cite{Banerjee:2015uta}. Note that the constitutive relations
themselves will not, in general, respect any kind of time reversal symmetry due
to dissipation. The effective field theory framework deals with these discrete
symmetries more carefully.

\subsection{\SK effective field theory}
\label{sec:GalSK-NC}

We can extend the null reduction philosophy above to the EFT as well. Compared
to our discussion in \cref{sec:GalSK}, we need to pick a coordinate system
$(\sigma^\sA) = (\sigma^-,\sigma^\alpha)$ on the fluid worldvolume such that
$\bbV^\sA(\sigma) = \delta^\sA_-$, and identify the reference mass chemical
shift field $\Lambda_\bbbeta(\sigma) = -\bbbeta^{-}(\sigma)$. Similarly on the
SK spacetimes we set $V^\sM_s(X_s) = \delta^\sM_-$ and identify the auxiliary
coordinate field with the U(1) mass phase $\varphi_s(s)$ via
$X^-_s(\sigma) = \sigma^- - \varphi_s(\sigma)$. This follows through to the
physical spacetime with the choice of average coordinate basis
$(x^\sM) = (x^-,x^\mu)$ such that $V^\mu(x) = \delta^\mu_-$; we can identify the
thermal shift field $\Lambda_\beta(x) = -\beta^-(x)$. The details will follow.

\subsubsection{Effective field theory on fluid worldvolume}

We start with a $(d+1)$-dimensional ``fluid worldvolume'' with coordinates
$\sigma^\alpha$, interpreted as a set of internal spacetime labels associated
with each ``fluid element''. On this spacetime lives the dynamical fields of the
theory: $X^\mu_s(\sigma)$ and $\varphi_s(\sigma)$ ($=\sigma^- - X^-_s(\sigma)$)
with $s=1,2$, which should be understood as the SK double copies of spacetime
coordinates and U(1) mass phases, respectively, of a given fluid
element. Decomposing $X^\mu_{1,2} = X^\mu_r \pm \hbar/2\, X^\mu_a$ and
$\varphi_{1,2} = \varphi_r \pm \hbar/2\,\varphi_a$, the average combinations
$X^\mu_r(\sigma)$ and $\varphi_r(\sigma)$ ($=\sigma^- - X^-_r(\sigma)$) are
understood as the true physical spacetime coordinates and U(1) phase of the
fluid elements, while $X^\mu_a(\sigma)$ and $\varphi_a(\sigma)$
($=-X_a^-(\sigma)$) as the associated stochastic noise. The worldvolume also
features a fixed thermal time vector field $\bbbeta^\alpha(\sigma)$ and a mass
chemical shift field $\Lambda_\bbbeta(\sigma)$ ($= - \bbbeta^-(\sigma)$), which
determine the global rest frame and reference chemical potential associated with
the global thermal state.

The EFT for Galilean hydrodynamics must be invariant under global Galilean
transformations of the coordinates $X^\mu_s(\sigma)$ and global U(1) shifts of
the phases $\varphi_s(\sigma)$, acting independently on the two SK
spacetimes. To probe the associated Noether currents, we can introduce double
copies of NC sources similar to \cref{eq:metric-decomposition}: clock forms
$n_{s\mu}(X_s)$, spatial metrics $h_{s\mu\nu}(X_s)$, and mass gauge fields
$A_{s\mu}(X_s)$. We can also introduce the inverse spatial metrics
$h_s^{\mu\nu}(X_s)$ and frame velocities $v^\mu_s(X_s)$ associated with these
sources using the normalisation conditions $v_s^\mu n_{s\mu} = 1$,
$h_{s\mu\nu} v^{\nu}_s = 0$, $h_s^{\mu\nu} n_{s\nu} = 0$, and
$h_{s\mu\nu}h_s^{\nu\rho} + n_{s\mu} v^\rho_s = \delta^\rho_\mu$. These sources
have a Milne redundancy \eqref{eq:MilneBoost} among them, acting independently
on the two copies. With the background sources in place, the system is required
to be invariant under local SK spacetime diffeomorphisms and gauge
transformations (following from \cref{eq:SK-phys-diffeo})
\begin{gather}
  X_s^\mu(\sigma) \to X'^\mu_s(X_s(\sigma)), \qquad \varphi_s(\sigma) \to
  \varphi_s(\sigma) - \Lambda_s(X_s(\sigma)),
  \label{eq:phys-diffeo-NC}
\end{gather}
which act on the background sources as usual according to
\cref{eq:NC.diffeo}. To make the symmetries \eqref{eq:phys-diffeo-NC} manifest,
we can perform a pullback onto the fluid worldvolume, along with a mass gauge
transformation for the gauge fields, to obtain
\begin{gather}
  \bbn_{s\alpha}(\sigma)
  = n_{s\mu}(X_s(\sigma))\,\dow_\alpha X^\mu_s(\sigma), \qquad
  \bbh_{s\alpha\beta}(\sigma)
  = h_{s\mu\nu}(X_s(\sigma))\,
    \dow_\alpha X^\mu_s(\sigma) \dow_\beta X^\nu_s(\sigma), \nn\\
  \bbA_{s\alpha}(\sigma)
  = A_{s\mu}(X_s(\sigma))\,\dow_\alpha X^\mu_s(\sigma)
  + \dow_\alpha\varphi_s(\sigma),
  % , \nn\\
  % \bbv^\alpha_s(\sigma) = v_s^\mu(X_s(\sigma))
  % \frac{\dow \sigma^\alpha(X_s(\sigma))}{\dow X_s^\mu(\sigma)}, \qquad
  % \bbh^{\alpha\beta}_s(\sigma) = h_s^{\mu\nu}(X_s(\sigma))
  % \frac{\dow \sigma^\alpha(X_s(\sigma))}{\dow X_s^\mu(\sigma)}
  % \frac{\dow \sigma^\beta(X_s(\sigma))}{\dow X_s^\nu(\sigma)}
  \label{eq:NCinvariants}
\end{gather}
and similarly for the inverse spatial metrics and the frame velocities. However,
these quantities still transform under the Milne boosts as
\begin{align}
  \bbh_{s\alpha\beta}(\sigma)
  &\to \bbh_{s\alpha\beta}(\sigma) - 2 \bbn_{(\alpha}(\sigma) \psi_{s\beta)}(\sigma)
    + \bbn_{s\alpha}(\sigma) \bbn_{s\beta}(\sigma) \psi_s^2(\sigma), \nn\\
  \bbA_{s\alpha}(\sigma)
  &\to \bbA_{s\alpha}(\sigma)
    + \psi_{s\alpha}(\sigma)
    - \half \bbn_{s\alpha}(\sigma)\psi^2_s(\sigma),
    \label{eq:milne-SK}
\end{align}
where $\psi_{s\alpha} = \bbh_{s\alpha\beta} \psi^\beta_s$ and
$\psi^2_{s} = \bbh_{s\alpha\beta} \psi^\alpha_s \psi^\beta_s$ for some vectors
$\psi^\alpha_s$ satisfying $\psi^\alpha_s \bbn_{s\alpha} = 0$.

The internal labelling scheme for fluid elements, i.e. the coordinates
$\sigma^\alpha$ on the worldvolume, can be arbitrarily redefined without
changing any physics. Similarly, the U(1) phases $\varphi_s(\sigma)$ can be
arbitrarily shifted among the fluid elements. This leads to a local invariance
of the theory under diffeomorphisms and U(1) gauge transformations on the fluid
worldvolume (following from \cref{eq:fluid-spacetime-symm-null})
\begin{subequations}
  \begin{gather}
    \sigma^\alpha \to \sigma'^\alpha(\sigma), \qquad \varphi_s(\sigma) \to
    \varphi_s(\sigma) + \lambda(\sigma).
  \end{gather}
  Note that the two phases are required to shift simultaneously. These
  transformations act naturally on all the fluid worldvolume objects
  \begin{gather}
    \bbn_{s\alpha}(\sigma) \to \bbn'_{s\alpha}(\sigma') =
    \frac{\dow\sigma^\beta}{\dow \sigma'^\alpha}\bbn_{s\beta}(\sigma), \qquad
    \bbh_{s\alpha\beta}(\sigma) \to \bbh'_{s\alpha\beta}(\sigma') =
    \frac{\dow\sigma^\gamma}{\dow \sigma'^\alpha} \frac{\dow \sigma^\delta}{\dow
      \sigma'^\beta}
    \bbh_{s\gamma\delta}(\sigma), \nn\\
    \bbA_{s\alpha}(\sigma) \to \bbA'_{s\alpha}(\sigma') =
    \frac{\dow\sigma^\beta}{\dow \sigma'^\alpha} \lb \bbA_{s\beta}(\sigma)
    + \dow_\beta \lambda(\sigma) \rb, \nn\\
    % \bbv^\alpha_s(\sigma) \to \bbv'^\alpha_s(\sigma') =
    % \frac{\dow\sigma'^\alpha}{\dow \sigma^\beta}\bbv^\beta_{s}(\sigma), \qquad
    % \bbh^{\alpha\beta}_s(\sigma) \to \bbh'^{\alpha\beta}_s(\sigma') =
    % \frac{\dow\sigma'^\alpha}{\dow \sigma^\gamma}
    % \frac{\dow \sigma'^\beta}{\dow \sigma^\delta}
    % \bbh^{\gamma\delta}_s(\sigma), \nn\\
    \bbbeta^\alpha(\sigma) \to \bbbeta'^\alpha(\sigma') =
    \frac{\dow\sigma'^\alpha(\sigma)}{\dow\sigma^\beta} \bbbeta^\beta(\sigma),
    \qquad \Lambda_\bbbeta(\sigma) \to \Lambda'_\bbbeta(\sigma') =
    \Lambda_\bbbeta(\sigma) - \bbbeta^\alpha(\sigma) \dow_\alpha
    \lambda(\sigma),
  \end{gather}%
  \label{eq:fluid-spacetime-symm-NC}%
\end{subequations}%
which can be implemented using the usual techniques. Note that the difference
combination of the pullback of mass gauge fields
$\bbA_{a\alpha} = (\bbA_{1\alpha}-\bbA_{2\alpha})/\hbar$ is gauge invariant,
while the dependence on the average combination
$\bbA_{r\alpha} = (\bbA_{1\alpha}+\bbA_{2\alpha})/2$ must only come via the
``time-component'' $\bbbeta^\alpha \bbA_{r\alpha} + \Lambda_\bbbeta$ or the
associated field strength $2\dow_{[\alpha}\bbA_{r\beta]}$. We can partially fix
the symmetries \eqref{eq:fluid-spacetime-symm-NC} to choose a basis
$\sigma^\alpha = (\tau,\sigma^i)$ and explicitly set
$\bbbeta^\alpha = \beta_0\delta^\alpha_\tau$ and
$\Lambda_\bbbeta = \beta_0\mu_0$, where $\beta_0 = (\kB T_0)^{-1}$ is the
constant inverse temperature and $\mu_0$ the chemical potential of the global
thermal state.  The residual symmetry transformations are then the arbitrary
spatial relabelling of fluid elements $\sigma^i\to\sigma'^i(\vec\sigma)$,
spatial redefinitions of the local time coordinate
$\tau \to \tau + f(\vec\sigma)$, and that of the U(1) phase fields
$\varphi_s \to \varphi_s + \lambda(\vec\sigma)$. Note that the arbitrary
redefinitions are not allowed to depend on $\tau$ itself, because the labelling
scheme chosen at one point in time cannot be arbitrarily changed as the fluid
evolves. If the fluid has some of its spacetime or U(1) symmetries spontaneously
broken, as in superfluids or crystals, the symmetries
\eqref{eq:fluid-spacetime-symm-NC} will need to be respectively lifted.

The SK effective action for Galilean hydrodynamics takes the schematic form
\begin{equation}
  S[\bbn_1,\bbh_1,\bbA_1,\bbn_2,\bbh_2,\bbA_2;\bbbeta,\Lambda_\bbbeta]
  = \int \df^{d+1}\sigma\sqrt{\bbgamma_r}\,
  \mathcal{L}[\bbn_1,\bbh_1,\bbA_1,\bbn_2,\bbh_2,\bbA_2;\bbbeta,\Lambda_\bbbeta].
  \label{eq:SK-action-NC}
\end{equation}
Here $\bbgamma_r = \det(\bbn_{r\alpha}\bbn_{r\beta} + \bbh_{r\alpha\beta})$ with
$\bbn_{r\alpha} = (\bbn_{1\alpha}+\bbn_{1\alpha})/2$ and
$\bbh_{r\alpha\beta} = (\bbh_{1\alpha\beta}+\bbh_{1\alpha\beta})/2$. The
Lagrangian $\mathcal{L}$ is a gauge and Milne-invariant scalar on the
worldvolume with appropriate contraction of $\alpha,\beta,\ldots$ indices. The
Milne invariance is hard to implement within the NC language.  In practise, it
is much easier to start from the null background effective action
\eqref{eq:SKaction-null} and identify
\begin{gather}
  \bbg_{s\sA\sB}(\sigma) \df \sigma^\sA \df\sigma^\sB
  = - 2\lb  \df \sigma^- - \bbA_{s\beta}(\sigma) \df \sigma^\beta \rb
  \bbn_{s\alpha}(\sigma) \df\sigma^\alpha
  + \bbh_{s\alpha\beta}(\sigma) \df \sigma^\alpha \df \sigma^\beta, \nn\\
  \bbV^\sA(\sigma) \dow_\sA = \dow_-, \qquad
  \bbbeta^\sA(\sigma)\dow_\sA = \bbbeta^\alpha(\sigma) \dow_\alpha
  - \Lambda_\bbbeta(\sigma) \dow_-.
  \label{eq:map-FS}
\end{gather}
We can define the SK double copies of Galilean currents by varying the
effective action with respect to various sources (according to
\cref{eq:doubleEMTensor-null}), leading to
\begin{gather}
  \rho^\mu_1(X_1)
  = \frac{\hbar}{\sqrt{\gamma_1(X_1)}}
  \frac{\delta S}{\delta A_{1\mu}(X_1)}, \qquad
  \rho^\mu_2(X_2)
  = \frac{-\hbar}{\sqrt{\gamma_2(X_2)}}
  \frac{\delta S}{\delta A_{2\mu}(X_2)}, \nn\\
  \epsilon^{\mu}_1(X_1)
  = \frac{-\hbar}{\sqrt{\gamma_1(X_1)}}
  \frac{\delta S}{\delta n_{1\nu}(X_1)}, \qquad
  \epsilon^{\mu}_2(X_2)
  = \frac{\hbar}{\sqrt{\gamma_2(X_2)}}
  \frac{\delta S}{\delta n_{2\nu}(X_2)}, \nn\\
  \tau^{\mu\nu}_1(X_1)
  = \frac{2\hbar\,h^\nu_{1\rho}(X_1) h^\mu_{1\sigma}(X_1) }{\sqrt{\gamma_1(X_1)}}
  \frac{\delta S}{\delta h_{1\rho\sigma}(X_1)}, \qquad
  \tau^{\mu\nu}_2(X_2)
  = \frac{-2\hbar\,h^\nu_{2\rho}(X_2) h^\mu_{2\sigma}(X_2)}{\sqrt{\gamma_2(X_2)}}
  \frac{\delta S}{\delta h_{2\rho\sigma}(X_2)},
  \label{eq:SK-currents12-defn-NC}
\end{gather}
with $\gamma_s = \det(n_{s\mu}n_{s\nu} + h_{s\mu\nu})$. The relative signs
between the two copies of the operators arises from the respective second copies
being inserted on the time-reversed part of the SK contour. We
have taken $S$ to be unitless, leading to the additional factors of $\hbar$ in
these formulae. Classical equations of motion for the dynamical fields $X^\mu_s$
and $\varphi_s$ imply that these operators satisfy the conservation equations
\eqref{eq:NC.Conservation} independently on the two SK spacetimes.

\subsubsection{Effective field theory on physical spacetime}

We can translate the above fluid worldvolume framework into a more useful
physical spacetime language. The idea is that we can use the average coordinates
$x^\mu = X^\mu_r(\sigma)$ as coordinates on the physical NC spacetime. Inverting
this map, we can express the spacetime labels as dynamical fields on the
physical spacetime $\sigma^\alpha = \sigma^\alpha(x)$.  Accordingly, pushforward
the remaining dynamical fields leads to the physical U(1) phase
$\varphi_r(x) = \varphi_r(\sigma(x))$ and the stochastic noise fields
$X_a^\mu(x) = X^\mu_a(\sigma(x))$ and $\varphi_a(x) = \varphi_a(\sigma(x))$.

Let us decompose the fluid worldvolume invariants \eqref{eq:NCinvariants} into
average and difference combinations
$\bbn_{1,2\,\alpha} = \bbn_{r\alpha} \pm \hbar/2\,\bbn_{a\alpha}$,
$\bbh_{1,2\,\alpha\beta} = \bbh_{r\alpha\beta} \pm
\hbar/2\,\bbh_{a\alpha\beta}$, and
$\bbA_{1,2\,\alpha} = \bbA_{r\alpha} \pm \hbar/2\,\bbA_{a\alpha}$. It should be
noted that while $\bbh_{1,2\alpha\beta}$ are degenerate, their linear
combinations $\bbh_{a\alpha\beta}$ are \emph{not} generically degenerate.  We
can define the pullbacks onto the physical spacetime using $\sigma^\alpha(x)$
and $\varphi_r(x)$ (similar to \cref{eq:semiclassical-expansion-null})
as\footnote{Please don't confuse the clock-form pullbacks $N_{r,a\,\mu}$ with
  the classical free energy current $N^\mu$.}
\begin{align}
  N_{r\mu}(x)
  &= \bbn_{r\alpha}(\sigma(x))\, \dow_\mu\sigma^\alpha(x)
    = n_{r\mu}(x) + \mathcal{O}(\hbar), \nn\\
  H_{r\mu\nu}(x)
  &= \bbh_{r\alpha\beta}(\sigma(x)) \, 
    \dow_\mu\sigma^\alpha(x) \dow_\nu\sigma^\beta(x)
    = h_{r\mu\nu}(x) + \mathcal{O}(\hbar), \nn\\
  B_{r\mu}(x)
  &= \bbA_{r\alpha}(\sigma(x)) \, \dow_{\mu}\sigma^\alpha(x)
    - \dow_{\mu}\varphi_r(x)
    = A_{r\mu}(x) + \mathcal{O}(\hbar), \nn\\
  N_{a\mu}(x)
  &= \bbn_{a\alpha}(\sigma(x))\, \dow_\mu\sigma^\alpha(x)
    = n_{a\mu}(x) + \lie_{X_a}n_{r\mu}(x) + \mathcal{O}(\hbar), \nn\\
  H_{a\mu\nu}(x)
  &= \bbh_{a\alpha\beta}(\sigma(x)) \, 
    \dow_\mu\sigma^\alpha(x) \dow_\nu\sigma^\beta(x)
    = h_{a\mu\nu}(x) + \lie_{X_a} h_{r\mu\nu} + \mathcal{O}(\hbar), \nn\\
  B_{a\mu}(x)
  &= \bbA_{a\alpha}(\sigma(x)) \, \dow_{\mu}\sigma^\alpha(x)
    = A_{a\mu}(x) + \dow_\mu \varphi_a(x)
    + \lie_{X_a}A_{r\mu}(x) + \mathcal{O}(\hbar).
\end{align}
Here $\lie_{X_a}$ denotes a Lie derivative along $X^\mu_a(x)$. These quantities
are invariant under the fluid worldvolume symmetries
\eqref{eq:fluid-spacetime-symm-NC}, however they transform quite non-trivially
under Milne boosts \eqref{eq:milne-SK}. Note that $\bbA_{r\alpha}$ is the only
worldvolume quantity that transforms under the fluid worldvolume gauge
transformations and hence is subjected to a gauge shift during the
pullback.\footnote{During the relativistic discussion
  in~\cite{Glorioso:2018wxw}, the authors chose to not to apply a gauge shift in
  the definition of $B_{r\mu}$. This does not make any practical difference
  except that the resultant $B_{r\mu}$ would transform under the fluid
  worldvolume gauge transformations and invariant under the physical spacetime
  ones. Also, in the statistical limit, it behaves as
  $B_{r\mu} \to A_{r\mu} + \dow_\mu \varphi_r +
  \mathcal{O}(\hbar)$.\label{foot:Br-def}} In the second equalities, we have
decomposed the respective background fields into average and difference
combinations according to $n_{1,2\mu} = n_{r\mu} \pm \hbar/2 n_{a\mu}$,
$h_{1,2\mu\nu} = h_{r\mu\nu} \pm \hbar/2 h_{a\mu\nu}$, and
$A_{1,2\mu} = A_{r\mu} \pm \hbar/2 A_{a\mu}$, and expanded the expressions up to
leading order in $\hbar$.\footnote{See \cref{foot:add-background}.}  We can
identify the average background fields with the classical single-copy NC
background fields introduced in \cref{sec:null-redn}. Since the SK background
frame velocities $v^\mu_{1,2}(x)$ are in our hand, we can always choose these to
be parallel to each other, i.e.
\begin{equation}
  v^{\mu}_{1,2}(x) = \frac{v^\mu(x)}{v^\mu(x) n_{1,2\mu}(x)}, \qquad
  v^\mu(x)
  \equiv \frac{v_1^\mu(x)}{v_1^\nu(x) n_{r\mu}(x)}
  = \frac{v_2^\mu(x)}{v_2^\nu(x) n_{r\mu}(x)}.
\end{equation}
Note that $v^\mu n_{r\mu} = 1$. This ensures that $h_{r,a\mu\nu}$ are degenerate
and satisfy $v^\mu h_{r,a\mu\nu} = 0$, and fixes the ``off-diagonal'' part of
the Milne redundancy. Note that the objects $H_{r,a\mu\nu}$ for finite $\hbar$
are still not guaranteed to be degenerate. The average inverse spatial metric
$h_r^{\mu\nu}$ can be defined in the usual manner via
$h_r^{\mu\nu} n_{r\nu} = 0$ and
$h_{r}^{\mu\lambda} h_{r\lambda\nu} + v^\mu n_{r\nu} =
\delta^\mu_\nu$.\footnote{The ``1/2'' inverse spatial metrics are non-trivially
  related as
  $h_{1,2}^{\mu\nu} = h_r^{\mu\nu} \mp \hbar/2\,
  h_{r}^{\mu\rho}h_{r}^{\nu\sigma} h_{a\rho\sigma} \mp \hbar\, v^{(\mu}
  h_r^{\nu)\rho} n_{a\rho} + \mathcal{O}(\hbar^2)$. Similarly
  $h_{1,2}{}^\mu_{~\nu} = h^\mu_{r\nu} - \hbar/2\, v^\mu h^\lambda_{r\nu}
  n_{a\lambda} + \mathcal{O}(\hbar^2)$.}

We can similarly obtain the ``hydrodynamic fields'' $\beta^\mu(x)$ and
$\Lambda_\beta(x)$ ($=-\beta^-(x)$) by pushforward of $\bbbeta^\alpha(\sigma)$
and $\Lambda_\bbbeta(\sigma)$ respectively (according to
\cref{eq:SK-hydro-fields-null})
\begin{equation}
  \beta^\mu(x) = \bbbeta^\alpha(\sigma(x)) \dow_\alpha X_r^\mu(\sigma(x)), \qquad
  \Lambda_\beta(x) = \Lambda_\bbbeta(\sigma(x))
  + \bbbeta^\alpha(\sigma(x)) \dow_\alpha \varphi_r(\sigma(x)).
 \label{eq:SK-beta-fields-NC}
\end{equation}
If we were to choose $\bbbeta^\alpha(\sigma) = \beta_0\delta^\alpha_\tau$ and
$\Lambda_\bbbeta = \beta_0\mu_0$, these definitions merely become
$\beta^\mu = \beta_0 \dow_\tau X^\mu_r$ and
$\Lambda_\beta = \beta_0(\mu_0 + \dow_\tau \varphi_r)$.  The conventional
Newton-Cartan hydrodynamic fields: fluid velocity $u^\mu(x)$ (normalised as
$u^\mu N_{r\mu} = 1$), local temperature $T(x)$, and mass chemical potential
$\mu(x)$ are defined as (see \cref{eq:SK-conventional-hydro-fields-null})
\begin{gather}
  \kB T(x) = \frac{1}{\beta^\mu(x) N_{r\mu}(x)}, \qquad
  u^\mu(x) = \frac{\beta^\mu(x)}{\beta^\lambda(x) N_{r\lambda}(x)}, \nn\\
  \mu(x)
  = \frac{\beta^\mu(x) B_{r\mu}(x) + \Lambda_\beta(x)}{\beta^\lambda(x) N_{r\lambda}(x)}
  + \half \frac{\beta^\mu(x)\beta^\nu(x)}{(\beta^\lambda(x) N_{r\lambda}(x))^2}
  H_{r\mu\nu}(x)
  + \frac{\hbar^2}{4} \frac{\beta^\mu(x)\beta^\nu(x)
    N_{a\mu}B_{a\nu}}{(\beta^\lambda(x) N_{r\lambda}(x))^2}.
  \label{eq:SK-hydro-fields-NC}
\end{gather}
Note that in the statistical limit ($\hbar\to0$), all the dependence on the
non-stochastic dynamical fields $\sigma^\alpha(x)$ and $\varphi_r(x)$ in the
physical spacetime formulation comes only via the hydrodynamic fields. This
justifies the validity of the choice of degrees of freedom in classical
hydrodynamics.

In the physical spacetime representation, all the fluid worldvolume symmetries
of the effective theory are explicitly realised. As a payoff, however, we need
to implement the ``average''\footnote{Explicitly in terms of
  \cref{eq:phys-diffeo-NC}, one finds
  \begin{align*}
    x'^\mu(x)
    &= \half \lb
      X'^\mu_1(x+\hbar/2\,X_a(x)) + X'^\mu_2(x- \hbar/2\,X_a(x)) \rb
      = \half  \lb X'^\mu_1(x) + X'^\mu_2(x) \rb + \mathcal{O}(\hbar), \nn\\
    \Lambda(x)
    &= \half \lb \Lambda_1(x+\hbar/2\,X_a(x))+ \Lambda_2(x-\hbar/2\,X_a(x))\rb
      = \half \lb \Lambda_1(x)+ \Lambda_2(x) \rb + \mathcal{O}(\hbar).
  \end{align*}
  In the statistical limit, they turn into the average combinations of the SK
  spacetime transformations.  } part of the SK spacetime symmetries
\eqref{eq:phys-diffeo-NC} translated onto the physical spacetime through the
maps $\sigma(x)$ and $\varphi_r(x)$, i.e.
\begin{subequations}
  \begin{equation}
    x^\mu \to x'^\mu(x), \qquad
    \varphi_r(x) \to \varphi_r(x) - \Lambda(x).
  \end{equation}
  Various fields transform under these as usual
  \begin{gather}
    N_{r\mu}(x) \to N'_{r\mu}(x') =
    \frac{\dow x^\nu}{\dow x'^\mu} N_{r\nu}(x), \qquad
    N_{a\mu\nu}(x) \to N'_{a\mu\nu}(x') =
    \frac{\dow x^\nu}{\dow x'^\mu} N_{a\nu}(x), \nn\\
    H_{r\mu\nu}(x) \to H'_{r\mu\nu}(x') =
    \frac{\dow x^\rho}{\dow x'^\mu} \frac{\dow x^\sigma}{\dow x'^\nu}
    H_{r\rho\sigma}(x), \qquad
    H_{a\mu\nu}(x) \to H'_{a\mu\nu}(x') =
    \frac{\dow x^\rho}{\dow x'^\mu} \frac{\dow x^\sigma}{\dow x'^\nu}
    H_{a\rho\sigma}(x), \nn\\
    B_{r\mu}(x) \to B'_{r\mu}(x') =
    \frac{\dow x^\nu}{\dow x'^\mu}
    \lb B_{r\nu}(x) + \dow_\nu \Lambda(x) \rb, \qquad
    B_{a\mu}(x) \to B'_{a\mu}(x') =
    \frac{\dow x^\nu}{\dow x'^\mu} B_{a\nu}(x), \nn\\
    \beta^\mu(x) \to \beta'^\mu(x') =
    \frac{\dow x'^\mu(\sigma)}{\dow x^\nu} \bbbeta^\nu(x), \qquad
    \Lambda_\beta(x) \to \Lambda'_\beta(x') =
    \Lambda_\beta(x) - \beta^\mu(x) \dow_ \mu \Lambda(x).
  \end{gather}%
  \label{eq:PS-symm-NC}
\end{subequations}%
Note that the chemical potential $\mu(x)$ defined through
\cref{eq:SK-hydro-fields-NC} is gauge invariant.

The hydrodynamic effective action \eqref{eq:SK-action-NC} can also be translated
to the physical spacetime language leading to
\begin{equation}
  S[N_r,H_r,B_r,N_a,H_a,B_a;\beta,\Lambda_\beta]
  = \int \df^{d+1}x\sqrt{\gamma_r}\,
  \mathcal{L}[N_r,H_r,B_r,N_a,H_a,B_a;\beta,\Lambda_\beta],
  \label{eq:SK-action-PS-NC}
\end{equation}
with $\gamma_r = \det(n_{r\mu}n_{r\nu} + h_{r\mu\nu})$. The action is manifestly
invariant under all the worldvolume and physical spacetime symmetries, with the
exception of Milne invariance. The latter can be implemented in practise by
starting from the null background effective action \eqref{eq:S-FS-null} with the
identification
\begin{gather}
  G_{r\sM\sN}(x) \df x^\sM \df x^\sN
  = - 2 \lb \df x^- - B_{r\nu}(x)\df x^\nu \rb N_{r\mu}(x) \df x^\mu
  + \lb H_{r\mu\nu}(x)
  + \frac{\hbar^2}{2} N_{a(\mu}(x) B_{a\nu)}(x) \rb \df x^\mu \df x^\nu, \nn\\
  G_{a\sM\sN}(x) \df x^\sM \df x^\sN
  = - 2 \lb \df x^- - B_{r\nu}(x) \df x^\nu \rb N_{a\mu}(x) \df x^\mu
  + \lb H_{a\mu\nu}(x)
  + 2 N_{r(\mu}(x) B_{a\nu)}(x) \rb \df x^\mu \df x^\nu, \nn\\
  V^\sM(x)\dow_\sM = \dow_-, \qquad
  \beta^\sM(x) \dow_\sM = \beta^\mu(x)\dow_\mu
  - \Lambda_\beta(x) \dow_-.
  \label{eq:map-PS}
\end{gather}
We can use \cref{eq:SK-currents12-defn-NC} to define ``$r/a$'' basis of various
Galilean observables. We obtain the physical mass current, energy current, and
stress tensor by varying with respect to ``$a$'' sources
\begin{subequations}
  \begin{align}
    \rho_r^\mu(x)
    &= \frac{1/\hbar}{\sqrt{\gamma_r(x)}} \frac{\delta S}{\delta A_{a\mu}(x)}
      = \frac{1/2}{\sqrt{\gamma_r(x)}} \lb \sqrt{\gamma_1(x)} \, \rho_1^\mu(x)
      + \sqrt{\gamma_2(x)}\, \rho_2^\mu(x) \rb, \nn\\
    \epsilon_r^\mu(x)
    &= \frac{-1/\hbar}{\sqrt{\gamma_r(x)}} \frac{\delta S}{\delta n_{a\mu}(x)}
      = \frac{1/2}{\sqrt{\gamma_r(x)}} \lb \sqrt{\gamma_1(x)} \, \epsilon_1^\mu(x)
      + \sqrt{\gamma_2(x)}\, \epsilon_2^\mu(x) \rb, \nn\\
    \tau_r^{\mu\nu}(x)
    &= \frac{2/\hbar\, h^\mu_{r\rho}h^\nu_{r\sigma}}{\sqrt{\gamma_r(x)}}
      \frac{\delta S}{\delta h_{a\rho\sigma}(x)}
      = \frac{1/2\,h^\mu_{r\rho}h^\nu_{r\sigma}}{\sqrt{\gamma_r(x)}}
      \lb \sqrt{\gamma_1(x)} \, \tau_1^{\rho\sigma}(x)
      + \sqrt{\gamma_2(x)}\, \tau_2^{\rho\sigma}(x) \rb
      + \mathcal{O}(\hbar).
  \end{align}
  while the associated stochastic noise is obtained by varying with respect to
  ``$r$'' sources
  \begin{align}
    \rho_a^\mu(x)
    &= \frac{1/\hbar}{\sqrt{\gamma_r(x)}} \frac{\delta S}{\delta A_{r\mu}(x)}
      = \frac{1/\hbar}{\sqrt{\gamma_r(x)}} \lb \sqrt{\gamma_1(x)} \, \rho_1^\mu(x)
      - \sqrt{\gamma_2(x)}\, \rho_2^\mu(x) \rb, \nn\\
    \epsilon_a^\mu(x)
    &= \frac{-1/\hbar}{\sqrt{\gamma_r(x)}} \frac{\delta S}{\delta n_{r\mu}(x)}
      = \frac{1/\hbar}{\sqrt{\gamma_r(x)}} \lb \sqrt{\gamma_1(x)} \, \epsilon_1^\mu(x)
      - \sqrt{\gamma_2(x)}\, \epsilon_2^\mu(x) \rb, \nn\\
    \tau_a^{\mu\nu}(x)
    &= \frac{2/\hbar\, h^\mu_{r\rho}h^\nu_{r\sigma}}{\sqrt{\gamma_r(x)}}
      \frac{\delta S}{\delta h_{r\rho\sigma}(x)}
      = \frac{1/\hbar\, h^\mu_{r\rho}h^\nu_{r\sigma}}{\sqrt{\gamma_r(x)}} \lb \sqrt{\gamma_1(x)} \, \tau_1^{\rho\sigma}(x)
      - \sqrt{\gamma_2(x)}\, \tau_2^{\rho\sigma}(x) \rb 
      + \mathcal{O}(\hbar).
  \end{align}%
  \label{eq:singleEMTensor-NC}%
\end{subequations}%
These can also be derived directly via null reduction of
\cref{eq:singleEMTensor-null}. While both set of quantities satisfy conservation
equations in flat spacetime, they are not individually conserved in the presence
of background sources. The physical operators are, however, conserved in small
$\hbar$ limit. Note that we have only simplified expressions for stress tensor
$\tau^{\mu\nu}_r$ and respective noise $\tau^{\mu\nu}_a$ within in small $\hbar$
expansion; expressions for finite $\hbar$ are quite involved.

\subsubsection{\SK constraints and explicit effective action}

The SK generating functional for Galilean hydrodynamics can be defined same as
\cref{eq:SK-Z-null-FS} in the worldvolume formulation or \cref{eq:SK-Z-null} in
the physical spacetime formulation. Out-of-equilibrium ordered correlations
functions (retarded, advanced, symmetric etc.) can be obtained by varying the
generating functional with respect to ``$r/a$'' combinations of background
sources similar to \cref{eq:corr-funs}. The generating functional must obey the
SK constraints given in \cref{eq:SKW1}, with the map between the NC and
higher-dimensional sources given in \cref{eq:metric-decomposition}. Within the
effective field theory, these constraints are implemented by requiring the
effective action to obey \cref{eq:SK-cons-FS} in the worldvolume formulation or
\cref{eq:SK-constraints-null} in the physical spacetime formulation; note the
maps \cref{eq:map-FS} and \cref{eq:map-PS} between the NC and null background ingredients.

For reference, we note the KMS conjugation properties of various NC
objects. These are similar to the ones proposed for the relativistic case
in~\cite{Crossley:2015evo}. Firstly, the background sources transform as
\begin{align}
  \tilde n_{1\mu}(x) = \Theta n_{1\mu}(x), \qquad
  \tilde n_{2\mu}(x) = \Theta n_{2\mu}(t+i\hbar\beta_0,\vec x), \nn\\
  \tilde h_{1\mu\nu}(x) = \Theta h_{1\mu\nu}(x), \qquad
  \tilde h_{2\mu\nu}(x) = \Theta h_{2\mu\nu}(t+i\hbar\beta_0,\vec x), \nn\\
  \tilde A_{1\mu}(x) = \Theta A_{1\mu}(x), \qquad
  \tilde A_{2\mu}(x) = \Theta A_{2\mu}(t+i\hbar\beta_0,\vec x).
\end{align}
This leads to their average and difference combinations transforming in the
statistical limit as
\begin{align}
  \tilde n_{r\mu}(x) = \Theta n_{r\mu}(x) + \mathcal{O}(\hbar), \qquad
  \tilde n_{a\mu}(x) = \Theta n_{a\mu}(x)
  + i\beta_0\Theta\dow_t n_{r\mu}(x) + \mathcal{O}(\hbar), \nn\\
  \tilde h_{r\mu\nu}(x) = \Theta h_{r\mu\nu}(x) + \mathcal{O}(\hbar), \qquad
  \tilde h_{a\mu\nu}(x) = \Theta h_{a\mu\nu}(x)
  + i\beta_0\Theta\dow_t h_{r\mu\nu}(x) + \mathcal{O}(\hbar), \nn\\
  \tilde A_{r\mu}(x) = \Theta A_{r\mu}(x) + \mathcal{O}(\hbar), \qquad
  \tilde A_{a\mu\nu}(x) = \Theta A_{a\mu}(x)
  + i\beta_0\Theta\dow_t A_{r\mu}(x) + \mathcal{O}(\hbar).
\end{align}
For the fluid worldvolume quantities we find\footnote{If we were to fix
  $\bbbeta^\alpha(\sigma) = \beta_0\delta^\alpha_\tau$ and
  $\Lambda_\bbbeta(\sigma) = \beta_0\mu_0$, the transformations in the ``2''
  sector will merely become
  \begin{gather*}
    \tilde X^\mu_2(\sigma) = \Theta X_2^\mu(\tau + i\hbar\beta_0,\vec\sigma) -
    i\hbar\beta_0 \delta^\mu_{t}, \qquad \tilde\varphi_2(\sigma) =
    \Theta\varphi_2(\tau + i\hbar\beta_0,\vec\sigma) +
    i\hbar\beta_0\mu_0, \nn\\
    \tilde\bbn_{2\alpha}(\sigma) = \Theta\bbn_{2\alpha}(\tau +
    i\hbar\beta_0,\vec\sigma), \qquad \tilde\bbh_{2\alpha\beta}(\sigma) =
    \Theta\bbh_{2\alpha\beta}(\tau + i\hbar\beta_0,\vec\sigma), \qquad
    \tilde\bbA_{2\alpha}(\sigma) = \Theta\bbA_{2\alpha}(\tau +
    i\hbar\beta_0,\vec\sigma),
  \end{gather*}
  similar to the ones proposed in~\cite{Glorioso:2018wxw} (up to a constant
  shift of phase).  }
\begin{gather}
  \tilde X_1^\mu(\sigma) = \Theta X_1^\mu(\sigma), \qquad
  \tilde X_2^\mu(\sigma) = \Theta X_2^\mu(\sigma + i\hbar\,\Theta\bbbeta(\sigma))
  - i\hbar\beta_0 \delta^\mu_{t}, \nn\\
  \tilde\varphi_1(\sigma) = \Theta\varphi_1(\sigma), \qquad
  \tilde\varphi_2(\sigma) = \Theta\varphi_2(\sigma +
  i\hbar\,\Theta\bbbeta(\sigma))
  + i\hbar\, \Theta\Lambda_\bbbeta(\sigma), \nn\\
  \tilde\bbbeta^\alpha(\sigma) = \Theta\bbbeta^\alpha(\sigma), \qquad
  \tilde\Lambda_\bbbeta(\sigma) = \Theta\Lambda_\bbbeta(\sigma), \nn\\
  \tilde\bbn_{1\alpha}(\sigma) = \Theta\bbh_{1\alpha}(\sigma), \quad
  \tilde\bbn_{2\alpha}(\sigma)
  = \dow_\alpha \Big(\sigma^\beta +i\hbar\,\Theta\bbbeta^\beta(\sigma)\Big)
  \Theta\bbn_{2\beta}(\sigma + i\hbar\,\Theta\bbbeta(\sigma)), \nn\\
  \tilde\bbh_{1\alpha\beta}(\sigma) = \Theta\bbh_{1\alpha\beta}(\sigma), \quad
  \tilde\bbh_{2\alpha\beta}(\sigma)
  = \dow_\alpha \Big(\sigma^\gamma +i\hbar\,\Theta\bbbeta^\gamma(\sigma)\Big)
  \dow_\beta \Big(\sigma^\delta+i\hbar\,\Theta\bbbeta^\delta(\sigma)\Big)
  \Theta\bbh_{2\gamma\delta}(\sigma + i\hbar\,\Theta\bbbeta(\sigma)), \nn\\
  \tilde\bbA_{1\alpha}(\sigma) = \Theta\bbA_{1\alpha}(\sigma), \quad
  \tilde\bbA_{2\alpha}(\sigma)
  = \dow_\alpha \lb \sigma^\beta +i\hbar\,\Theta\bbbeta^\beta(\sigma)\rb
  \Theta\bbA_{2\beta}(\sigma + i\hbar\,\Theta\bbbeta(\sigma))
  - i\hbar\,\Theta \dow_\alpha \Lambda_\bbbeta(\sigma).
\end{gather}
The argument $(\sigma + i\hbar\,\Theta\bbbeta(\sigma))$ should be understood as
a vector, i.e. $\sigma^\alpha + i\hbar\,\Theta\bbbeta^\alpha(\sigma)$. The KMS
conjugation in the ``1'' sector is merely a $\Theta$-conjugation, while in the
``2'' sector it is given by a $\Theta$-conjugation followed by a diffeomorphism
along $i\hbar\,\Theta\bbbeta^\alpha(\sigma)$ and a gauge shift along
$i\hbar\,\Theta\Lambda_\bbbeta(\sigma)$. On the physical spacetime we find (in
the statistical limit)
\begin{gather}
  \tilde\sigma^\alpha(x) = \Theta\sigma^\alpha(x), \qquad
  \tilde X_a^\mu(x) = \Theta X^\mu_a(x) - i\Theta\beta^\mu(x)
  + i\beta_0\delta^\mu_t  + \mathcal{O}(\hbar), \nn\\
  \tilde\varphi_r(x) = \Theta\varphi_r(x), \qquad
  \tilde\varphi_a(x) = \Theta\varphi_a(x) - i\Theta\Lambda_\beta(x)
  + \mathcal{O}(\hbar), \nn\\
  \tilde\beta^\mu(x) = \Theta\beta^\mu(x) + \mathcal{O}(\hbar), \qquad
  \tilde\Lambda_\beta(x) = \Theta\Lambda_\beta(x) + \mathcal{O}(\hbar), \nn\\
  \tilde N_{r\mu}(x) = \Theta N_{r\mu}(x) + \mathcal{O}(\hbar), \qquad
  \tilde N_{a\mu}(x) = \Theta N_{a\mu}(x)
  + i\Theta\delta_\scB N_{r\mu}(x) + \mathcal{O}(\hbar), \nn\\
  \tilde H_{r\mu\nu}(x) = \Theta H_{r\mu\nu}(x) + \mathcal{O}(\hbar), \qquad
  \tilde H_{a\mu\nu}(x) = \Theta H_{a\mu\nu}(x)
  + i\Theta\delta_\scB H_{r\mu\nu}(x) + \mathcal{O}(\hbar), \nn\\
  \tilde B_{r\mu}(x) = \Theta B_{r\mu}(x) + \mathcal{O}(\hbar), \qquad
  \tilde B_{a\mu\nu}(x) = \Theta B_{a\mu}(x)
  + i\Theta\delta_\scB B_{r\mu}(x) + \mathcal{O}(\hbar).
  \label{eq:KMS-conjugation-G-rel}
\end{gather}
The operator $\delta_\scB$ combines a Lie derivative along $\beta^\mu$ and a
gauge transformation along $\Lambda_\beta$, i.e.
$\delta_{\scB}N_{r\mu} = \lie_\beta N_{r\mu}$,
$\delta_{\scB}H_{r\mu\nu} = \lie_\beta H_{r\mu\nu}$, and
$\delta_\scB B_{r\mu} = \lie_\beta B_{r\mu} + \dow_\mu \Lambda_\beta$.

In the statistical limit, the KMS transformation acts on the building blocks of
the effective action simply as $\scB \to\Theta\scB$, $\Phi_r \to \Theta\Phi_r$,
and $\Phi_a \to \Theta\Phi_a + i\Theta\delta_\scB \Phi_r$ for the hydrodynamic
fields $\scB = (\beta^\mu,\Lambda_\beta)$ and the invariants
$\Phi_{r,a} = (N_{r,a\mu}, 1/2\,H_{r,a\mu\nu}, B_{r,a\mu})$. The construction of
the explicit effective action allowed by the KMS constraints follows similar to
\cref{sec:classicalLimit-null} with a trivial substitution of
$1/2\,G_{r,a}\to \Phi_{r,a}$ and $\lie_\beta \to \delta_\scB$. In particular,
the most general effective for Galilean hydrodynamics is given in terms of a set
of real totally-symmetric multi-linear operators $\mathcal{D}_m(\circ,\ldots)$
made out of $\Phi_r$ and $\scB$, allowing $m$ number of arguments from the
vector space spanned by $i\delta_\scB\Phi_r$ and $\Phi_a$; to wit
\begin{align}
  \mathcal{L}
  &= \cD_1(\Phi_a)
    + i \sum_{n=1}^\infty
    \cD_{2n}(\underbrace{\Phi_a,\ldots}_{\times n},
    \underbrace{\Phi_a {+} i\delta_\scB\Phi_r,
    \ldots}_{\times n}) \nn\\
  &\qquad
    + \sum_{n=1}^\infty
    \cD_{2n+1}(\Phi_a{+}{\textstyle\frac{i}{2}}\delta_\scB \Phi_r,
    \underbrace{\Phi_a,\ldots}_{\times n},
    \underbrace{\Phi_a {+} i \delta_\scB \Phi_r,\ldots}_{\times n})
    + \mathcal{O}(\hbar).
    \label{eq:L-G-NC}%
\end{align}
The operators satisfy three constraints
\begin{subequations}
  \begin{gather}
    \mathcal{D}_1(\delta_\scB \Phi_r) = \frac{1}{\sqrt{\gamma_r}} \dow_\mu \lb
    \sqrt{\gamma_r}\, \mathcal{N}_0^\mu\rb
    \quad \text{for some vector~} \mathcal{N}_0^\mu, \\
    \mathcal{D}_m(\Phi_a,\Phi_a,\ldots)
    \quad \text{is $\Theta$-even}~\forall m, \\
    \mathcal{D}_2(\Phi,\Phi)\big|_{\text{leading order}} \geq 0
    \quad \text{for arbitrary~} \Phi = (N_\mu,1/2\, H_{\mu\nu}, B_\mu).
  \end{gather}
\end{subequations}
The classical constitutive relations only get contributions from the first three
operators $\mathcal{D}_{1,2,3}$. These agree with the adiabaticity equation
\eqref{eq:adiabaticity-NC}, which leads to the emergent local second law of
thermodynamics. The proof of the same follows from our discussion in
\cref{sec:classicalLimit-null}.

For example, the effective action for one-derivative Galilean hydrodynamics can
be reduced from its null fluid incarnation in \eqref{eq:1der-action}. We find a
slightly cumbersome expression
\begin{align}
  \mathcal{L}
  &= 
    \rho\, u^\mu \lb A_{a\mu} + \dow_{\mu}\varphi_a + \lie_{X_a} A_{r\mu} \rb \nn\\
  &\qquad
    - \Big( \lb \varepsilon + p
    + {\textstyle\half} \rho\, h_{r\rho\sigma}u^\rho u^\sigma \rb u^\mu
    - p\, v^\mu \Big)
    \lb n_{a\mu} + \lie_{X_a}n_{r\mu} \rb
    + \frac12 \lb \rho\, u^\mu u^\nu + p\, h_r^{\mu\nu} \rb
    \lb h_{a\mu\nu} + \lie_{X_a} h_{r\mu\nu} \rb \nn\\
  &\qquad
    + \lb - \frac{a_0}{2} \epsilon^{\mu\nu\rho\sigma} n_\nu \Omega_{\rho\sigma}
    + \frac{a_2}{2} \epsilon^{\mu\nu\rho\sigma} n_\nu F^n_{r\rho\sigma}
    \rb \nn\\
  &\qquad\qquad
    \lb
    \lb A_{a\mu} + \dow_{\mu} \varphi_a + \lie_{X_a} A_{r\mu} \rb
    + u^\lambda \lb h_{a\lambda\mu} + \lie_{X_a} h_{r\lambda\mu} \rb
    - \lb \vec u_\mu u^\lambda + \half \vec u^2 \delta^\lambda_\mu \rb
    \lb n_{a\lambda} + \lie_{X_a} n_{r\lambda} \rb \rb \nn\\
  &\qquad
    - \lb - \frac{a_2}{2} \epsilon^{\mu\nu\rho\sigma} n_\nu \Omega_{\rho\sigma}
    + \frac{a_1}{2} \epsilon^{\mu\nu\rho\sigma} n_\nu F^n_{r\rho\sigma}
    \rb
    \lb n_{a\mu} + \lie_{X_a} n_{\mu} \rb \nn\\
  &\qquad
    + \frac{i\kB T}{4}\lb 2\eta\, h_r^{\mu(\rho} h_r^{\sigma)\nu}
    + \lb \zeta - {\textstyle\frac{2}{d}}\eta \rb h_r^{\mu\nu}h_r^{\rho\sigma} \rb
    \lb h_{a\mu\nu} + \lie_{X_a}h_{r\mu\nu}
    - 2\lb n_{a(\mu} + \lie_{X_a}n_{r(\mu}\rb h_{r\nu)\lambda} u^\lambda \rb \nn\\
  &\qquad\qquad\qquad
    \lb h_{a\rho\sigma} 
    + \lie_{(X_a+i\beta)}h_{r\rho\sigma}
    - 2\lb n_{a(\mu} + \lie_{(X_a+i\beta)}n_{r(\mu}\rb h_{r\nu)\lambda} u^\lambda \rb \nn\\
  &\qquad
    + i\kB T^2\kappa\, h^{\mu\nu}_r \lb n_{a\mu} + \lie_{X_a}n_{r\mu} \rb
    \lb n_{a\nu} + \lie_{(X_a + i\beta)}n_{r\nu} \rb \nn\\
  &= 
    \rho^\mu \lb A_{a\mu} + \dow_{\mu}\varphi_a + \lie_{X_a} A_{r\mu} \rb
    - \epsilon^\mu \lb n_{a\mu} + \lie_{X_a}n_{r\mu} \rb
    + \lb v^\mu \pi^\nu
    + \half\tau^{\mu\nu} \rb \lb h_{a\mu\nu} + \lie_{X_a} h_{r\mu\nu} \rb
    \nn\\
  &\qquad
    + \frac{i\kB T}{4}\lb 2\eta\, h_r^{\mu(\rho} h_r^{\sigma)\nu}
    + \lb \zeta - {\textstyle\frac{2}{d}}\eta \rb h_r^{\mu\nu}h_r^{\rho\sigma} \rb
    \lb h_{a\mu\nu} + \lie_{X_a}h_{r\mu\nu}
    - 2\lb n_{a(\mu} + \lie_{X_a}n_{r(\mu}\rb h_{r\nu)\lambda} u^\lambda \rb \nn\\
  &\qquad\qquad\qquad
    \lb h_{a\rho\sigma} 
    + \lie_{X_a}h_{r\rho\sigma}
    - 2\lb n_{a(\mu} + \lie_{X_a}n_{r(\mu}\rb h_{r\nu)\lambda} u^\lambda \rb \nn\\
  &\qquad
    + i\kB T^2\kappa\, h^{\mu\nu}_r \lb n_{a\mu} + \lie_{X_a}n_{r\mu} \rb
    \lb n_{a\nu} + \lie_{X_a}n_{r\nu} \rb.
    \label{eq:Gal1derAction-NC}
\end{align}
Here $\lie_{X_a}$ denotes a Lie derivative along $X_a^\mu$ and
$\lie_{(X_a+i\beta)}$ along $X^\mu_a + i\beta^\mu$. In the second step, we have
substituted the thermodynamic mass frame constitutive relations from
\cref{eq:NC-consti-thermo}. This is the generalisation of the effective action
\eqref{eq:1der-action-NC-intro} presented in the introduction to include the
parity-violating terms. It can be explicitly checked that varying the action
with respect to difference sources $A_{a\mu}$, $n_{a\mu}$, and $h_{a\mu\nu}$, in
a configuration with zero ``$a$'' type fields
$X^\mu_a = \varphi_a = A_{a\mu} = n_{a\mu} = h_{a\mu\nu}$, leads to the
classical constitutive relations \cref{eq:NC-consti-thermo}, while extremising
with respect to $X^\mu_a$ and $\varphi_a$ leads to the conservation equations
\eqref{eq:NC.Conservation}.

\section{Outlook}
\label{sec:discussion}

In this work, we constructed a \SK effective field theory for dissipative
Galilean (non-relativistic) hydrodynamics. We used the null background formalism
of Galilean field theories to recast Galilean fluids in $(d+1)$ spacetime
dimensions in terms of relativistic null fluids in $(d+2)$ dimensions. This
allowed us to write down an EFT for Galilean fluids along the lines of its
relativistic cousin developed recently~\cite{Crossley:2015evo}, with appropriate
modifications to account for a null Killing vector. We explicitly constructed
the most generic effective action allowed by the formalism in the statistical
limit, where the quantum fluctuations are suppressed compared to thermal
fluctuations. We also inspected how the field theory gives rise to an emergent
second law of thermodynamics in the classical limit. As a concrete example, we
looked at the EFT for one-derivative dissipative Galilean fluids. We also looked
at the linearised limit of this theory, truncated to three point interactions,
which is appropriate to describe stochastic fluctuations in a Galilean fluid
around its equilibrium state.

The $(d+2)$-dimensional null fluid formalism is also convenient as it makes the
Galilean (Milne) boost symmetry manifest within the EFT, however the presence of
an auxiliary coordinate direction does obscure the underlying physics to some
degree. Therefore, we also spent some time translating our results to the more
widely used $(d+1)$-dimensional Newton-Cartan formulation of Galilean field
theories. While the two formulations are entirely equivalent at the end of the
day, the latter is technically more challenging as it involves a non-invertible
spatial metric, a torsional connection, and requires one to introduce a Galilean
observer to distinguish between energy and momentum. In the NC language, the
underlying structure of the EFT looks quite similar to that of charged
relativistic hydrodynamics from~\cite{Crossley:2015evo}, but with relativistic
symmetries and background sources swapped for the Galilean ones. The Galilean
EFT can also be arrived at by taking a non-relativistic limit of the
relativistic theory, as we illustrate in\cref{sec:nonrelLimit}.

Several generalisations of this work are immediately possible. We have assumed
the Galilean symmetry group underlying the fluid to be unbroken, but the EFT
requirements can also be relaxed to describe fluid phases with spontaneously
broken symmetries. For example, in the null background language, to describe
Galilean superfluids with spontaneously broken U(1) we can provide each fluid
element with a reference phase $\phi(\sigma)$. We must restrict the derivatives
of the phase as $\bbbeta^\sA \dow_\sA\phi = 0$ (since the reference phase must
be fixed in the comoving frame of the fluid elements) and
$\bbV^\sA \dow_\sA\phi = 1$ (required for spontaneously breaking the
U(1)).\footnote{Choosing
  $\bbbeta^\sA = \beta_0(\delta^\sA_\tau - \mu_0 \delta^\sA_-)$ and
  $\bbV^\sA = \delta^\sA_-$, these conditions merely say that $\dow_-\phi = 1$
  and $\dow_\tau\phi = \mu_0$. Recall that choosing this basis fixed the
  worldvolume diffeomorphisms \eqref{eq:fluid-spacetime-symm-null} down to
  spatial $\sigma^i$ diffeomorphisms, and residual spatial reparametrisations of
  $\sigma^\tau$ and $\sigma^-$. We can further fix the
  $\sigma^-$-reparametrisations explicitly by choosing
  $\phi(\sigma) = \sigma^- + \mu_0\sigma^\tau$. In the NC language, this amounts
  to lifting the worldvolume phase shift symmetry in
  \cref{eq:fluid-spacetime-symm-NC}. This is similar to what was proposed for
  relativistic superfluids in~\cite{Glorioso:2018wxw}. One must be careful in
  the Galilean case, however, because although $\bbA_{r\sigma}$ is now gauge
  invariant, it still transforms under Milne boosts.} On the physical spacetime,
this results in the reference phase satisfying $\beta^\sM\dow_\sM \phi = 0$ and
$V^\sM\dow_\sM\phi = 1$, leading to the Josephson's equation for Galilean
superfluids $u^\sM\dow_\sM \phi = \mu$. In the NC language, the Josephson's
equation is equivalently stated as $u^\mu\xi_\mu = \mu - \half \vec u^2$, where
$\xi_\mu = \dow_\mu \phi + A_{r\mu}$ is identified as the superfluid velocity;
see~\cite{Banerjee:2016qxf,Jain:2018jxj} for more discussion in case of
classical Galilean superfluids. Similarly, we can construct an EFT for
viscoelastic hydrodynamics with spontaneously broken translations by including
reference ``crystal coordinate fields'' $\phi^I(\sigma)$ with $I=1,\ldots,d$
satisfying $\bbbeta^\sA \dow_\sA \phi^I = 0$ and $\bbV^\sA \dow_\sA\phi^I = 0$;
see~\cite{Armas:2019sbe, Armas:2020bmo} for a classical analogue in the
relativistic case. More complicated symmetry breaking patterns are also
possible, describing a various phases of liquid crystals. The reference fields
to be included for such phases can be inferred by employing a coset construction
similar to~\cite{Landry:2019iel}.

The EFT can also be extended to the cases where the underlying global symmetries
are anomalous. In the simplest case that we are considering, the U(1) mass
symmetry does not admit any anomalies~\cite{Jain:2015jla}. The system can admit
gravitational/rotational anomalies, but these only show up in $d=4n+1$ spatial
dimensions at ($4n+2$)th derivative order in the parity-violating sector. In
particular, there are no such anomalies in $d=3$ and none at one-derivative
order in general. If one were to include additional symmetries in the Galilean
hydrodynamic setup, such as additional U(1) or non-Abelian charges
(e.g. electromagnetic or flavor), the anomaly structure can be much more
non-trivial; see~\cite{Jensen:2014hqa,Jain:2015jla}. The inclusion of anomalies
into the EFT, in such cases, would follow similar to the relativistic discussion
in e.g.~\cite{Jensen:2018hse}. Generalisations are also possible to more exotic
theories of hydrodynamics with higher-form symmetries (see
e.g.~\cite{Grozdanov:2018ewh, Grozdanov:2016tdf, Armas:2018ibg, Armas:2018zbe,
  Armas:2018atq}). So far, such dissipative EFTs have not been constructed even
in the relativistic case; see~\cite{Glorioso:2018kcp} for a non-dissipative
discussion of relativistic hydrodynamics with one-form symmetries.

In a more pragmatic direction, one can use the EFT for stochastic fluctuations
from \cref{sec:linear-Gal} to compute stochastic loop corrections to classical
hydrodynamic correlation functions. A similar analysis was done for the
simplified case of energy diffusion (with no momentum or density modes)
in~\cite{Chen-Lin:2018kfl}. Authors wrote down the most generic effective action
describing energy diffusion at one-derivative order and quartic order in
fluctuations, and worked out one-loop stochastic corrections to energy-energy
two-point functions. The discussion in \cref{sec:linear-Gal} can, in principle,
be extended to include quartic and higher interactions as well. However, as far
as one-loop corrections to two-point functions are concerned, even in full
hydrodynamics, the quartic interactions only contribute to renormalise the
hydrodynamic parameters (shear and energy diffusion constants, susceptibility,
etc.) and do not lead to any cutoff independent predictions.

Formally, the EFT for hydrodynamics has been setup for finite $\hbar$, capable
of describing both stochastic and quantum fluctuations within the same effective
framework. However, to concretely implement the KMS condition, we were forced to
work in the small $\hbar$ (statistical) limit. It is still an open question,
whether or not the effective KMS symmetry can be implemented at finite
$\hbar$. We should note that, in this work, we have focused on the hydrodynamic
EFT framework of~\cite{Crossley:2015evo}. There is another contending framework
due to~\cite{Haehl:2018lcu}, where the discrete KMS symmetry is replaced by a
local U(1)$_\sfT$ symmetry; see~\cite{Haehl:2017zac} for a comparison. It will
be interesting to revisit our Galilean results in the
context~\cite{Haehl:2018lcu}.

\acknowledgements

The author would like to thank Jay Armas, Pavel Kovtun, Kristan Jensen, and Hong
Liu for various helpful discussions and comments. This work is supported in part
by the NSERC Discovery Grant program of Canada.

\appendix

\section{Second law constraints in Galilean hydrodynamics}
\label{sec:second}

In this appendix we give a derivation of the second law constraints in Galilean
hydrodynamics. For ease, we will work in the null fluid framework. Translation
to NC language is straight-forward. We start with the decomposition of null
fluid energy-momentum tensor as in \cref{eq:null-fluid-EM} and write down
expressions for $p^{\sM\sN}$ and $q^\sM$ satisfying
$p^{\sM\sN}u_\sN = p^{\sM\sN}V_\sN = 0$ and $q^\sM u_\sM = q^\sM V_\sM = 0$
\begin{align}
  p^{\sM\sN}
  &= p\, \Delta^{\sM\sN}
  - \eta\, \sigma^{\sM\sN}
  - \zeta\, \Delta^{\sM\sN}\nabla_\sR u^\sR, \nn\\
  q^{\sM}
  &= - \kappa\, \Delta^{\sM\sN}\nabla_\sN T
    + T\tilde\kappa\, 2\nabla^{[\sM} V^{\sN]} u_\sN
    + \xi_\Omega \epsilon^{\sM\sN\sR\sS\sT} V_\sN u_\sR \nabla_\sS u_\sT
    + \xi_H \epsilon^{\sM\sN\sR\sS\sT} V_\sN u_\sR \nabla_\sS V_\sT.
\end{align}
At this point, all the coefficients appearing above are completely arbitrary. We
have introduced two new coefficients $\tilde\kappa$ and $\xi_H$ coupled to
$2\nabla_{[\sM}V_{\sN]} = 2\partial_{[\sM}g_{\sN]-}$ which do not contribute on
a flat background. These are similar to terms coupled to gauge field strength
$F_{\mu\nu}$ in relativistic theories. Similarly, for the non-canonical part of
the entropy current in \cref{eq:null-fluid-EM} we write down
\begin{align}
  \Upsilon^\sM
  &= \frac{a_2}{T} \epsilon^{\sM\sN\sR\sS\sT} V_\sN u_\sR \nabla_\sS u_\sT
    + \frac{a_1}{2T} \epsilon^{\sM\sN\sR\sS\sT} V_\sN u_\sR \nabla_\sS V_\sT
    - \frac{a_0}{2T} \epsilon^{-\sM\sN\sR\sS} u_\sN \nabla_\sR u_\sS.
\end{align}
The final term does not transform nicely under symmetries, but the respective
contribution to entropy current divergence can be made to behave nicely with a
judicious choice of the coefficient $a_0$. We have only written down
parity-violating terms in $\Upsilon^\sM$, because divergence of any
one-derivative parity-preserving term always contains a term with a
double-derivative acting on a quantity, which cannot be made positive
semi-definite. Computing the entropy-current divergence, we find
\begin{align}
  \nabla_\sM S^\sM
  &= 
    \frac{\eta}{2T} \sigma^{\sM\sN}\sigma_{\sM\sN}
    + \frac{\zeta}{T} (\nabla_\sM u^\sM)^2
    + \kappa\, \Delta^{\sM\sN}
    \lb\frac1T  \nabla_\sM T - 2\nabla_{[\sM}V_{\sR]} u^\sR \rb
    \lb \frac1T \nabla_\sN T - 2\nabla_{[\sN}V_{\sS]} u^\sS \rb
     \nn\\
  &\qquad
    - (\tilde\kappa - \kappa) H^{\sM\sR} u_\sR \lb\frac1T  \nabla_\sM T -
    2\nabla_{[\sM}V_{\sS]} u^\sS \rb \nn\\
  &\qquad
    - \frac1T \lb \xi_\Omega - a_2 + a_0\frac{\varepsilon+p}{\rho} \rb
    \epsilon^{\sM\sN\sR\sS\sT} V_\sN u_\sR \nabla_\sS u_\sT 
    \lb\frac1T  \nabla_\sM T - 2\nabla_{[\sM}V_{\sP]} u^\sP \rb \nn\\
  &\qquad
    - \frac{1}{T}
    \lb \xi_H - a_1 + a_2\frac{\epsilon+p}{\rho} \rb
    \epsilon^{\sM\sN\sR\sS\sT} V_\sN u_\sR \nabla_\sS V_\sT
    \lb\frac1T  \nabla_\sM T - 2\nabla_{[\sM}V_{\sP]} u^\sP \rb \nn\\
  &\qquad
    + T \nabla_\sM \bfrac{a_2 - a_0\mu}{T^2}
    \epsilon^{\sM\sN\sR\sS\sT} V_\sN u_\sR \nabla_\sS u_\sT
    + \frac{T^2}{2} \lb \nabla_\sM \frac{a_1}{T^3}
    - \frac{2a_2}{T^2} \nabla_\sM \frac{\mu}{T} \rb
    \epsilon^{\sM\sN\sR\sS\sT} V_\sN u_\sR \nabla_\sS V_\sT
      \nn\\
  &\qquad
    - \nabla_\sM \bfrac{a_0}{2T} \epsilon^{-\sM\sN\sR\sS} u_\sN \nabla_\sR u_\sS
    + \mathcal{O}(\dow^3).
    \label{eq:ECdiv-1der-app}
\end{align}
In deriving this expression, we used the first order equations of
motion. Terms in the first line can be made positive semi-definite by requiring
\begin{equation}
  \eta\geq 0, \qquad
  \zeta\geq 0, \qquad
  \kappa\geq 0.
\end{equation}
All the remaining terms, unfortunately, must be required to vanish. This leads
to equality constraints
\begin{gather}
  \tilde\kappa = \kappa \\
  a_0 = 2K_0T, \qquad
  a_1 = 2K_1 T^3 + 2K_2 T^2\mu + 2K_0T\mu^2, \qquad
  a_2 = K_2 T^2 + 2K_0T\mu, \nn\\
  \xi_\Omega = a_2 - \frac{\varepsilon+p}{\rho} a_0, \qquad
  \xi_H = a_1 - \frac{\epsilon+p}{\rho} a_2,
  \label{eq:transconstraints-app}
\end{gather}
for some constants $K_0$, $K_1$, $K_2$. Note that the constant $K_1$ only
appears in $\lambda_H$ and $a_1$ and does not contribute in the absence of
sources. 

\section{Effective field theory via non-relativistic limit}
\label{sec:nonrelLimit}

In this section we outline a systematic procedure to perform the
non-relativistic limit of relativistic hydrodynamics, to arrive at the Galilean
results presented in the main text. This essentially entails judiciously
introducing factors of $c$ (speed of light) so as to segregate spatial and
temporal components of various fields, and take $c\to\infty$ in the relevant
equations.

\subsection{Relativistic hydrodynamics}
\label{sec:relEFT}

\subsubsection{Classical hydrodynamics}

Let us start with a lightening recap of relativistic hydrodynamics. Classical
hydrodynamics is formulated in terms of conserved currents. However, unlike the
Galilean case discussed in \cref{sec:ClassicalGalilean}, the conserved currents
for charged relativistic hydrodynamics are the covariant energy-momentum tensor
$T^{\mu\nu}$ and charge/particle number current $J^\mu$, associated with
Poincar\'e and U(1) invariance of the underlying microscopic theory
respectively. The indices $\mu,\nu,\ldots$ run over the $(d+1)$-dimensional
spacetime coordinates. Coupling the setup to a background metric $g_{\mu\nu}$
with signature $(-1,1,1,\ldots)$ and gauge field $A_\mu$, the conservation
equations are expressed as
\begin{equation}
  \nabla_\mu T^{\mu\nu} = F^{\nu\rho} J_\rho, \qquad
  \nabla_\mu J^\mu = 0.
  \label{eq:conservation-rel}
\end{equation}
Here $\nabla_\mu$ is the covariant derivative associated with $g_{\mu\nu}$ and
$F_{\mu\nu} = 2\dow_{[\mu}A_{\nu]}$ is the antisymmetric field strength
tensor. We pick an appropriate set of degrees of freedom, called the
hydrodynamic fields, to be solved for using the conservation equations: fluid
velocity $u^\mu$ normalised as $u^\mu u_\mu = - c^2$, temperature $T$, and
chemical potential $\mu$. We leave the factors of speed of light $c$ explicit to
make contact with Galilean physics later. Normally, one needs to pick a
hydrodynamic frame to fix the inherent redefinition freedom in these variables;
for now we leave this freedom unfixed.

Hydrodynamic constitutive relations are the most generic expressions for
$T^{\mu\nu}$ and $J^\mu$, in accordance with symmetries, written in terms of the
dynamical fields $u^\mu$, $T$, $\mu$, and the background fields $g_{\mu\nu}$,
$A_\mu$, arranged in a derivative expansion. These constitutive relations are
required to satisfy the local second law of thermodynamics. Similar to our
discussion in \cref{sec:GalAdiabaticity}, the second law can be stated offshell
in terms of the adiabaticity equation~\cite{Loganayagam:2011mu}, i.e. there must
exist a free energy current $N^\mu$ and a quadratic form $\Delta$ such that
\begin{equation}
  \nabla_\mu N^\mu
  = \half T^{\mu\nu} \delta_\scB g_{\mu\nu}
  + J^\mu \delta_\scB A_\mu
  + \Delta, \qquad
  \Delta \geq 0,
  \label{eq:adiabaticity-rel}
\end{equation}
is satisfied offshell. Here $\delta_\scB g_{\mu\nu} = \lie_\beta g_{\mu\nu}$ and
$\delta_\scB A_\mu = \lie_\beta A_\mu + \dow_\mu \Lambda_\beta$ denotes a Lie
derivative along $\beta^\mu = u^\mu/(\kB T)$ and gauge shift along
$\Lambda_\beta = (\mu - u^\mu A_\mu)/(\kB T)$. The entropy current is defined as
$S^\mu = \kB N^\mu - (\mu J^\mu + T^{\mu\nu}u_\nu)/T$, whose divergence
satisfies $\nabla_\mu S^\mu = \kB\Delta \geq 0$ onshell, leading to the sign
definiteness of entropy production. By expressing the adiabaticity equation in
the form \eqref{eq:adiabaticity-rel}, we have partially fixed the redefinition
freedom in $u^\mu$, $T$, and $\mu$ noted above, formally known as a
``thermodynamic frame''. The residual freedom can be fixed by requiring certain
tensor structures to be eliminated from the constitutive relations using
equations of motion. We will make the choice to eliminate
$u^\mu\delta_\scB A_\mu$ and $u^\mu \delta_\scB g_{\mu\nu}$, which leads to the
``thermodynamic Landau frame'' wherein the non-hydrostatic (nhs) part of the
constitutive relations are transverse to the fluid velocity
$T^{\mu\nu}_{\text{nhs}} u_\nu = J^\mu_{\text{nhs}} u_\mu = 0$. This is distinct
from the conventional ``Landau frame'' employed in~\cite{landau1959fluid}, where
the full constitutive relations are chosen to obey
$T^{\mu\nu} u_\nu = -\varepsilon(T,\mu) u^\mu$ and $J^\mu u_\mu = -n(T,\mu)$,
with $\varepsilon$ and $n$ being the thermodynamic energy and charge densities
respectively.

For example, truncated to one-derivative order, the constitutive relations of a
charged relativistic fluid in thermodynamic Landau frame are given
as~\cite{Banerjee:2012iz} (see~\cite{Jain:2018jxj} for a
review)\footnote{\label{foot:rel-Landau}In the conventional Landau frame, the
  constitutive relations are equivalently given by
  \begin{align}
    T^{\mu\nu}
    &= \frac{1}{c^2}(\epsilon+p)u^\mu u^\nu
      + p\, g^{\mu\nu} - \eta\, \sigma^{\mu\nu} -
      \zeta\,P^{\mu\nu} \nabla_\lambda u^\lambda, \nn\\
    J^\mu
    &= n\, u^\mu - \frac{\sigma}{c^4} P^{\mu\nu}
      \lb T\dow_\nu \frac{\mu}{T} - E_{\nu} \rb
      + \frac{1}{2c}\xi_B \epsilon^{\mu\nu\rho\sigma} u_\nu F_{\rho\sigma}
      + \frac{1}{c^3} \xi_\Omega
      \epsilon^{\mu\nu\rho\sigma}u_\nu\dow_\rho u_\sigma.
  \end{align}
  In this frame, the constitutive relations do not satisfy the adiabaticity
  equation offshell.
}
\begin{align}
  T^{\mu\nu}
  &= \frac{1}{c^2}(\varepsilon +p)u^\mu u^\nu
    + p\, g^{\mu\nu} - \eta\, \sigma^{\mu\nu} -
    \zeta\,\Delta^{\mu\nu} \nabla_\lambda u^\lambda
    + \frac{\alpha_1}{c^5} 2u^{(\mu}\epsilon^{\nu)\rho\sigma\lambda}
    u_\rho \dow_\sigma u_\lambda
    + \frac{\alpha_2}{2c^3} 2u^{(\mu} \epsilon^{\nu)\rho\sigma\lambda}
    u_\rho F_{\sigma\lambda}, \nn\\
  J^\mu
  &= n\, u^\mu - \sigma \Delta^{\mu\nu}
    \lb T\dow_\nu \frac{\mu}{T} - F_{\nu\lambda}u^\lambda \rb
    + \frac{\alpha_0}{2c} \epsilon^{\mu\nu\rho\sigma} u_\nu F_{\rho\sigma}
    + \frac{\alpha_2}{c^3} \epsilon^{\mu\nu\rho\sigma} u_\nu \dow_\rho u_\sigma,
    \label{eq:rel-consti}
\end{align}
where $\Delta^{\mu\nu} = g^{\mu\nu} + u^\mu u^\nu/c^2$ is the projector
transverse to fluid velocity. The convention for the Levi-Civita tensor is
$\epsilon_{0123} = \sqrt{-g}$. Thermodynamic pressure $p(T,\mu)$ is related to
energy density $\varepsilon(T,\mu)$, charge/particle number density $n(T,\mu)$,
and entropy density $s(T,\mu)$ via the thermodynamic relations
$\df p = s\df T + n\df \mu$ and $\varepsilon = Ts+\mu n-p$. In the parity-even
sector, $\eta(T,\mu)$, $\zeta(T,\mu)$, and $\sigma(T,\mu)$ are non-negative
dissipative transport coefficients identified as shear viscosity, bulk
viscosity, and electric conductivity respectively.
$\sigma^{\mu\nu} =
2\Delta^{\mu\rho}\Delta^{\nu\sigma}(\nabla_{(\rho}u_{\sigma)}-1/d\,
\Delta_{\rho\sigma} \nabla_\lambda u^\lambda)$ is the fluid shear tensor. While
the parity-even part of the constitutive relations is applicable for any number
of dimensions, the parity-odd part is dimension specific. The above expressions
are valid in $d=3$ spatial dimensions, with the transport coefficients
\begin{gather}
  \alpha_0 =
  % \frac{6\mu C}{c^6} +
  2C_0T, \qquad
  \alpha_1 =
  % \frac{2\mu^3 C}{c^6} +
  2C_1T^3 + 2C_2 T^2\mu +  2C_0 T \mu^2, \qquad
  \alpha_2 = % \frac{3\mu^2 C}{c^6} +
  C_2T^2 + 2C_0T\mu,
  \label{eq:relC}
\end{gather}
for three arbitrary constants $C_0$, $C_1$, $C_2$.\footnote{If the U(1) symmetry
  is anomalous, \cref{eq:relC} will also include contributions from the anomaly
  coefficient.} These are related to the physical parity-odd conductivities as
$\xi_B = \alpha_0 - n/(\varepsilon+p) \alpha_2$ and
$\xi_\Omega = \alpha_2 - n/(\varepsilon+p) \alpha_1$ (see
\cref{foot:rel-Landau}). The associated free-energy density and quadratic form
are given as
\begin{align}
  \kB N^\mu
  &= \frac{p}{T} u^\mu
    + \frac{\alpha_1}{2Tc^3} \epsilon^{\mu\nu\rho\sigma}u_\nu \dow_\rho u_\sigma
    + \frac{\alpha_2}{2Tc} \epsilon^{\mu\nu\rho\sigma}u_\nu F_{\rho\sigma}
    + \frac{c}{2} C_0 \epsilon^{\mu\nu\rho\sigma} A_\nu F_{\rho\sigma}, \nn\\
  \kB\Delta
  &= \frac{\eta}{2T} \sigma^{\mu\nu} \sigma_{\mu\nu}
    + \frac{\zeta}{T} (\nabla_\lambda u^\lambda)^2
    + \frac{\sigma}{T} P^{\mu\nu}
    \lb T\dow_\mu \frac{\mu}{T} - F_{\mu\lambda} u^\lambda \rb
    \lb T\dow_\nu \frac{\mu}{T} - F_{\nu\lambda} u^\lambda \rb.
    \label{eq:N-rel}
\end{align}
Note that there is a gauge-non-invariant term in $N^\mu$ coupled to $C_0$. This
is not an issue because the free-energy current $N^\mu$ itself is not a physical
observable, only the constitutive relations are.

% where  is the projector transverse to
% $u^\mu$. 
% \begin{gather}
%   \tilde b_1 = \frac{3\mu^2}{Tc^6} C + 2\mu C_0 + T C_2, \qquad
%   \tilde b_2 = \frac{\mu^3}{Tc^6} C + \mu^2 C_0 + T^2 C_1 + \mu T C_2, \nn\\
%   \tilde a_1 = 2\mu C_0 + 2T C_2, \qquad
%   \tilde a_2 = \mu^2 C_0 + 3T^2 C_1 + 2\mu T C_2, \nn\\
% \end{gather}

\begin{table}[t]
  \centering
  \begin{tabular}{c|ccc|c|c}
    \toprule
    & C & P & T & PT & CPT \\
    \midrule
    $X^0$, $ct = x^0$, $c\tau = \sigma^0$  & $+$ & $+$ & $-$ & $-$ & $-$ \\
    $X^i$, $x^i$, $\sigma^i$ & $+$ & $-$ & $+$ & $-$ & $-$ \\
    $\varphi$ & $-$ & $+$ & $-$ & $-$ & $+$ \\
    \midrule

    $u^i$, $\beta^i$, $\bbbeta^i$ & $+$ & $-$ & $-$ & $+$ & $+$ \\
    $T$, $\beta^0$, $\bbbeta^0$ & $+$ & $+$ & $+$ & $+$ & $+$ \\
    $\mu$, $\Lambda_\beta$, $\Lambda_\bbbeta$ & $-$ & $+$ & $+$ & $+$ & $-$ \\
    
    \midrule
    $T^{00}$, $g_{00}$ & $+$ & $+$ & $+$ & $+$ & $+$ \\
    $T^{0i}$, $g_{0i}$ & $+$ & $-$ & $-$ & $+$ & $+$ \\
    $T^{ij}$, $g_{ij}$ & $+$ & $+$ & $+$ & $+$ & $+$ \\
    $J^0$, $A_0$ & $-$ & $+$ & $+$ & $+$ & $-$ \\
    $J^i$, $A_i$ & $-$ & $-$ & $-$ & $+$ & $-$ \\
    \bottomrule
  \end{tabular}
  \caption{Action of parity (P), time-reversal (T), and charge conjugation (C)
    of various quantities in classical relativistic hydrodynamics and effective
    field theory. \SK double copies of various quantities in the effective
    theory, ``$1/2$'' or ``$r/a$'', behave the same as their unlabelled
    versions.\label{tab:CPT-rel}}
\end{table}

We can also demand the hydrodynamic theory to obey additional constraints like
the discrete time-reversal symmetry T, spacetime parity PT, or the full CPT
including a charge conjugation, denoted collectively as $\Theta$; see
\cref{tab:CPT-rel}. These are slightly subtle as they need to be implemented at
the microscopic level and not at the level of constitutive relations; we expect
the dissipative effects to explicitly violate any kind of
$\Theta$-symmetry. Nonetheless, constitutive relations evaluated on a
hydrostatic configuration, characterised by
$\delta_\scB g_{\mu\nu} = \delta_\scB A_\mu = 0$ in a thermodynamic frame, must
respect $\Theta$~\cite{Banerjee:2012iz,Jensen:2012jh}. In the non-hydrostatic
sector, discrete symmetries lead to Onsager's reciprocity relations among
hydrodynamic retarded two-point functions; see~\cite{Kovtun:2012rj}. For
instance, taking $\Theta$ to be T does not lead to any constraints on the
constitutive relations \eqref{eq:rel-consti}, while PT forces us to set
$C_{0,1,2} = 0$. On the other hand, choosing $\Theta$ to be CPT sets
$C_{0,1} = 0$ leaving only $C_2$ in the parity-violating
sector~\cite{Banerjee:2012iz}. The effective field theory framework deals with
these discrete symmetries more systematically, combining them with the thermal
KMS condition.

\subsubsection{\SK effective field theory}

An effective field theory framework for relativistic hydrodynamics has been
proposed in~\cite{Crossley:2015evo, Glorioso:2017fpd, Glorioso:2016gsa}. We will
not delve in the technicalities; a comprehensive review can be found
in~\cite{Glorioso:2018wxw}. We directly note the effective action for
one-derivative relativistic hydrodynamics in the statistical limit. The
ingredients we need to introduce are the (physical spacetime formulation)
dynamical fields: fluid spacetime labels $\sigma^\alpha(x)$, phase field
$\varphi_r(x)$, associated stochastic noise fields $X_a^\mu(x)$, $\varphi_a(x)$,
two sets of background fields $g_{r\mu\nu}(x) = g_{\mu\nu}(x)$,
$A_{r\mu}(x) = A_{\mu}(x)$, $g_{a\mu\nu}(x)$, and $A_{a\mu}(x)$, and reference
thermal vector $\bbbeta^\alpha(\sigma) = \beta_0\delta^\alpha_\tau$ and chemical
shift field $\Lambda_\bbbeta(\sigma) = \beta_0\mu_0$. The hydrodynamic fields
$\beta^\mu(x)$, $\Lambda_\beta(x)$ are related to these as same as
\cref{eq:SK-beta-fields-NC}. The effective action is found to be
\begin{align}
  \mathcal{L}
  &= \half \lb \frac{\varepsilon+p}{c^2} u^\mu u^\nu
    + p\,\Delta^{\mu\nu}
    + \frac{\alpha_1}{c^5} 2u^{(\mu}\epsilon^{\nu)\rho\sigma\lambda}
    u_\rho \dow_\sigma u_\lambda
    + \frac{\alpha_2}{2c^3} 2u^{(\mu} \epsilon^{\nu)\rho\sigma\lambda}
    u_\rho F_{\sigma\lambda}
    \rb G_{a\mu\nu} \nn\\
  &\qquad
    + \lb n\, u^\mu
    + \frac{\alpha_0}{2c} \epsilon^{\mu\nu\rho\sigma} u_\nu F_{\rho\sigma}
    + \frac{\alpha_2}{c^3} \epsilon^{\mu\nu\rho\sigma} u_\nu \dow_\rho u_\sigma
    \rb B_{a\mu} \nn\\
  &\qquad 
    + \frac{i\kB T}{4} \lb 2\eta \Delta^{\mu(\rho}\Delta^{\sigma)\nu}
    + \lb\zeta-{\textstyle\frac{d}{2}}\eta\rb \Delta^{\mu\nu}\Delta^{\rho\sigma}
    \rb G_{a\mu\nu}
    \lb G_{a\rho\sigma} + i\delta_\sB G_{r\rho\sigma} \rb \nn\\
  &\qquad
    + i\kB T\sigma P^{\mu\nu} B_{a\mu} \lb B_{a\mu} + i\delta_\scB B_{r\mu} \rb.
    \label{eq:rel-action}
\end{align}
Here $G_{r\mu\nu} = g_{r\mu\nu}$, $B_{r\mu} = A_\mu$,
$G_{a\mu\nu} = g_{a\mu\nu} + \lie_{X_a} g_{\mu\nu}$, and
$B_{a\mu} = A_{a\mu} + \dow_\mu \varphi_a + \lie_{X_a} A_\mu$. The operator
$\delta_\scB$ acts same as defined below \cref{eq:adiabaticity-rel}. It is easy
to see that varying with respect to the sources $g_{a\mu\nu}$ and $A_{a\mu}$,
and switching off all ``$a$'' type fields, one recovers the classical
constitutive relations \eqref{eq:rel-consti}. The classical conservation
equations, on the other hand, are obtained by varying with respect to $X^\mu_a$
and $\varphi_a$.

We note that the effective action \eqref{eq:rel-action} conforms with the
general structure of the effective action advertised in \cref{eq:L-G-null} with
$1/2\,G_a$ replaced with $\Phi_{r,a} = (1/2\,G_{r,a\mu\nu}, B_{r,a\mu})$ and
$\lie_\beta$ with $\delta_\scB$. In fact, with these trivial substitutions, the
entire discussion of \cref{eq:rel-action} follows through to relativistic
hydrodynamics line by line and can be used to extend the effective field theory
to arbitrarily high orders in the derivative expansion and noise expansion.

\subsection{Non-relativistic limit}
\label{sec:nr-limit}

We are now ready to perform the non-relativistic limit. A general discussion on
the procedure can be found in~\cite{Jensen:2014wha}. The limit proceeds in three
steps: (1) identify the dynamical and background fields between Galilean
hydrodynamics and the relativistic parent, (2) choose $c$-scaling for various
transport coefficients, and (3) take $c\to\infty$. For the first step, we take
the prescription from~\cite{Jensen:2014wha}. The second step is hit-and-trial;
choosing the $c$-scaling for a transport coefficient to be too high will lead to
a divergent $c\to\infty$ limit and is not good, while choosing it to be too low
will cause it to vanish in $c\to\infty$ limit and we will miss out on some
transport coefficients.

Briefly in this section, we use the label ``rel'' for relativistic quantities in
case of notational tension.

\subsubsection{Newton-Cartan sources via non-relativistic limit}

A non-relativistic background is characterised by the presence of a preferred
one-form $n_\mu$ characterising the non-relativistic notion of absolute time. It
is convenient to introduce a local Galilean frame velocity $v^\mu$ such that
$v^\mu n_\mu = 1$. We can use this to decompose the background fields in a
series expansion in $c^2$ according to
\begin{gather}
  g_{\mu\nu} = - c^2 n_\mu n_\nu + h_{\mu\nu}  + \mathcal{O}(1/c^2), \qquad
  g^{\mu\nu} = - \frac{1}{c^2} v^\mu v^\nu + h^{\mu\nu}   + \mathcal{O}(1/c^2), \nn\\
  A^\rel_\mu = m c^2 n_\mu + m A_\mu  + \mathcal{O}(1/c^2),
  \label{eq:rel-nonrel-background-map}
\end{gather}
where $h_{\mu\nu}v^\nu = h^{\mu\nu} n_\nu = 0$ and
$n_\mu v^\nu + h_{\mu\lambda}h^{\lambda\nu} = \delta^\nu_\mu$. Here $m$ is a
constant representing mass per particle. The constituent fields above can be
identified with the Newton-Cartan structure discussed in
\cref{sec:NC-formulation}. The relativistic diffeomorphism and U(1) invariance
maps equivalently to the Galilean ones given in \cref{eq:NC.diffeo}. On the
other hand, the ambiguity in the choice of the frame velocity $v^\mu$ leads to a
redundancy, for some $\psi^\mu$ satisfying $\psi^\mu n_\mu = 0$,
\begin{gather}
  v^\mu \to v^\mu + \psi^\mu, \qquad
  h_{\mu\nu} \to h_{\mu\nu} - 2 n_{(\mu}\psi_{\nu)} + n_\mu n_\nu \psi^2, \qquad
  A_\mu \to A_\mu + \psi_{\mu} - \half n_\mu \psi^2, \nn\\
  n_\mu \to n_\mu - \frac{1}{c^2}\lb \psi_{\mu} - \half n_\mu \psi^2\rb, \qquad
  h^{\mu\nu} \to
  h^{\mu\nu} + \frac{1}{c^2} \lb 2 v^{(\mu}\psi^{\nu)} + \psi^\mu \psi^\nu \rb,
\end{gather}
identified as Galilean Milne invariance in $c\to\infty$ limit. Since the metric
has a $c^2$ piece, the associated Levi-Civita connection
$\Gamma^\lambda_{\mu\nu}$ is not well defined in $c\to\infty$
limit. Nonetheless, we can define a torsional connection
$\tilde\Gamma^\lambda_{\mu\nu}$ using $g_{\mu\nu}$ and $F^\rel_{\mu\nu}$ which
behaves regularly. We have
\begin{align}
  \tilde\Gamma^\lambda_{\mu\nu}
  &= \Gamma^\lambda_{\mu\nu}
    - \frac{1}{2m} g^{\lambda\rho} \lb n_\rho F^\rel_{\mu\nu}
    - 2n_{(\mu} F^\rel_{\nu)\rho} \rb
    \nn\\
  &= v^\lambda \dow_{\mu} n_{\nu}
    + \half h^{\lambda\rho} \lb 2\dow_{(\mu} h_{\nu)\rho} - \dow_\rho h_{\mu\nu} \rb
    + n_{(\mu} F_{\nu)\rho} h^{\rho\lambda} + \mathcal{O}(1/c).
\end{align}
Note that the connection still preserves the relativistic metric $g_{\mu\nu}$.
This yields the Newton-Cartan connection \eqref{eq:NC-connection} in
$c\to\infty$ limit. Finally, the Levi-Civita tensor scales as
$\epsilon_\rel^{\mu\nu\rho\sigma} = 1/c\, \epsilon^{\mu\nu\rho\sigma}$, with the
$1/c$ factor coming from the measure.

We can define the non-relativistic conserved currents as 
\begin{align}
  \rho^\mu
  &= \lim_{c\to\infty} m J^\mu, \qquad
  \epsilon^\mu
  = - \lim_{c\to\infty} \lb T^{\mu}_{~~\nu} v^\nu + mc^2 J^\mu \rb, \qquad
  \tau^{\mu\nu}
    = \lim_{c\to\infty} h^\mu_{~\rho} h^{\nu\sigma} T^\rho_{~\sigma}.
    \label{eq:rel-nonrel-current-map}
\end{align}
These limits must be well defined for the system to admit a non-relativistic
limit. The Galilean conservation equations follow simply from the relativistic
ones\footnote{In case the U(1) symmetry in the relativistic theory is anomalous,
  i.e.
  $\nabla_\mu J^\mu = - 3C/(4c^5) \epsilon^{\mu\nu\rho\sigma}_\rel
  F^\rel_{\mu\nu}F^\rel_{\rho\sigma}$, the anomaly gets washed away in the
  non-relativistic limit. It only leads to a shift of the non-relativistic
  energy current
  $\epsilon^\mu \to \epsilon^\mu - 2Cm^3\epsilon^{\mu\nu\rho\sigma}
  n_\nu\dow_\rho n_\sigma$.}
\begin{align}
  m \nabla_\mu J^\mu
  &= \lb \tilde\nabla_\mu + n_{\mu\lambda} v^\lambda \rb \rho^\mu
    + \mathcal{O}(1/c) = 0, \nn\\
  - v^\nu \bigg( \nabla_\mu T^\mu_{~\,\nu}
  &- F^\rel_{\nu\lambda} J^\lambda
  + m c^2 n_\nu \nabla_\mu J^\mu \bigg) \nn\\
  &= \lb \tilde\nabla_\mu + n_{\mu\lambda} v^\lambda \rb
    \epsilon^\mu
    - v^\lambda n_{\lambda\mu} \epsilon^\mu
    + \lb v^{\mu} \pi^\lambda + \tau^{\mu\lambda} \rb
    h_{\lambda\rho} \tilde\nabla_\mu v^\rho
    + \mathcal{O}(1/c) = 0, \nn\\
  h^{\rho\nu}\bigg( \nabla_\mu T^\mu_{~\,\nu}
  &- F^\rel_{\nu\lambda} J^\lambda \bigg) \nn\\
  &= \lb \tilde\nabla_\mu + n_{\mu\lambda} v^\lambda \rb
    \lb v^{\mu} \pi^\rho + \tau^{\mu\rho} \rb
    + h^{\rho\lambda} n_{\lambda\mu} \epsilon^\mu
    + \rho^\mu \tilde\nabla_\mu v^\rho
    + \mathcal{O}(1/c) = 0.
\end{align}
These are same as the ones derived in \cref{eq:NC.Conservation}.

\subsubsection{Non-relativistic limit of hydrodynamics}

Normalisation of the fluid velocity $u^\mu u^\nu g_{\mu\nu} = -c^2$ implies that
\begin{gather}
  u^\mu n_\mu = 1 + \frac{1}{2c^2} \vec u^2 + \mathcal{O}(1/c^{4}), \qquad
  u_\mu = - c^2 n_\mu - \frac{1}{2} \vec u^2 n_\mu + \vec u_\mu
  + \mathcal{O}(1/c^{2}),
\end{gather}
where $\vec u_\mu = h_{\mu\nu} u^\nu$ and $\vec u^2 = h_{\mu\nu} u^\mu
u^\nu$. Having set this, through a straight-forward calculation we can deduce
that the Galilean fluid constitutive relations \eqref{eq:NC-consti} follow by
taking $c\to\infty$ in the relativistic fluid constitutive relations
\eqref{eq:rel-consti} with the mapping \eqref{eq:rel-nonrel-current-map},
provided that we choose
\begin{gather}
  n = \frac{1}{m}\rho, \qquad
  \varepsilon_\rel = \rho c^2 + \varepsilon, \qquad
  \mu_\rel = mc^2 + m\mu, \qquad
  \sigma = \frac{T\kappa}{m^2c^4}, \nn\\
  C_0 = \frac{1}{m^2} K_0, \qquad
  C_1 = K_1, \qquad
  C_2 = \frac{1}{m} K_2.
\end{gather}
Most of these identifications are expected on physical grounds: the number
density is given by mass density divided by $m$, relativistic energy density has
a rest mass contribution and so does the relativistic chemical potential. The
mapping of electric conductivity to thermal conductivity is slightly unnatural,
but it follows from dimensional analysis. Similarly, mapping of $C_{0,1,2}$ to
$K_{0,1,2}$ also follows on dimensional grounds. The remaining relativistic
coefficients and quantities map to their namesake on the Galilean side.

Similar mapping rules can be applied to the effective field theory. The
dynamical fields $\sigma^\alpha$, $\varphi_r$, and $X_a^\mu$, $\varphi_a$
directly map to their respective Galilean versions, while the average background
fields $g_{r\mu\nu}$, $A_{r\mu}^\rel$ map to $n_{r\mu}$, $h_{r\mu\nu}$,
$A_{r\mu}$ according to \cref{eq:rel-nonrel-background-map}. We just need the
mapping between the difference background fields, which we propose
\begin{gather}
  g_{a\mu\nu} = - 2c^2 n_{r(\mu} n_{a\nu)} + h_{a\mu\nu}  + \mathcal{O}(1/c^2), \qquad
  A^\rel_{a\mu} = m c^2 n_{a\mu} + m A_{a\mu}  + \mathcal{O}(1/c^2).
\end{gather}
We leave it to the reader to verify that this identification maps the
relativistic one-derivative order effective action \eqref{eq:rel-action} to the
Galilean one in \cref{eq:Gal1derAction-NC}. In principle, the same procedure can
be iterated to arbitrarily high orders in the derivative expansion.

\makereferences


\providecommand{\href}[2]{#2}\begingroup\raggedright\begin{thebibliography}{10}

\bibitem{Martin:1973zz}
P.~Martin, E.~Siggia and H.~Rose, \emph{{Statistical Dynamics of Classical
  Systems}}, \href{https://doi.org/10.1103/PhysRevA.8.423}{\emph{Phys.\ Rev.\
  A} {\bfseries 8} (1973) 423}.

\bibitem{Pomeau:1974hg}
Y.~Pomeau and P.~Resibois, \emph{{Time Dependent Correlation Functions and
  Mode-Mode Coupling Theories}}, {\emph{Phys. Rept.} {\bfseries 19} (1974) 63}.

\bibitem{1974Phy....75....1D}
I.~M. {De Schepper}, H.~{Van Beyeren} and M.~H. {Ernst}, \emph{{The
  nonexistence of the linear diffusion equation beyond Fick's law}},
  \href{https://doi.org/10.1016/0031-8914(74)90290-0}{\emph{Physica} {\bfseries
  75} (1974) 1}.

\bibitem{Kovtun:2012rj}
P.~Kovtun, \emph{{Lectures on hydrodynamic fluctuations in relativistic
  theories}}, \href{https://doi.org/10.1088/1751-8113/45/47/473001}{\emph{J.
  Phys. A} {\bfseries 45} (2012) 473001}
  [\href{https://arxiv.org/abs/1205.5040}{{\ttfamily 1205.5040}}].

\bibitem{Landau:1980mil}
L.~Landau and E.~Lifshitz, \emph{{Statistical Physics, Part 1}}, vol.~5 of
  \emph{Course of Theoretical Physics}. Butterworth-Heinemann, Oxford, 1980.

\bibitem{landauHydroFluctuations1957}
L.~Landau and E.~Lifshitz, \emph{Hydrodynamic fluctuations}, {\emph{Sov. Phys.
  JETP} {\bfseries 5} (1957) 512}.

\bibitem{Hohenberg:1977ym}
P.~Hohenberg and B.~Halperin, \emph{{Theory of Dynamic Critical Phenomena}},
  \href{https://doi.org/10.1103/RevModPhys.49.435}{\emph{Rev.\ Mod.\ Phys.}
  {\bfseries 49} (1977) 435}.

\bibitem{DeDominicis:1977fw}
C.~De~Dominicis and L.~Peliti, \emph{{Field Theory Renormalization and Critical
  Dynamics Above t(c): Helium, Antiferromagnets and Liquid Gas Systems}},
  \href{https://doi.org/10.1103/PhysRevB.18.353}{\emph{Phys. Rev. B} {\bfseries
  18} (1978) 353}.

\bibitem{Khalatnikov:1983ak}
I.~Khalatnikov, V.~Lebedev and A.~Sukhorukov, \emph{{Diagram Technique for
  Calculating Long Wave Fluctuation Effects}},
  \href{https://doi.org/10.1016/0375-9601(83)90716-8}{\emph{Phys. Lett. A}
  {\bfseries 94} (1983) 271}.

\bibitem{Dubovsky:2011sj}
S.~Dubovsky, L.~Hui, A.~Nicolis and D.~T. Son, \emph{{Effective field theory
  for hydrodynamics: thermodynamics, and the derivative expansion}},
  \href{https://doi.org/10.1103/PhysRevD.85.085029}{\emph{Phys.\ Rev.\ D}
  {\bfseries 85} (2012) 085029}
  [\href{https://arxiv.org/abs/1107.0731}{{\ttfamily 1107.0731}}].

\bibitem{Grozdanov:2013dba}
S.~Grozdanov and J.~Polonyi, \emph{{Viscosity and dissipative hydrodynamics
  from effective field theory}},
  \href{https://doi.org/10.1103/PhysRevD.91.105031}{\emph{Phys. Rev. D}
  {\bfseries 91} (2015) 105031}
  [\href{https://arxiv.org/abs/1305.3670}{{\ttfamily 1305.3670}}].

\bibitem{Crossley:2015evo}
M.~Crossley, P.~Glorioso and H.~Liu, \emph{{Effective field theory of
  dissipative fluids}},
  \href{https://doi.org/10.1007/JHEP09(2017)095}{\emph{JHEP} {\bfseries 09}
  (2017) 095} [\href{https://arxiv.org/abs/1511.03646}{{\ttfamily
  1511.03646}}].

\bibitem{Glorioso:2017fpd}
P.~Glorioso, M.~Crossley and H.~Liu, \emph{{Effective field theory of
  dissipative fluids (II): classical limit, dynamical KMS symmetry and entropy
  current}}, \href{https://doi.org/10.1007/JHEP09(2017)096}{\emph{JHEP}
  {\bfseries 09} (2017) 096}
  [\href{https://arxiv.org/abs/1701.07817}{{\ttfamily 1701.07817}}].

\bibitem{Glorioso:2016gsa}
P.~Glorioso and H.~Liu, \emph{{The second law of thermodynamics from symmetry
  and unitarity}},  \href{https://arxiv.org/abs/1612.07705}{{\ttfamily
  1612.07705}}.

\bibitem{Gao:2017bqf}
P.~Gao and H.~Liu, \emph{{Emergent Supersymmetry in Local Equilibrium
  Systems}}, \href{https://doi.org/10.1007/JHEP01(2018)040}{\emph{JHEP}
  {\bfseries 01} (2018) 040}
  [\href{https://arxiv.org/abs/1701.07445}{{\ttfamily 1701.07445}}].

\bibitem{Glorioso:2017lcn}
P.~Glorioso, H.~Liu and S.~Rajagopal, \emph{{Global Anomalies, Discrete
  Symmetries, and Hydrodynamic Effective Actions}},
  \href{https://doi.org/10.1007/JHEP01(2019)043}{\emph{JHEP} {\bfseries 01}
  (2019) 043} [\href{https://arxiv.org/abs/1710.03768}{{\ttfamily
  1710.03768}}].

\bibitem{Gao:2018bxz}
P.~Gao, P.~Glorioso and H.~Liu, \emph{{Ghostbusters: Unitarity and Causality of
  Non-equilibrium Effective Field Theories}},
  \href{https://doi.org/10.1007/JHEP03(2020)040}{\emph{JHEP} {\bfseries 03}
  (2020) 040} [\href{https://arxiv.org/abs/1803.10778}{{\ttfamily
  1803.10778}}].

\bibitem{Jensen:2017kzi}
K.~Jensen, N.~Pinzani-Fokeeva and A.~Yarom, \emph{{Dissipative hydrodynamics in
  superspace}}, \href{https://doi.org/10.1007/JHEP09(2018)127}{\emph{JHEP}
  {\bfseries 09} (2018) 127}
  [\href{https://arxiv.org/abs/1701.07436}{{\ttfamily 1701.07436}}].

\bibitem{Jensen:2018hse}
K.~Jensen, R.~Marjieh, N.~Pinzani-Fokeeva and A.~Yarom, \emph{{A panoply of
  Schwinger-Keldysh transport}},
  \href{https://doi.org/10.21468/SciPostPhys.5.5.053}{\emph{SciPost Phys.}
  {\bfseries 5} (2018) 053} [\href{https://arxiv.org/abs/1804.04654}{{\ttfamily
  1804.04654}}].

\bibitem{Haehl:2018lcu}
F.~M. Haehl, R.~Loganayagam and M.~Rangamani, \emph{{Effective Action for
  Relativistic Hydrodynamics: Fluctuations, Dissipation, and Entropy Inflow}},
  \href{https://doi.org/10.1007/JHEP10(2018)194}{\emph{JHEP} {\bfseries 10}
  (2018) 194} [\href{https://arxiv.org/abs/1803.11155}{{\ttfamily
  1803.11155}}].

\bibitem{Glorioso:2018wxw}
H.~Liu and P.~Glorioso, \emph{{Lectures on non-equilibrium effective field
  theories and fluctuating hydrodynamics}},
  \href{https://doi.org/10.22323/1.305.0008}{\emph{PoS} {\bfseries TASI2017}
  (2018) 008} [\href{https://arxiv.org/abs/1805.09331}{{\ttfamily
  1805.09331}}].

\bibitem{Wang:1998wg}
E.~Wang and U.~W. Heinz, \emph{{A Generalized fluctuation dissipation theorem
  for nonlinear response functions}},
  \href{https://doi.org/10.1103/PhysRevD.66.025008}{\emph{Phys.\ Rev.\ D}
  {\bfseries 66} (2002) 025008}
  [\href{https://arxiv.org/abs/hep-th/9809016}{{\ttfamily hep-th/9809016}}].

\bibitem{Chen-Lin:2018kfl}
X.~Chen-Lin, L.~V. Delacretaz and S.~A. Hartnoll, \emph{{Theory of diffusive
  fluctuations}},
  \href{https://doi.org/10.1103/PhysRevLett.122.091602}{\emph{Phys.\ Rev.\
  Lett.} {\bfseries 122} (2019) 091602}
  [\href{https://arxiv.org/abs/1811.12540}{{\ttfamily 1811.12540}}].

\bibitem{Glorioso:2020loc}
P.~Glorioso, L.~V. Delacretaz, X.~Chen, R.~M. Nandkishore and A.~Lucas,
  \emph{{Hydrodynamics in lattice models with continuous non-Abelian
  symmetries}},  \href{https://arxiv.org/abs/2007.13753}{{\ttfamily
  2007.13753}}.

\bibitem{Landry:2019iel}
M.~J. Landry, \emph{{The coset construction for non-equilibrium systems}},
  \href{https://doi.org/10.1007/JHEP07(2020)200}{\emph{JHEP} {\bfseries 07}
  (2020) 200} [\href{https://arxiv.org/abs/1912.12301}{{\ttfamily
  1912.12301}}].

\bibitem{Glorioso:2018kcp}
P.~Glorioso and D.~T. Son, \emph{{Effective field theory of
  magnetohydrodynamics from generalized global symmetries}},
  \href{https://arxiv.org/abs/1811.04879}{{\ttfamily 1811.04879}}.

\bibitem{Gralla:2018kif}
S.~E. Gralla and N.~Iqbal, \emph{{Effective Field Theory of Force-Free
  Electrodynamics}},
  \href{https://doi.org/10.1103/PhysRevD.99.105004}{\emph{Phys.\ Rev.\ D}
  {\bfseries 99} (2019) 105004}
  [\href{https://arxiv.org/abs/1811.07438}{{\ttfamily 1811.07438}}].

\bibitem{Cartan:1924yea}
E.~Cartan, \emph{{Sur les varietes a connexion affine et la theorie de la
  relativite generalisee. (premiere partie) (Suite).}}, {\emph{Annales Sci.
  Ecole Norm. Sup.} {\bfseries 41} (1924) 1}.

\bibitem{Cartan:1923zea}
E.~Cartan, \emph{{Sur les varietes a connexion affine et la theorie de la
  relativite generalisee. (premiere partie)}}, {\emph{Annales Sci. Ecole Norm.
  Sup.} {\bfseries 40} (1923) 325}.

\bibitem{Jensen:2014aia}
K.~Jensen, \emph{{On the coupling of Galilean-invariant field theories to
  curved spacetime}},
  \href{https://doi.org/10.21468/SciPostPhys.5.1.011}{\emph{SciPost Phys.}
  {\bfseries 5} (2018) 011} [\href{https://arxiv.org/abs/1408.6855}{{\ttfamily
  1408.6855}}].

\bibitem{Jensen:2014ama}
K.~Jensen, \emph{{Aspects of hot Galilean field theory}},
  \href{https://doi.org/10.1007/JHEP04(2015)123}{\emph{JHEP} {\bfseries 04}
  (2015) 123} [\href{https://arxiv.org/abs/1411.7024}{{\ttfamily 1411.7024}}].

\bibitem{Jensen:2014wha}
K.~Jensen and A.~Karch, \emph{{Revisiting non-relativistic limits}},
  \href{https://doi.org/10.1007/JHEP04(2015)155}{\emph{JHEP} {\bfseries 04}
  (2015) 155} [\href{https://arxiv.org/abs/1412.2738}{{\ttfamily 1412.2738}}].

\bibitem{Jensen:2014hqa}
K.~Jensen, \emph{{Anomalies for Galilean fields}},
  \href{https://doi.org/10.21468/SciPostPhys.5.1.005}{\emph{SciPost Phys.}
  {\bfseries 5} (2018) 005} [\href{https://arxiv.org/abs/1412.7750}{{\ttfamily
  1412.7750}}].

\bibitem{Festuccia:2016awg}
G.~Festuccia, D.~Hansen, J.~Hartong and N.~A. Obers, \emph{{Torsional
  Newton-Cartan Geometry from the Noether Procedure}},
  \href{https://doi.org/10.1103/PhysRevD.94.105023}{\emph{Phys.\ Rev.\ D}
  {\bfseries 94} (2016) 105023}
  [\href{https://arxiv.org/abs/1607.01926}{{\ttfamily 1607.01926}}].

\bibitem{Bergshoeff:2017dqq}
E.~Bergshoeff, A.~Chatzistavrakidis, L.~Romano and J.~Rosseel,
  \emph{{Newton-Cartan Gravity and Torsion}},
  \href{https://doi.org/10.1007/JHEP10(2017)194}{\emph{JHEP} {\bfseries 10}
  (2017) 194} [\href{https://arxiv.org/abs/1708.05414}{{\ttfamily
  1708.05414}}].

\bibitem{Bergshoeff:2014uea}
E.~A. Bergshoeff, J.~Hartong and J.~Rosseel, \emph{{Torsional Newton--Cartan
  geometry and the Schrodinger algebra}},
  \href{https://doi.org/10.1088/0264-9381/32/13/135017}{\emph{Class.\ Quant.\
  Grav.} {\bfseries 32} (2015) 135017}
  [\href{https://arxiv.org/abs/1409.5555}{{\ttfamily 1409.5555}}].

\bibitem{Christensen:2013rfa}
M.~H. Christensen, J.~Hartong, N.~A. Obers and B.~Rollier, \emph{{Boundary
  Stress-Energy Tensor and Newton-Cartan Geometry in Lifshitz Holography}},
  \href{https://doi.org/10.1007/JHEP01(2014)057}{\emph{JHEP} {\bfseries 01}
  (2014) 057} [\href{https://arxiv.org/abs/1311.6471}{{\ttfamily 1311.6471}}].

\bibitem{Christensen:2013lma}
M.~H. Christensen, J.~Hartong, N.~A. Obers and B.~Rollier, \emph{{Torsional
  Newton-Cartan Geometry and Lifshitz Holography}},
  \href{https://doi.org/10.1103/PhysRevD.89.061901}{\emph{Phys.\ Rev.\ D}
  {\bfseries 89} (2014) 061901}
  [\href{https://arxiv.org/abs/1311.4794}{{\ttfamily 1311.4794}}].

\bibitem{Hartong:2016nyx}
J.~Hartong, N.~A. Obers and M.~Sanchioni, \emph{{Lifshitz Hydrodynamics from
  Lifshitz Black Branes with Linear Momentum}},
  \href{https://doi.org/10.1007/JHEP10(2016)120}{\emph{JHEP} {\bfseries 10}
  (2016) 120} [\href{https://arxiv.org/abs/1606.09543}{{\ttfamily
  1606.09543}}].

\bibitem{Hartong:2014pma}
J.~Hartong, E.~Kiritsis and N.~A. Obers, \emph{{Schrodinger Invariance from
  Lifshitz Isometries in Holography and Field Theory}},
  \href{https://doi.org/10.1103/PhysRevD.92.066003}{\emph{Phys.\ Rev.\ D}
  {\bfseries 92} (2015) 066003}
  [\href{https://arxiv.org/abs/1409.1522}{{\ttfamily 1409.1522}}].

\bibitem{Geracie:2015xfa}
M.~Geracie, K.~Prabhu and M.~M. Roberts, \emph{{Fields and fluids on curved
  non-relativistic spacetimes}},
  \href{https://doi.org/10.1007/JHEP08(2015)042}{\emph{JHEP} {\bfseries 08}
  (2015) 042} [\href{https://arxiv.org/abs/1503.02680}{{\ttfamily
  1503.02680}}].

\bibitem{Geracie:2016dpu}
M.~Geracie, K.~Prabhu and M.~M. Roberts, \emph{{Physical stress, mass, and
  energy for non-relativistic matter}},
  \href{https://doi.org/10.1007/JHEP06(2017)089}{\emph{JHEP} {\bfseries 06}
  (2017) 089} [\href{https://arxiv.org/abs/1609.06729}{{\ttfamily
  1609.06729}}].

\bibitem{Duval:2009vt}
C.~Duval and P.~A. Horvathy, \emph{{Non-relativistic conformal symmetries and
  Newton-Cartan structures}},
  \href{https://doi.org/10.1088/1751-8113/42/46/465206}{\emph{J.\ Phys.\ A}
  {\bfseries 42} (2009) 465206}
  [\href{https://arxiv.org/abs/0904.0531}{{\ttfamily 0904.0531}}].

\bibitem{Banerjee:2017rch}
R.~Banerjee and P.~Mukherjee, \emph{{Milne boost from galilean gauge theory}},
  \href{https://doi.org/10.1016/j.physletb.2018.01.033}{\emph{Phys.\ Lett.\ B}
  {\bfseries 778} (2018) 303}
  [\href{https://arxiv.org/abs/1710.10882}{{\ttfamily 1710.10882}}].

\bibitem{Banerjee:2016laq}
R.~Banerjee and P.~Mukherjee, \emph{{Torsional Newton--Cartan geometry from
  Galilean gauge theory}},
  \href{https://doi.org/10.1088/0264-9381/33/22/225013}{\emph{Class. Quant.
  Grav.} {\bfseries 33} (2016) 225013}
  [\href{https://arxiv.org/abs/1604.06893}{{\ttfamily 1604.06893}}].

\bibitem{Banerjee:2015hra}
N.~Banerjee, S.~Dutta and A.~Jain, \emph{{Null Fluids - A New Viewpoint of
  Galilean Fluids}},
  \href{https://doi.org/10.1103/PhysRevD.93.105020}{\emph{Phys. Rev. D}
  {\bfseries 93} (2016) 105020}
  [\href{https://arxiv.org/abs/1509.04718}{{\ttfamily 1509.04718}}].

\bibitem{Maldacena:2008wh}
J.~Maldacena, D.~Martelli and Y.~Tachikawa, \emph{{Comments on string theory
  backgrounds with non-relativistic conformal symmetry}},
  \href{https://doi.org/10.1088/1126-6708/2008/10/072}{\emph{JHEP} {\bfseries
  10} (2008) 072} [\href{https://arxiv.org/abs/0807.1100}{{\ttfamily
  0807.1100}}].

\bibitem{Adams:2008wt}
A.~Adams, K.~Balasubramanian and J.~McGreevy, \emph{{Hot Spacetimes for Cold
  Atoms}}, \href{https://doi.org/10.1088/1126-6708/2008/11/059}{\emph{JHEP}
  {\bfseries 11} (2008) 059} [\href{https://arxiv.org/abs/0807.1111}{{\ttfamily
  0807.1111}}].

\bibitem{Herzog:2008wg}
C.~P. Herzog, M.~Rangamani and S.~F. Ross, \emph{{Heating up Galilean
  holography}},
  \href{https://doi.org/10.1088/1126-6708/2008/11/080}{\emph{JHEP} {\bfseries
  11} (2008) 080} [\href{https://arxiv.org/abs/0807.1099}{{\ttfamily
  0807.1099}}].

\bibitem{Duval:1984cj}
C.~Duval, G.~Burdet, H.~Kunzle and M.~Perrin, \emph{{Bargmann Structures and
  Newton-cartan Theory}},
  \href{https://doi.org/10.1103/PhysRevD.31.1841}{\emph{Phys.\ Rev.\ D}
  {\bfseries 31} (1985) 1841}.

\bibitem{Duval:1990hj}
C.~Duval, G.~W. Gibbons and P.~Horvathy, \emph{{Celestial mechanics, conformal
  structures and gravitational waves}},
  \href{https://doi.org/10.1103/PhysRevD.43.3907}{\emph{Phys.\ Rev.\ D}
  {\bfseries 43} (1991) 3907}
  [\href{https://arxiv.org/abs/hep-th/0512188}{{\ttfamily hep-th/0512188}}].

\bibitem{Julia:1994bs}
B.~Julia and H.~Nicolai, \emph{{Null Killing vector dimensional reduction and
  Galilean geometrodynamics}},
  \href{https://doi.org/10.1016/0550-3213(94)00584-2}{\emph{Nucl.\ Phys.\ B}
  {\bfseries 439} (1995) 291}
  [\href{https://arxiv.org/abs/hep-th/9412002}{{\ttfamily hep-th/9412002}}].

\bibitem{Jain:2015jla}
A.~Jain, \emph{{Galilean Anomalies and Their Effect on Hydrodynamics}},
  \href{https://doi.org/10.1103/PhysRevD.93.065007}{\emph{Phys. Rev. D}
  {\bfseries 93} (2016) 065007}
  [\href{https://arxiv.org/abs/1509.05777}{{\ttfamily 1509.05777}}].

\bibitem{Banerjee:2015uta}
N.~Banerjee, S.~Dutta and A.~Jain, \emph{{Equilibrium partition function for
  nonrelativistic fluids}},
  \href{https://doi.org/10.1103/PhysRevD.92.081701}{\emph{Phys. Rev. D}
  {\bfseries 92} (2015) 081701}
  [\href{https://arxiv.org/abs/1505.05677}{{\ttfamily 1505.05677}}].

\bibitem{Banerjee:2016qxf}
N.~Banerjee, S.~Dutta and A.~Jain, \emph{{First Order Galilean Superfluid
  Dynamics}}, \href{https://doi.org/10.1103/PhysRevD.96.065004}{\emph{Phys.
  Rev. D} {\bfseries 96} (2017) 065004}
  [\href{https://arxiv.org/abs/1612.01550}{{\ttfamily 1612.01550}}].

\bibitem{Banerjee:2017ouw}
N.~Banerjee, S.~Atul~Bhatkar and A.~Jain, \emph{{Second order Galilean fluids
  and Stokes' law}},
  \href{https://doi.org/10.1103/PhysRevD.97.096018}{\emph{Phys. Rev. D}
  {\bfseries 97} (2018) 096018}
  [\href{https://arxiv.org/abs/1711.09076}{{\ttfamily 1711.09076}}].

\bibitem{Jain:2018jxj}
A.~Jain, \emph{{A universal framework for hydrodynamics}}, Ph.D. thesis, Durham
  U., CPT, 6, 2018.

\bibitem{Hassaine:1999hn}
M.~Hassaine and P.~Horvathy, \emph{{Field dependent symmetries of a
  nonrelativistic fluid model}},
  \href{https://doi.org/10.1006/aphy.1999.6002}{\emph{Annals Phys.} {\bfseries
  282} (2000) 218} [\href{https://arxiv.org/abs/math-ph/9904022}{{\ttfamily
  math-ph/9904022}}].

\bibitem{Horvathy:2009kz}
P.~Horvathy and P.-M. Zhang, \emph{{Non-relativistic conformal symmetries in
  fluid mechanics}},
  \href{https://doi.org/10.1140/epjc/s10052-009-1221-x}{\emph{Eur.\ Phys.\ J.\
  C} {\bfseries 65} (2010) 607}
  [\href{https://arxiv.org/abs/0906.3594}{{\ttfamily 0906.3594}}].

\bibitem{Bellac:2011kqa}
M.~L. Bellac, \emph{{Thermal Field Theory}}, Cambridge Monographs on
  Mathematical Physics. Cambridge University Press, 3, 2011,
  \href{https://doi.org/10.1017/CBO9780511721700}{10.1017/CBO9780511721700}.

\bibitem{landau1959fluid}
L.~Landau and E.~Lifshitz, \emph{{Fluid Mechanics}}, Teoreticheskaia fizika.
  Pergamon Press, 1959.

\bibitem{Loganayagam:2011mu}
R.~Loganayagam, \emph{{Anomaly Induced Transport in Arbitrary Dimensions}},
  \href{https://arxiv.org/abs/1106.0277}{{\ttfamily 1106.0277}}.

\bibitem{Kovtun:2019hdm}
P.~Kovtun, \emph{{First-order relativistic hydrodynamics is stable}},
  \href{https://doi.org/10.1007/JHEP10(2019)034}{\emph{JHEP} {\bfseries 10}
  (2019) 034} [\href{https://arxiv.org/abs/1907.08191}{{\ttfamily
  1907.08191}}].

\bibitem{Banerjee:2012iz}
N.~Banerjee, J.~Bhattacharya, S.~Bhattacharyya, S.~Jain, S.~Minwalla and
  T.~Sharma, \emph{{Constraints on Fluid Dynamics from Equilibrium Partition
  Functions}}, \href{https://doi.org/10.1007/JHEP09(2012)046}{\emph{JHEP}
  {\bfseries 09} (2012) 046} [\href{https://arxiv.org/abs/1203.3544}{{\ttfamily
  1203.3544}}].

\bibitem{Jensen:2012jh}
K.~Jensen, M.~Kaminski, P.~Kovtun, R.~Meyer, A.~Ritz and A.~Yarom,
  \emph{{Towards hydrodynamics without an entropy current}},
  \href{https://doi.org/10.1103/PhysRevLett.109.101601}{\emph{Phys.\ Rev.\
  Lett.} {\bfseries 109} (2012) 101601}
  [\href{https://arxiv.org/abs/1203.3556}{{\ttfamily 1203.3556}}].

\bibitem{Arzt:1993gz}
C.~Arzt, \emph{{Reduced effective Lagrangians}},
  \href{https://doi.org/10.1016/0370-2693(94)01419-D}{\emph{Phys. Lett. B}
  {\bfseries 342} (1995) 189}
  [\href{https://arxiv.org/abs/hep-ph/9304230}{{\ttfamily hep-ph/9304230}}].

\bibitem{Bhattacharyya:2014bha}
S.~Bhattacharyya, \emph{{Entropy Current from Partition Function: One
  Example}}, \href{https://doi.org/10.1007/JHEP07(2014)139}{\emph{JHEP}
  {\bfseries 07} (2014) 139} [\href{https://arxiv.org/abs/1403.7639}{{\ttfamily
  1403.7639}}].

\bibitem{Bhattacharyya:2013lha}
S.~Bhattacharyya, \emph{{Entropy current and equilibrium partition function in
  fluid dynamics}}, \href{https://doi.org/10.1007/JHEP08(2014)165}{\emph{JHEP}
  {\bfseries 08} (2014) 165} [\href{https://arxiv.org/abs/1312.0220}{{\ttfamily
  1312.0220}}].

\bibitem{Armas:2019sbe}
J.~Armas and A.~Jain, \emph{{Viscoelastic hydrodynamics and holography}},
  \href{https://doi.org/10.1007/JHEP01(2020)126}{\emph{JHEP} {\bfseries 01}
  (2020) 126} [\href{https://arxiv.org/abs/1908.01175}{{\ttfamily
  1908.01175}}].

\bibitem{Armas:2020bmo}
J.~Armas and A.~Jain, \emph{{Hydrodynamics for charge density waves and their
  holographic duals}},
  \href{https://doi.org/10.1103/PhysRevD.101.121901}{\emph{Phys. Rev. D}
  {\bfseries 101} (2020) 121901}
  [\href{https://arxiv.org/abs/2001.07357}{{\ttfamily 2001.07357}}].

\bibitem{Grozdanov:2018ewh}
S.~s. Grozdanov and N.~Poovuttikul, \emph{{Generalized global symmetries in
  states with dynamical defects: The case of the transverse sound in field
  theory and holography}},
  \href{https://doi.org/10.1103/PhysRevD.97.106005}{\emph{Phys.\ Rev.\ D}
  {\bfseries 97} (2018) 106005}
  [\href{https://arxiv.org/abs/1801.03199}{{\ttfamily 1801.03199}}].

\bibitem{Grozdanov:2016tdf}
S.~s. Grozdanov, D.~M. Hofman and N.~Iqbal, \emph{{Generalized global
  symmetries and dissipative magnetohydrodynamics}},
  \href{https://doi.org/10.1103/PhysRevD.95.096003}{\emph{Phys.\ Rev.\ D}
  {\bfseries 95} (2017) 096003}
  [\href{https://arxiv.org/abs/1610.07392}{{\ttfamily 1610.07392}}].

\bibitem{Armas:2018ibg}
J.~Armas, J.~Gath, A.~Jain and A.~V. Pedersen, \emph{{Dissipative hydrodynamics
  with higher-form symmetry}},
  \href{https://doi.org/10.1007/JHEP05(2018)192}{\emph{JHEP} {\bfseries 05}
  (2018) 192} [\href{https://arxiv.org/abs/1803.00991}{{\ttfamily
  1803.00991}}].

\bibitem{Armas:2018zbe}
J.~Armas and A.~Jain, \emph{{One-form superfluids \& magnetohydrodynamics}},
  \href{https://doi.org/10.1007/JHEP01(2020)041}{\emph{JHEP} {\bfseries 01}
  (2020) 041} [\href{https://arxiv.org/abs/1811.04913}{{\ttfamily
  1811.04913}}].

\bibitem{Armas:2018atq}
J.~Armas and A.~Jain, \emph{{Magnetohydrodynamics as superfluidity}},
  \href{https://doi.org/10.1103/PhysRevLett.122.141603}{\emph{Phys. Rev. Lett.}
  {\bfseries 122} (2019) 141603}
  [\href{https://arxiv.org/abs/1808.01939}{{\ttfamily 1808.01939}}].

\bibitem{Haehl:2017zac}
F.~M. Haehl, R.~Loganayagam and M.~Rangamani, \emph{{Two roads to hydrodynamic
  effective actions: a comparison}},
  \href{https://arxiv.org/abs/1701.07896}{{\ttfamily 1701.07896}}.

\end{thebibliography}\endgroup
\end{document}